\UseRawInputEncoding
\documentclass[numberedappendix,appendixfloats]{emulateapj}

\usepackage{psfig}
\usepackage{amsmath}
\usepackage{amssymb}
\usepackage{graphicx}
\usepackage{appendix}
\usepackage{color}
\usepackage{multirow}
\usepackage[normalem]{ulem}

\newcommand{\lSect}[1]{{\label{sec:#1}}}
\newcommand{\lFig}[1]{{\label{fig:#1}}}
\newcommand{\lEq}[1]{{\label{eq:#1}}}
\newcommand{\lTab}[1]{{\label{tab:#1}}}
\def\gtaprx {\lower .1ex\hbox{\rlap{\raise .6ex\hbox{\hskip .3ex
	{\ifmmode{\scriptscriptstyle >}\else
		{$\scriptscriptstyle >$}\fi}}}
	\kern -.4ex{\ifmmode{\scriptscriptstyle \sim}\else
		{$\scriptscriptstyle\sim$}\fi}}}
\def\ltaprx {\lower .1ex\hbox{\rlap{\raise .6ex\hbox{\hskip .3ex
	{\ifmmode{\scriptscriptstyle <}\else
		{$\scriptscriptstyle <$}\fi}}}
	\kern -.4ex{\ifmmode{\scriptscriptstyle \sim}\else
		{$\scriptscriptstyle\sim$}\fi}}}
\newcommand{\FIGFF}[2]{{\ref{fig:#2}{#1}}}

\newcommand{\FIG}[2]{{Fig.~\FIGFF{#1}{#2}}}
\newcommand{\Fig}[1]{{\FIG{}{#1}}}

\newcommand{\Sectff}[1]{{\ref{sec:#1}}}
\newcommand{\Sect}[1]{{\S\Sectff{#1}}}

\newcommand{\Eqref}[1]{{\ref{eq:#1}}}
\newcommand{\Eqff}[1]{{(\Eqref{#1})}}

\newcommand{\Eq}[1]{{eq.~\Eqff{#1}}}

\newcommand{\Msun}{\ensuremath{\mathrm{M}_\odot}}

\newcommand{\Rsun}{\ensuremath{\mathrm{R}_\odot}}

\newcommand{\Tab}[1]{{Table~\ref{tab:#1}}}

\def\gtaprx {\lower .1ex\hbox{\rlap{\raise .6ex\hbox{\hskip .3ex
	{\ifmmode{\scriptscriptstyle >}\else
		{$\scriptscriptstyle >$}\fi}}}
	\kern -.4ex{\ifmmode{\scriptscriptstyle \sim}\else
		{$\scriptscriptstyle\sim$}\fi}}}
\def\ltaprx {\lower .1ex\hbox{\rlap{\raise .6ex\hbox{\hskip .3ex
	{\ifmmode{\scriptscriptstyle <}\else
		{$\scriptscriptstyle <$}\fi}}}
	\kern -.4ex{\ifmmode{\scriptscriptstyle \sim}\else
		{$\scriptscriptstyle\sim$}\fi}}}

\begin{document}


\title{Model Light Curves for Type Ib and Ic Supernovae}

\author{S.\ E.\ Woosley\altaffilmark{1}, Tuguldur
  Sukhbold\altaffilmark{2,3,4}, and D. N. Kasen\altaffilmark{5,6}}
\altaffiltext{1}{Department of Astronomy and Astrophysics, University
  of California, Santa Cruz, CA 95064}
\altaffiltext{2}{Department of Astronomy, Ohio State University,
  Columbus, Ohio, 43210}
\altaffiltext{3}{Center for Cosmology and AstroParticle Physics, Ohio
  State University, Columbus OH 43210}
\altaffiltext{4}{NASA Hubble Fellow}
\altaffiltext{5}{Department of Astronomy, University of California,
  Berkeley, CA 94720}
\altaffiltext{6}{Lawrence Berkeley National Laboratory, Berkeley, CA, 94720}

\begin{abstract} 
  Using the Monte Carlo code, SEDONA, multiband photometry and spectra
  are calculated for supernovae derived from stripped helium stars
  with presupernova masses from 2.2 to 10.0 \Msun. The models are
  representative of evolution in close binaries and have
    previously been exploded using a parametrized one-dimensional
    model for neutrino-transport. A subset, those with presupernova
  masses in the range 2.2 - 5.6 \Msun, have many properties in common
  with observed Type Ib and Ic supernovae, including a median ejected
  mass near 2 \Msun, explosion energies near $1 \times 10^{51}$ erg,
  typical $^{56}$Ni masses 0.07 - 0.09 \Msun, peak times of about 20
  days, and a narrow range for the $V$-$R$ color index 10 days post
  $V$-maximum near 0.3 mag. The median peak bolometric luminosity,
  near 10$^{42.3}$ erg s$^{-1}$, is fainter, however, than for several
  observational tabulations and the brightest explosion has a
  bolometric luminosity of only 10$^{42.50}$ erg s$^{-1}$. The
  brightest absolute $B$, $V$, and $R$ magnitudes at peak are $-17.2$,
  $-17.8$, and $-18.0$. These limits are fainter than some allegedly
  typical Type Ib and Ic supernovae and could reflect problems in our
  models or the observational analysis. Helium stars with lower and
  higher masses also produce interesting transients that may have been
  observed including fast, faint, blue transients and long, red, faint
  Type Ic supernovae.  New models are specifically presented for SN
  2007Y, SN 2007gr, SN 2009jf, LSQ13abf, SN 2008D, and SN 2010X.
 \end{abstract}

\section{Introduction}
\lSect{intro}

In a series of papers \citep{Woo19,Ert20,Woo20}, we have explored the
observational consequences of helium stars evolved with mass
loss. The intent of these papers has been to capture some of the
essential elements of massive stellar evolution in binary systems with
sufficiently close separations that the envelope of one or both stars
is lost to its companion when the star starts to become a red
supergiant. While the treatment of the binary interaction itself is
crude - the entire envelope is assumed to be lost at central helium
ignition - these studies have the advantage of including a broad range
of masses and following the evolution through the final stages of core
collapse and explosion in a realistic way.

Because of the limitations of the KEPLER code \citep{Wea78} and the
interesting upper bound it gave on supernova luminosities, it is
worthwhile to repeat the calculations of light curves in \citet{Ert20}
with a better treatment of the radiation transport. KEPLER does not
give reliable estimates of the photometry and spectra that are of
great interest to observers.  The present paper uses the SEDONA, an
implicit Monte-Carlo code that includes the effects of tens of
millions of lines, to compute the light curves and spectra. While the
version of SEDONA employed here also has limitations, such as assuming
a homologously coasting state for the ejecta and local thermodynamic
equilibrium (LTE), using it can make a better connection between the
explosion models and several recent compilations of observational
data.

The paper is observationally oriented. Bolometric luminosities are
calculated, but also absolute magnitudes for various filters, color
indices, light curve shapes, peak magnitude distributions, and
spectra. In six cases, these results are compared with specific
historic supernovae. In the end, the case is compelling that most
stripped envelope supernovae are best understood in a traditional
scenario of binary mass exchange, neutrino-powered explosions without
rotation, and radioactivity-illuminated light curves. There remains,
however, a puzzling minority of observed events that are too bright.
This could reflect deficiencies in the one-dimensional explosion
models, errors in estimating bolometric luminosities from its
components by the observers, errors in extinction corrections or
distances, or the existence of more than one mechanism operating to
explode and illuminate common stripped-envelope supernovae. We return
to these issues in the conclusions (\Sect{conclude}).

\section{Models and Codes}
\lSect{models}

\subsection{Explosion Models}

The presupernova evolution of mass-losing helium stars was studied by
\citet{Woo19} for star masses, at helium ignition, from 1.6
\Msun \ to 120 \Msun. This range corresponds to main sequence masses
from approximately 10 to 250 \Msun. For standard assumptions regarding
mass loss \citep[][with $f_{\rm WR} = 1$]{Yoo17}, helium stars with
initial masses below 2.5 \Msun \ were found to produce white dwarfs or
electron-capture supernova and their final evolution was not followed.
For the other stars, mass loss resulted in presupernova masses in the range
2 to 60 \Msun. The effect of a larger mass loss rate, closer to that
favored by Yoon, $f_{\rm WR}$ = 1.5, was also examined.

The explosion of a large subset of these models, those that did not
end up as as pulsational-pair instability supernovae or collapse
  directly to black holes, was simulated in a one-dimensional (1D)
neutrino-transport code, PHOTB, by \citet{Ert20}. The collapse
  to a black hole was determined by following the accretion onto the
  proto-neutron star long enough to assure that no outgoing shock
  emerged. In some cases not considered here, a black hole was formed
  by fall back even though the outer part of the star was ejected.
This reduced model set consisted of 133 helium stars with initial
masses from 2.5 to 40 \Msun, corresponding to presupernova masses
from 2.1 to 19.6 \Msun. The efficiency of the neutrino-powered
explosions was calibrated to SN 1987A and the Crab.  The ejected mass,
fall-back mass, final kinetic energy, and remnant mass were determined
for each star in a self-consistent way that depended upon the
progenitor structure. Considerable attention was paid to limits on the
amount of $^{56}$Ni each explosion produced, and lower and upper
bounds consistent with the (1D) neutrino-powered assumption and
nucleosynthesis were established. An additional set of models with
mass loss rates 1.5 times the standard Yoon values ($f_{\rm WR}$ =
1.5) was also considered.  More recently, \citet{Woo20} used the
remnant mass distributions calculated by \citet{Ert20} to estimate the
initial mass-function averaged birth function and mean masses for
black holes and neutron stars in close binary systems and found good
agreement with present observables.

\citet{Ert20} used a variety of central engines to simulate their
explosions. Their standard model was W18, but four other cases were
also considered that used different presupernova models for SN
  1987A. The average outcomes for these such as explosion energy, mass
  cut, and $^{56}$Ni synthesis did not vary greatly for the five
  different central engines. See Figs. 2 and 3 and Table 2 of
  \citet{Ert20} and Figs. 8, 9, 13, and 17 and Table 4 of
  \citet{Suk16}. Since we will find (\Sect{mix}; \Fig{kepmix}) that
bolometric light curves calculated using SEDONA and KEPLER agree quite
well, we will, with one exception, use only the models from
\citet{Ert20} with their W18 central engine for the present study. The
exception is an 8.00 \Msun \ helium star evolved with twice the
standard mass loss rate ($f_{\rm WR}$ = 2) and exploded manually
without the benefit of a neutrino-transport calculation. This was
necessary to examine the effects of the composition on the outcome of
a star with presupernova mass low enough to be common yet composed
mostly of carbon and oxygen and not helium \citep[see \Sect{colors}
  and][]{Yoo19}. The evolution of this star gave a presupernova mass
of 3.63 \Msun \ and an iron core of 1.48 \Msun, similar to a standard
model ($f_{\rm WR} = 1$) with initial mass 4.7 \Msun.  An explosion
with final kinetic energy $1.28 \times 10^{51}$ erg was generated
using a piston situated near the edge of the silicon core at 1.64
\Msun.

Except for this single model, the results of \citet{Ert20} for the
bolometric properties using the other central engines remain valid and
sufficient.  In addition to the standard model set with $f_{\rm WR} =
1$, we continued to carry the parallel set with $f_{\rm WR} = 1.5$
which was also exploded by \citet{Ert20}.

\begin{deluxetable}{cccccc}
\tablecaption{Properties of main models}
\tablehead{ \colhead{${M_{\rm He}}$}  &
            \colhead{${M_{\rm preSN}}$}  &
            \colhead{${\rm Ni+Tr}$}  &
            \colhead{$3/4 \times{M_{\rm NSE}}$}  &
            \colhead{${E_{\rm exp}}$}  &
            \colhead{${M_{\rm ej}}$}
            \\
            \colhead{[\Msun]}  &
            \colhead{[\Msun]}  &
            \colhead{[\Msun]}  &
            \colhead{[\Msun]}  &
            \colhead{[B]}      &
            \colhead{[\Msun]}
            }\\
\startdata
\multicolumn{6}{c}{$f_{\rm WR}=1$}\\
\multicolumn{6}{c}{}\\
  2.70   &   2.21   &   0.012   &   0.024   &   0.21   &   0.85 \\
  2.80   &   2.30   &   0.011   &   0.024   &   0.23   &   0.89 \\
  2.90   &   2.37   &   0.018   &   0.035   &   0.38   &   0.93 \\
  3.20   &   2.59   &   0.037   &   0.060   &   0.67   &   1.14 \\
  3.30   &   2.67   &   0.035   &   0.055   &   0.59   &   1.20 \\
  3.40   &   2.74   &   0.024   &   0.039   &   0.40   &   1.22 \\
  3.50   &   2.81   &   0.026   &   0.040   &   0.42   &   1.27 \\
  3.60   &   2.88   &   0.024   &   0.035   &   0.35   &   1.32 \\
  3.70   &   2.95   &   0.046   &   0.064   &   0.69   &   1.45 \\
  3.80   &   3.02   &   0.049   &   0.070   &   0.80   &   1.51 \\
  3.90   &   3.09   &   0.044   &   0.059   &   0.59   &   1.59 \\
  4.00   &   3.15   &   0.045   &   0.061   &   0.64   &   1.62 \\
  4.10   &   3.22   &   0.042   &   0.058   &   0.61   &   1.70 \\
  4.20   &   3.29   &   0.051   &   0.069   &   0.77   &   1.77 \\
  4.30   &   3.36   &   0.068   &   0.087   &   0.95   &   1.85 \\
  4.40   &   3.42   &   0.064   &   0.085   &   0.97   &   1.92 \\
  4.50   &   3.49   &   0.070   &   0.099   &   1.28   &   1.90 \\
  4.62   &   3.57   &   0.081   &   0.109   &   1.39   &   1.98 \\
  4.75   &   3.65   &   0.076   &   0.108   &   1.46   &   2.10 \\
  4.88   &   3.73   &   0.076   &   0.109   &   1.47   &   2.16 \\
  5.00   &   3.81   &   0.080   &   0.112   &   1.49   &   2.22 \\
  5.13   &   3.89   &   0.074   &   0.105   &   1.39   &   2.28 \\
  5.25   &   3.97   &   0.065   &   0.095   &   1.26   &   2.33 \\
  5.38   &   4.05   &   0.094   &   0.126   &   1.65   &   2.55 \\
  5.50   &   4.13   &   0.082   &   0.117   &   1.58   &   2.60 \\
  5.63   &   4.21   &   0.067   &   0.096   &   1.26   &   2.41 \\
  5.75   &   4.29   &   0.070   &   0.098   &   1.28   &   2.55 \\
  5.88   &   4.36   &   0.083   &   0.117   &   1.57   &   2.79 \\
  6.00   &   4.44   &   0.058   &   0.084   &   1.07   &   2.82 \\
  6.13   &   4.52   &   0.088   &   0.122   &   1.62   &   2.94 \\
  6.25   &   4.59   &   0.085   &   0.119   &   1.58   &   3.00 \\
  6.38   &   4.67   &   0.078   &   0.113   &   1.52   &   3.05 \\
  6.50   &   4.75   &   0.077   &   0.111   &   1.49   &   3.14 \\
  6.63   &   4.82   &   0.081   &   0.115   &   1.53   &   3.21 \\
  6.75   &   4.89   &   0.066   &   0.076   &   0.77   &   2.99 \\
  6.88   &   4.97   &   0.088   &   0.113   &   1.35   &   3.29 \\
  7.00   &   5.04   &   0.088   &   0.114   &   1.37   &   3.33 \\
  7.13   &   5.12   &   0.103   &   0.135   &   1.70   &   3.56 \\
  7.25   &   5.19   &   0.088   &   0.103   &   1.06   &   3.58 \\
  7.38   &   5.26   &   0.093   &   0.119   &   1.43   &   3.57 \\
  7.50   &   5.34   &   0.091   &   0.116   &   1.38   &   3.61 \\
  7.63   &   5.41   &   0.094   &   0.119   &   1.42   &   3.70 \\
  7.75   &   5.48   &   0.097   &   0.122   &   1.45   &   3.76 \\
  7.88   &   5.55   &   0.094   &   0.120   &   1.42   &   3.82 \\
  8.00   &   5.63   &   0.044   &   0.061   &   0.70   &   3.95 \\
\cline{1-6}\\
\multicolumn{6}{c}{$f_{\rm WR}=1.5$}\\
\multicolumn{6}{c}{}\\
5.00   &   3.43   &   0.094   &   0.124   &   1.60   &   1.85 \\
5.50   &   3.70   &   0.087   &   0.115   &   1.47   &   2.08 \\
6.00   &   3.96   &   0.087   &   0.118   &   1.56   &   2.32 \\
6.50   &   4.21   &   0.068   &   0.098   &   1.30   &   2.50 \\
7.00   &   4.45   &   0.077   &   0.109   &   1.45   &   2.84 \\
7.50   &   4.69   &   0.077   &   0.108   &   1.39   &   3.02 \\
8.00   &   4.92   &   0.076   &   0.110   &   1.47   &   3.26 \\
8.50   &   4.90   &   0.089   &   0.114   &   1.35   &   3.18 \\
9.00   &   4.87   &   0.052   &   0.078   &   1.01   &   3.14 \\
9.50   &   4.88   &   0.093   &   0.118   &   1.40   &   3.16 \\
10.0   &   4.96   &   0.084   &   0.106   &   1.24   &   3.21 \\
10.5   &   5.08   &   0.077   &   0.092   &   0.96   &   3.45 \\
11.0   &   5.19   &   0.090   &   0.105   &   1.13   &   3.44 \\
11.5   &   5.32   &   0.079   &   0.095   &   1.02   &   3.60 \\
12.0   &   5.43   &   0.076   &   0.092   &   0.99   &   3.71 \\
12.5   &   5.53   &   0.101   &   0.124   &   1.41   &   3.82 \\
13.0   &   5.64   &   0.094   &   0.115   &   1.30   &   3.98
\enddata
\tablecomments{Explosive properties are all based on \texttt{W18}
  engine \citep{Ert20}.}  \lTab{main}
\end{deluxetable}

The supernova models used are summarized in \Tab{main} and
\Tab{spcl}. Except for the 8.0 \Msun \ model mentioned above,
quantities in the tables were adopted without modification from
\citet{Ert20}. ``${\rm Ni+Tr}$'' and ``$3/4 \times{M_{\rm NSE}}$''
refer to a typical and a maximum $^{56}$Ni mass and their meanings are
discussed extensively in that publication. Here, in keeping with the
goal of producing the brightest possible supernovae this model set
allows, the ``$3/4 \times{M_{\rm NSE}}$'' value is adopted here as
standard.  These values can reasonably be reduced in any light curve
simulation, but not increased.  $M_{\rm presn}$ and $M_{\rm ej}$ are
the masses of the presupernova star and the ejecta, after any
fallback is over, and $E_{\rm exp}$ is its terminal kinetic
energy. For the exceptional 8.00 \Msun \ model that was manually
exploded, a $^{56}$Ni mass of 0.099 \Msun \ was enforced by adjusting
the mass cut. The reason for this will become apparent in
\Sect{colors} when the colors of this model are compared with
another (4.50 \Msun; $f_{\rm WR}$ = 1) model with this mass
of $^{56}$Ni.

The standard models that used $f_{\rm WR}$ = 1 are labeled by their
initial helium star mass and the prefix ``He''.  Model He5.00 thus
began its computational lifetime as a 5 \Msun \ helium star, lost mass
using $f_{\rm WR} =1$, and exploded with a final mass of 3.81 \Msun
\ using the W18 central engine. The other explosion models in
\Tab{main} that used $f_{\rm WR} = 1.5$ will have an ``x1.5'' attached
to their name. Model He5.00x1.5 thus also began as a 5 \Msun \ helium
star, but ended with a mass of 3.43 \Msun \ and was exploded using the
W18 central engine. All of these models in \Tab{main} are candidates
for normal Type Ib and Ic supernovae (\Sect{sn1bc}).  The files are
available to those wanting to use these explosion models for
their own studies.

An additional 4 low and high mass models using the same explosion and
mass-loss assumptions, but not considered ``normal'' are given in
\Tab{spcl}. Model He8.00x2 had an initial mass of 8.00 \Msun \ but
twice the standard mass loss rate.

\begin{deluxetable}{cccccc}
\tablecaption{Properties of special case models}
\tablehead{ \colhead{${M_{\rm He}}$}  &
            \colhead{${M_{\rm preSN}}$}  &
            \colhead{${\rm Ni+Tr}$}  &
            \colhead{$3/4 \times{M_{\rm NSE}}$}  &
            \colhead{${E_{\rm exp}}$}  &
            \colhead{${M_{\rm ej}}$}
            \\
            \colhead{[\Msun]}  &
            \colhead{[\Msun]}  &
            \colhead{[\Msun]}  &
            \colhead{[\Msun]}  &
            \colhead{[B]}      &
            \colhead{[\Msun]}
            }\\
\startdata

\multicolumn{2}{c}{$f_{\rm WR}=2.0$} & \multicolumn{4}{c}{artificial explosion}\\
\multicolumn{6}{c}{}\\
8.00x2   &   3.63   &   -   &   0.099   &   1.28   &   1.99 \\
\\
\multicolumn{2}{c}{$f_{\rm WR}=1.0$} & \multicolumn{4}{c}{${\rm engine}=\texttt{W18}$}\\
\multicolumn{6}{c}{}\\
2.50   &   2.07   &   0.007   &   0.015   &   0.11   &   0.74 \\
2.60   &   2.15   &   0.008   &   0.018   &   0.15   &   0.79 \\
12.0   &   7.24   &   0.071   &   0.079   &   0.81   &   5.33 \\
19.0   &   9.98   &   0.093   &   0.102   &   1.00   &   8.32 \\
\multicolumn{6}{c}{}
\enddata
\lTab{spcl}
\end{deluxetable}

\subsection{Use of the SEDONA Code}
\lSect{sedona}

SEDONA \citep{Kas06a,Kas08,Rot15} is a multi-dimensional implicit
Monte Carlo code, used here in one dimension. Given the ejecta
properties i.e., the density, composition, and velocity profile of the
freely expanding supernova material, SEDONA does a detailed treatment
of the gamma-ray transport and self-consistently determines the
emergent broadband light curves and spectra. The temperature structure
is solved assuming radiative equilibrium. Lines are treated in the
expansion opacity formalism and assumed to be purely
absorbing. Otherwise no free parameters need be adjusted in the
radiative transfer calculation. It is assumed that the excited states
of the various ions are in thermal equilibrium. Opacities include the
effects of over 20 million iron-group lines and about 800,000 lines of
hydrogen through calcium
\citep[http://kurucz.harvard.edu/vitabib.html;][]{Kur94,Kur95,Kur09}. The
ions included here were He, C, N, O, Ne, Mg, Si, Ar, Ca, Ti, Cr, Fe,
Co, and Ni. The only energy source was the decay of $^{56}$Ni and
$^{56}$Co. Each ion except helium carried 7 stages of ionization and
from 300 to 2000 levels. The initial thermal energy from the
  KEPLER explosion model was included.

To make a link with the explosion simulations, KEPLER models were
evolved to a coasting configuration, typically 1 day, but 2 days in
some of the more massive models with slower expansion speeds, and
mapped into SEDONA. The original zoning was preserved except that few
bottom zones with coasting speeds less than 500 km s$^{-1}$ and
surface zones moving faster than 30,000 km s$^{-1}$ were excised. The
$^{56}$Ni mass and velocity of the remaining zones was then adjusted
slightly so as to preserve the final total $^{56}$Ni and kinetic
energy given by \citet[][see also \Tab{main} and
  \Tab{spcl}]{Ert20}. In a few cases where sensitivity to $^{56}$Ni
mass was studied, the $^{56}$Ni mass was multiplied by a constant and
the other mass fractions renormalized so as to keep the sum of the
mass fractions equal to one. Typical spatial resolution varied from
200 zones for the models with small ejected mass to 1220 zones for the
heaviest model. The models were then mixed as discussed in \Sect{mix}.

Most runs were initialized with 1 million particles and added 140,000
particles per time step. Particles that escaped were allowed to leave
the grid. Typical runs took about 350 time steps and terminated after
100 days. Finer time steps were used early in the calculation. The
typical time step near maximum light was 0.3 days. The typical maximum
number of particles on the grid at that time was 15 million. Several
runs were carried out with four times the number of particles in order
to obtain well converged spectra. These showed that the broad-band
light curves were well converged for the smaller value.  Calculations
were run in the LUX supercomputer at UCSC using 4 to 8 nodes (160 to
320 CPU). Run times were shorter for the lighter models with higher
velocity and fewer zones, but ranged from several hours to 15 hours.

The principal output of SEDONA consists of a time series of spectra
from the far UV to infrared (10$^{14}$ to $2 \times 10^{16}$ Hz; 100
to 20,000 \AA). In order to generate multiband light curves, these
spectra were then folded with different filter response functions
(\Sect{colors}).

\subsection{Mixing}
\lSect{mix}

One of the greatest uncertainties in any modeling of the multi-color
light curves or spectra of Type Ib and Ic supernovae is the extent to
which the ejecta, especially $^{56}$Ni and $^4$He, have been mixed.
See \citet{Des12}, \citet{Des15}, and \citet{Yoo19} for
discussions. The mixing of iron-group elements to large velocity
provides a source of opacity that blankets the blue and ultraviolet
emission making the supernova redder at peak. The decay of $^{56}$Ni
provides heat that maintains ionization to a larger radius in a mixed
model hence supporting a larger photosphere and cooler effective
temperature. The early color evolution of the supernova is thus
especially sensitive to mixing \citet{Yoo19}. If $^{56}$Ni and $^4$He
are mixed together, the decay of the former can aid in exciting lines
of the latter \citep{Luc91,Des12,Woo97}, dramatically affecting the
spectrum and even the classification of the supernova as Type Ib or
Ic. Above some mass, the small residual helium may be ejected with
such high velocity that, depending on mixing, no $^{56}$Ni and $^4$He
commingle. Hence no strong, non-thermal lines of helium will be
present in the spectrum and it will be Type Ic. But the supernova can
also be Type Ic if there is a thick buffer of carbon, oxygen, and
heavier elements between the $^{56}$Ni and even a thick shell of
helium. When $^{56}$Ni is mixed to larger radii, the light curve rises
earlier and may decline earlier. The peak bolometric luminosity is
thus affected, though the change is generally not large.

Despite its importance, mixing in Type Ib and Ic supernova has not
received nearly as much attention in multi-dimensional simulation as
Type IIp. Focus in the latter has frequently been on the
Rayleigh-Taylor instability created as the outgoing shock decelerates
in regions of increasing $\rho r^3$, where $\rho$ is the local density
and $r$ the shock radius. Major mixing occurs when the helium core
encounters the hydrogen envelope and is forced to decelerate. This
kind of mixing is present in exploding helium stars as well, but
lacking a hydrogen envelope, the consequences are not as
dramatic.

A second kind of mixing occurs in all varieties of core-collapse
supernovae and is driven by the essential asymmetry of the central
engine. In neutrino-powered explosions, the ``hot bubble'' is unstable
and expands at different rates for different angles. See for example,
Figs 10 and 11 of \citet{Won17}. If a magnetar powers the explosion,
similar anisotropies will exist \citep{Che17}. For explosions that
impart their energy in a time short compared with the shock crossing
time for the presupernova star, about a minute here, the final
velocity of the ejected plumes depends on their initial speed and how
much matter they interact with on the way out. There is usually an
upper bound to their speed \citep{Won17}.

As is often noted, this mixing is not microscopic. It does not lead to
the homogenization of the composition. Rather material is ejected in
plumes and clumps that extend to higher final velocity than some of
the other ejecta that initially was at smaller radius. Mixing is thus
intrinsically a multi-dimensional process that cannot be properly
replicated in 1D. For a clear demonstration in nature, see the
supernova remnant for Cas-A and its well studied ``fast moving
knots''. Whatever is done here in 1D will be a gross approximation.

Many artificial prescriptions exist for mixing in one-dimensional
simulations of supernovae. Most have, at their root, early attempts to
model mixing in SN 1987A, and involve moving a ``boxcar average''
through the supernova ejecta multiple times. A boxcar with an interval of
specified mass, is moved outwards through the supernova model, zone by
zone until either some maximum mass or the surface of the ejecta is
reached. At each zone the overlying interval is completely mixed while
conserving mass. The mixing operation is often performed through the
supernova several times with a decreasing mixing interval so that the
final curve for abundance as a function of mass is smooth.  In our
previous studies \citep[e.g.][]{Ert20}, an initial boxcar width of
0.15 $M_{\rm ej}$ was employed , where $M_{\rm ej}$ is the mass
ejected in the supernova (presupernova mass minus remnant mass).  This
mixing region was moved through all the ejecta three times and then a
final fourth mixing was applied using an interval half as great, i.e.,
0.075 $M_{\rm ej}$. Importantly, this mixing was continued to the
surface of the star, so some $^{56}$Ni was mixed to arbitrarily high
velocity.

A similar prescription was employed by \citet{Des15}, but instead of
mixing in mass, they mixed in velocity intervals. Specifically, for
the study of Type Ib and Ic supernovae, they recommended a 1000 km
s$^{-1}$ mixing interval as characteristic of weak mixing and 2000 km
s$^{-1}$ as representative of strong mixing and explored the
consequences. The Dessart et al mixing and Ertl et al mixing are shown
for Model He6.00, which ejected 2.32 \Msun \ (see \Tab{main}) in
\Fig{mix}. Note the tail extending to very high velocity for the
Dessart major mixing case and the Ertl et al prescription. Since the
abundance of $^{56}$Ni in these high velocity zones exceeds the
abundance of solar iron in the initial model, 0.0014, this will have a
major effect on the spectrum.

\begin{figure}
\includegraphics[width=0.48\textwidth]{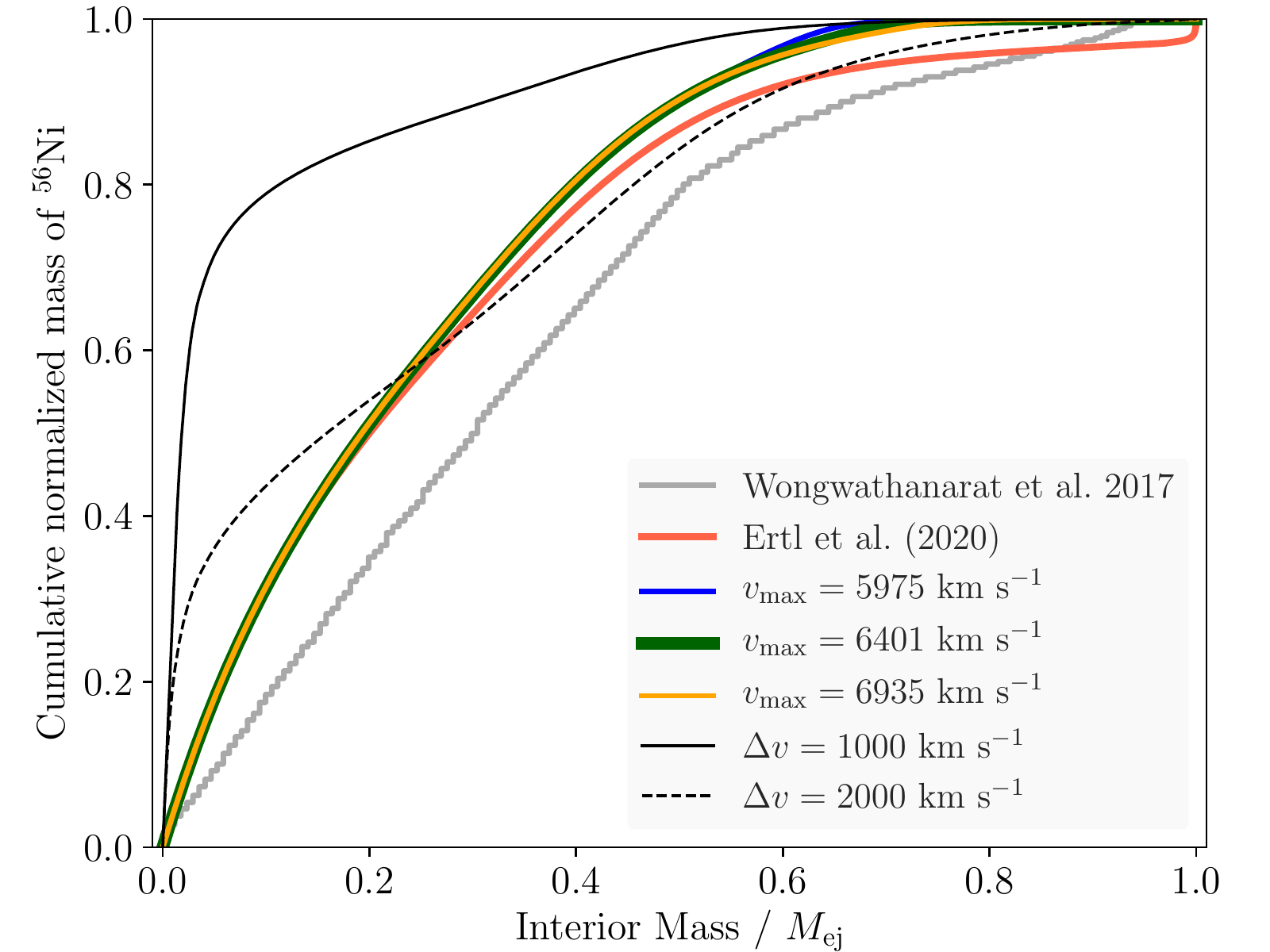}
\includegraphics[width=0.48\textwidth]{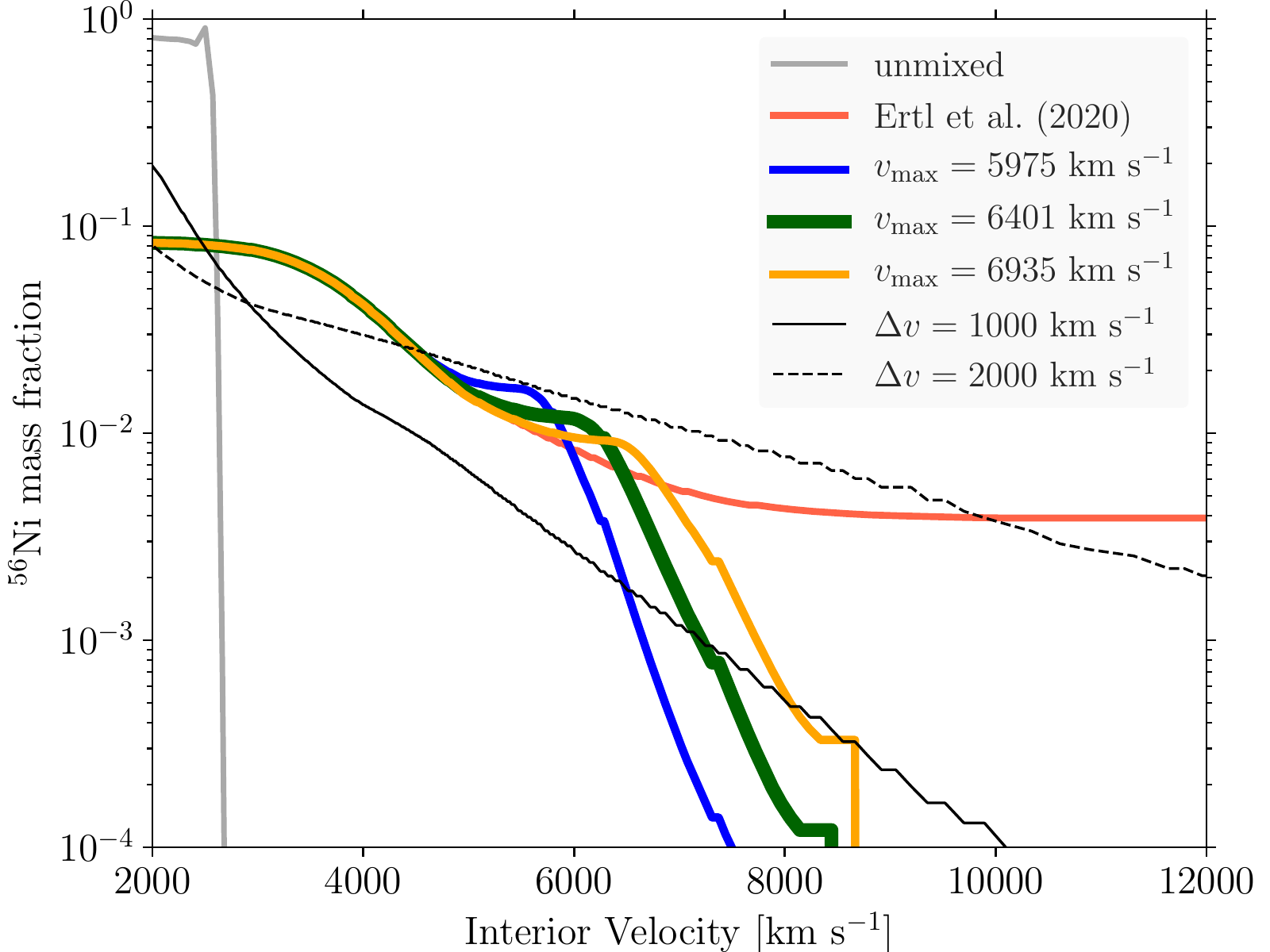}
\caption{Mixing in Model He6.00. The top panel shows several possible
  prescriptions for the mixing of $^{56}$Ni in mass compared with a
  multi-dimensional simulation of a Type IIb supernova with a similar helium
  core mass by \citet[][solid gray curve]{Won17}. Three of the
  prescriptions involve mixing to a maximum velocity using a moving
  boxcar method based on mass coordinate (blue, green, orange). Two others 
  (solid and dashed black) invoke a similar
  mixing in velocity space as suggested by \citet{Des15}. The solid
  red line is the mixing used in the previous KEPLER study of light
  curves \citep{Ert20}. The thick dark green line labeled 6400 km
  s$^{-1}$ is the standard mixing used in this paper. $B$ and especially
  $U$ magnitudes are very sensitive to the extent to which $^{56}$Ni is
  mixed to high velocity. The bottom panel shows this same mixing
  plotted as a function of terminal velocity. \lFig{mix}}
\end{figure}

While we could find no published three-dimensional studies of
post-explosive mixing in completely stripped massive stars exploded
with neutrino transport, the results should resemble closely what
\citet{Won17} calculated for a Type IIb supernova with a very low mass
hydrogenic envelope. Their initial model had a helium and heavy
element core of 4.4 \Msun\ capped by 0.3 \Msun\ of low density
hydrogen envelope. The ejected mass was 3.0 \Msun\ of core material
plus the 0.3 \Msun\ of envelope, and the explosion energy was $1.47
\times 10^{51}$ erg. Though intended as a model for Cas-A, the helium
core mass and structure of their Model W15-2-cw-IIb is very similar to
Model He6.0 here (\Tab{main}).  \citet{Won17} found that mixing in
their ejecta was characterized by a maximum speed for $^{56}$Ni of
about 7000 km s$^{-1}$ with only a small amount, about 1\%, being
mixed out to 7000 - 9000 km s$^{-1}$. They point out that this is
consistent with the maximum speed seen for $^{44}$Ti in Cas-A of 6300
$\pm$ 1250 km s$^{-1}$ (Grefenstette et al. 2017). They also point out
that this speed is consistent with dimensional arguments for the bulk
speed $(2 E/M_{\rm ej})^{1/2} \approx 7000$ km s$^{-1}$. \citet{Des16}
say that this bulk speed is characteristic of the Doppler velocity
measured for He I (5875 \AA) at maximum in supernovae of Type IIb and
Ib. It thus seems quite doubtful that substantial $^{56}$Ni will be
mixed to speeds in excess of 10,000 km s$^{-1}$ in models like
He6.00. Dessart's high mixing and Ertl et al. standard mixing are thus
unrealistic at high speeds where the ejected material has been
accelerated by a shock going down a density gradient.

With admittedly large residual uncertainty, we adopt a mixing
formalism where the bulk of the $^{56}$Ni and other heavy elements is
mixed out to a speed given by the bulk velocity. That is the mixing is
characterized by a maximum speed of
\begin{equation}
v_{\rm mix} \ = \ 6000 \left(\frac{E_{51}}{M_{\rm ej}/3 \Msun}\right)^{1/2} \ {\rm km\ s^{-1}}
\lEq{vmix}
\end{equation}
where $E_{51}$ is the kinetic energy of the ejecta in units of
10$^{51}$ erg.  To account for a small amount of material at greater
speeds, a small amount of matter is mixed out to a value 30\% greater.
This formula would not describe mixing in a red or blue supergiant and is
only for stripped helium stars.

\begin{figure}
\includegraphics[width=0.48\textwidth]{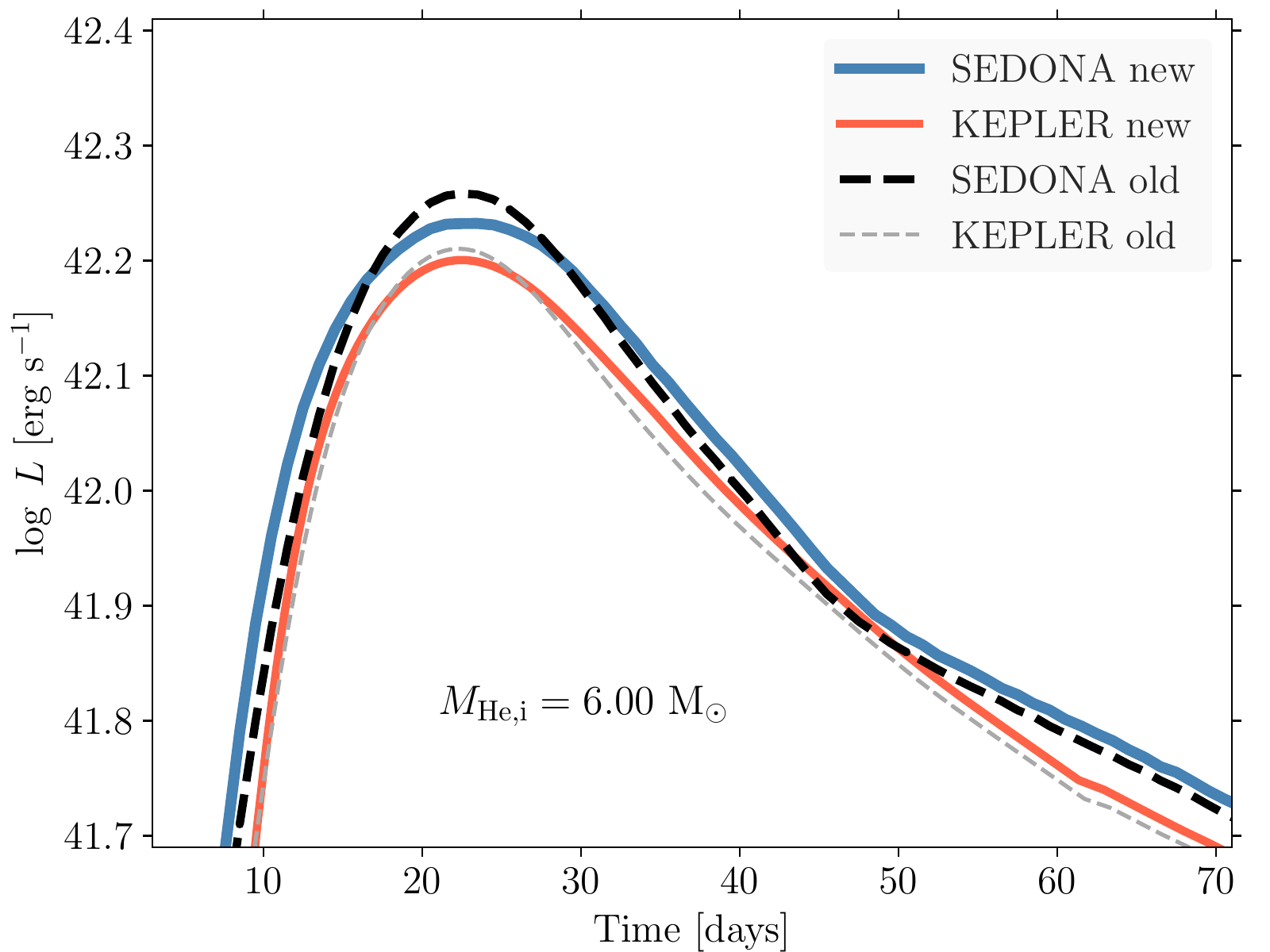}
\caption{The effect of two mixing prescriptions on the light curve is
  shown for Model He6.00 calculated with two codes, KEPLER and SEDONA.
  The curves labeled ``old'' use the mixing defined in \citet{Ert20},
  including the unrealistic mixing of $^{56}$Ni to arbitrarily high
  velocities. The ``new'' mixing is described in the text and
  \Fig{mix}. In both cases, the KEPLER and SEDONA calculations are in
  good agreement for the bolometric light curve. The difference at
  peak is about 9\%. \lFig{kepmix}}
\end{figure}

This mixing is accomplished by the usual boxcar averaging
technique.  Two zones, $n_1$ and $n_2$ are determined that have
velocities equal to $v_{\rm mix}$ and 1.3 $v_{\rm mix}$. Two mixing
masses (boxcar sizes) are taken to be $m_1 = 0.1 M_{\rm ej}$ and $m_2
= 0.02 M_{\rm ej}$. The boxcar $m_1$ is passed 7 times from the center
of the star to zone $n_1$ and then the boxcar $m_2$ is passed 4 times from
the center of the star to zone $n_2$. The results for three different
choices for $v_{\rm mix}$ are shown for Model He6.00 in \Fig{mix}. The 
choice of 6400 km s$^{-1}$ corresponds to the characteristic value in
\Eq{vmix}.

Calculations using SEDONA show that neither the bolometric luminosity
nor $V$ magnitude at peak is very sensitive to the mixing
prescription, but the $B$ and especially $U$ magnitudes are. Were the
mixing assumed in \citet{Ert20} to be employed here, the colors of the
supernova at peak, as characterized by ($B-V$), would be too red
compared with observations.

\section{Normal Type Ib and Ic Supernovae}
\lSect{sn1bc}

It is first necessary to define just which observations and models
will be considered ``normal'' Type Ib and Ic supernovae. We exclude
any observed events displaying gross asymmetry, broad lines, or
extreme energy (much greater than $2 \times 10^{51}$ erg). Our
observational sample will thus not include broad-lined Type Ic
supernovae or gamma-ray burst supernovae.  Present modeling does not
support the hypothesis that these are non-rotating, neutrino-powered,
radioactivity-illuminated explosions. The stars that uniformly
collapse to black holes in our survey or pulsational pair-instability
form a gross upper boundary mass to normal Type Ib and Ic
supernovae. Stars that experience strong silicon flashes or produce
very little $^{56}$Ni form a lower one. Such stars exist, are
abundant, and probably explode, but the early evolution of the typical
Ib light curve excludes their inclusion in the ``normal'' sample. The
width of the light curve also excludes additional models on the upper
end. These cases are thus discussed separately in \Sect{fastblue} and
\Sect{heavy}.

Applying these criteria, \citet{Ert20} claimed that normal Type Ib and
Ic supernovae resulted from presupernova masses 2.7 to 5.6 \Msun. For
the standard mass loss rates this implied initial helium star masses
from 3.3 to 8.0 \Msun \ (\Tab{main}). For $f_{\rm WR}$ = 1.5, the
corresponding initial helium star mass range increased to 3.7 to 13
\Msun.  This larger mass loss rate was actually preferred by
\citet{Yoo17} and \citet{Woo20}. We will find later (\Sect{colors})
that the results, especially near peak, are most sensitive to the
presupernova mass, not the initial mass and mass loss that was used to
get there, so these models with higher mass loss rates are included in
our analysis, but not emphasized.

Here, in order to include what could potentially be a large number of
lower mass explosions for which the silicon flash is not very strong,
we extend the definition of ordinary Type Ib and Ic supernova to
include smaller presupernova masses down to 2.2 \Msun. For standard
mass loss rates, this means we include initial helium core masses as
low as 2.7 \Msun.  The new lower bound still excludes a few stars with
still smaller masses that have substantial radius expansion even
without a strong silicon flash \citep{Woo19} and make little $^{56}$Ni,
and models with such a strong flash that the light curve would be
dominated by circumstellar shock interaction. The larger set thus
continues to exclude Models He3.00 and He3.10 for which we have no
explosion calculation from \citet{Ert20} that did not also experience
a strong silicon flash. They also exclude He2.50 and He2.60 which made
very little ${56}$Ni (less than 0.02 \Msun).

The flash in Model He3.20 was weak enough that the resulting light
curve was dominated by radioactivity and not circumstellar
interaction, so it is included. \Tab{main} lists the full range of
models treated as normal Type Ib and Ic in the new survey.

When averages are to be calculated, two ranges will be considered: a
high one, 3.3 - 8 \Msun \ consistent with \citet{Ert20}, and a new low
one that we now favor, 2.7 - 8.0 \Msun \ (excluding 3.1 and 3.2),
which includes a substantial fraction of potentially observable low
mass events. Obviously the inclusion of lower mass models will
decrease the averages for $^{56}$Ni produced, kinetic energy,
luminosity, and ejected mass. 

Except for Model He8.00x2, no new explosion models have been
calculated, so all previous conclusions that depended only on
stellar evolution and hydrodynamics and not on detailed radiation
transport carry over from \citet{Ert20}. That includes ejected masses,
compositions, explosion kinetic energies, and $^{56}$Ni production for
individual supernovae (\Tab{main}). These quantities are summarized
for other central engines in Table 5 and Fig. 24 of \citet{Ert20}. The
greatest $^{56}$Ni mass produced for the W18 parametrization was 0.135
\Msun \ (as upper limit, $3/4\times M_{\rm NSE}$) in Model He7.13,
which also had the greatest explosion energy, $1.70 \times 10^{51}$
erg. (\Tab{main}). When the more energetic S19.6 central engine was
employed, the maximum $^{56}$Ni production and explosion energy rose
only slightly to 0.148 \Msun \ and $1.94 \times 10^{51}$ erg (also for
the 7.13 \Msun \ model).

\begin{figure}
\includegraphics[width=0.48\textwidth]{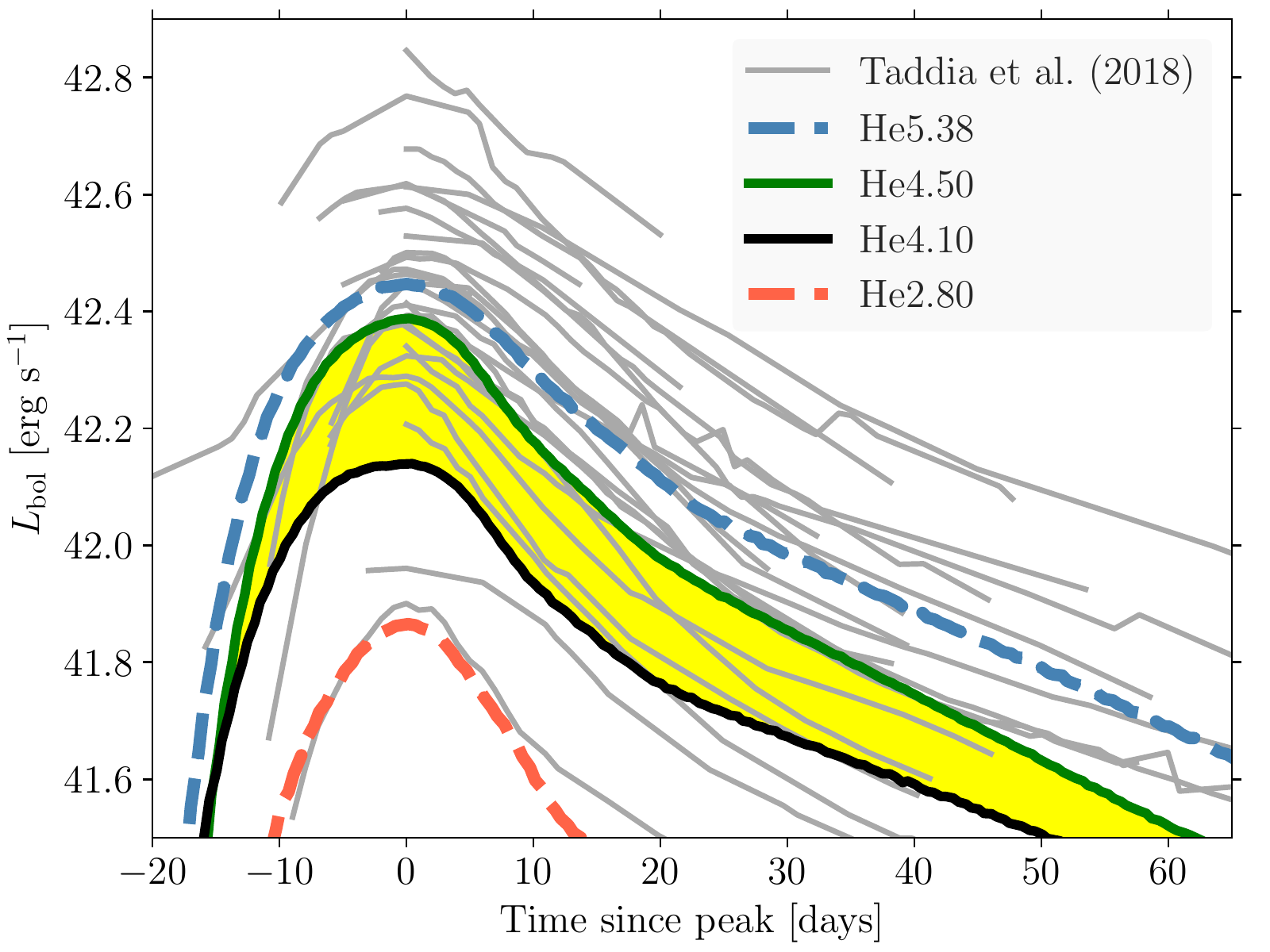}
\caption{Bolometric light curves for the Carnegie sample of 32 Type
  Ib, Ic and IIb events \citep{Tad18} compared with models. The dark
  black and green lines are Models He4.10 and He4.50 respectively. The
  yellow band bounded by these two models roughly represent the
  typical range of the survey. The red and blue dashed lines are our
  faintest (He2.80) and brightest (He5.38) ones. A substantial set of
  observed normal supernovae are brighter than the most luminous
  model. Fainter supernova pose a lesser problem since the $^{56}$Ni
  masses used here are upper bounds
  (\Tab{main};\Sect{LSQ13}). \lFig{taddia}}
\end{figure}

\begin{figure}
\includegraphics[width=0.48\textwidth]{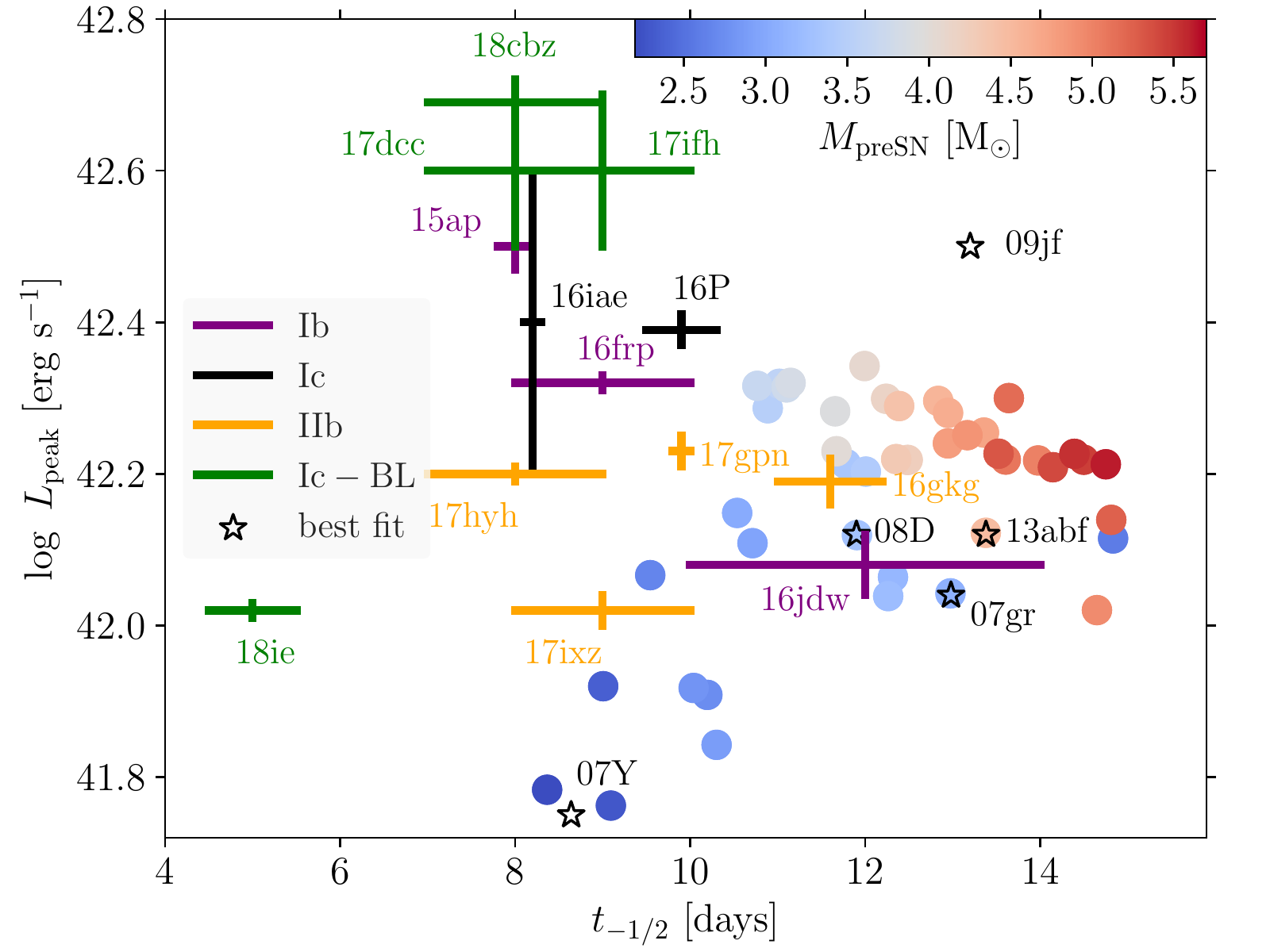}
\includegraphics[width=0.48\textwidth]{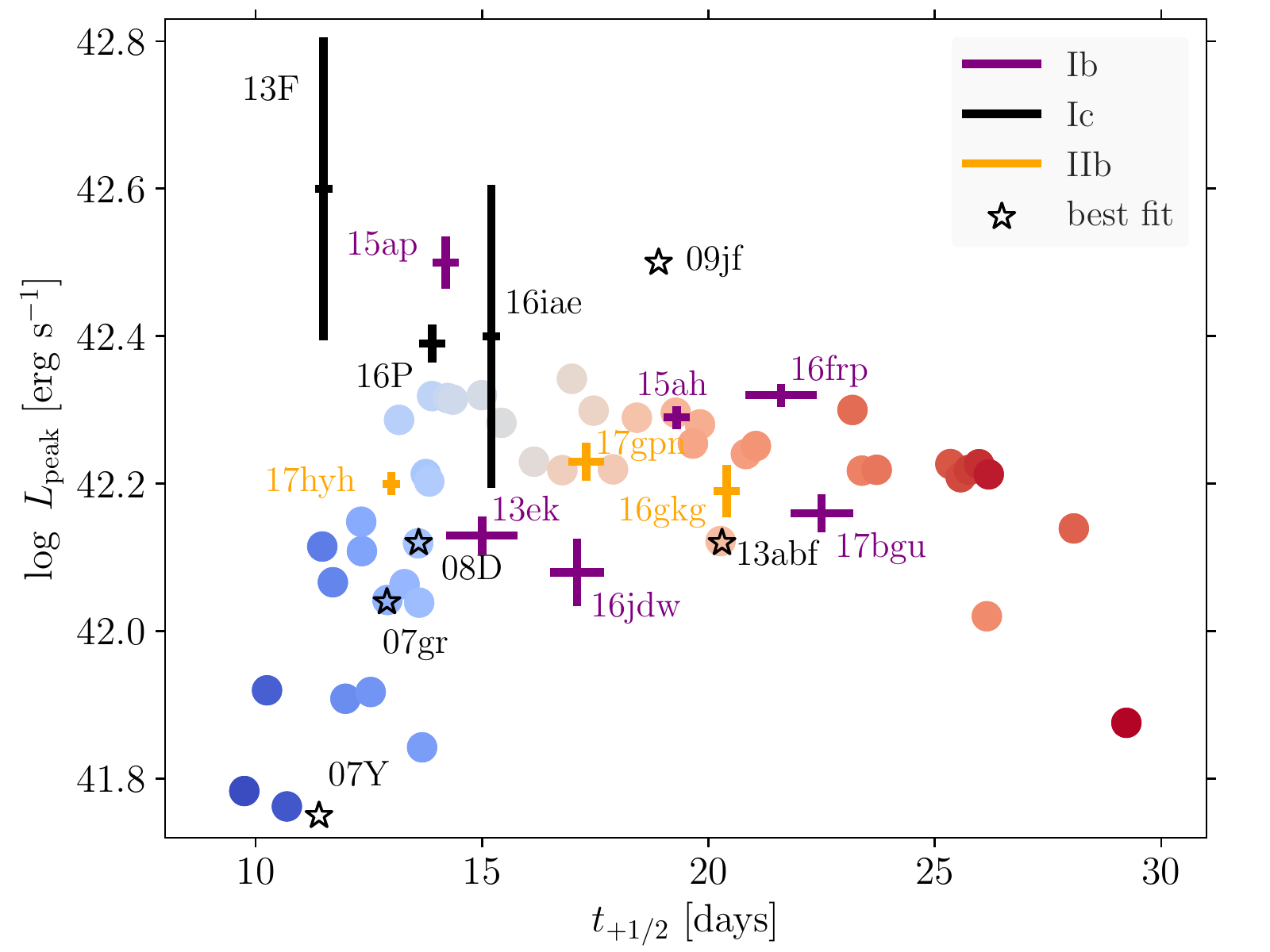}
\caption{Peak luminosity in the 4000--10000 \AA\ range vs time to rise
  to peak (top) or time to decline (bottom) for a baseline of 50\%
  $L_{\rm peak}$.  Filled circles are model results color coded
    by their presupernova mass while points with error bars are observations
  \citep[Table 3 of][]{Pre19}.  Five points indicated with
    asterisks, 07Y, 07gr, 09jf, 13abf, and 08D, are supernovae
    discussed in \Sect{observations}. The points are models that fit
    the observed light curves very well. In two cases, 09jf and 07Y,
    the $^{56}$Ni abundance was adjusted to give these good fits (see
    \Fig{2009jf} and \Fig{2007Y}).  In the top panel, some outliers
  attributed by \citet{Pre19} to Type Ic-BL and a very low mass event
  (SN 2018ie) are included, while some supernovae that lacked
  data for $t_{-1/2}$ in the Prentice et al. table (SNe 2013F,
    2013bb, 2013ek, 2015ah, and 2017bgu) are absent. In the bottom
  panel, the outliers are omitted and most of the previously missing
  supernovae now have data.  The models predict a slightly longer rise
  to peak ($t_{-1/2}$) than these observations, but a comparable time
  to decay to 50\% of peak luminosity ($t_{+1/2}$). For these
  observations, there is a deficiency of bright models with short
  time scales, but the observed luminosity of SN 2013F is uncertain.
  \lFig{prentice}}
\end{figure}

\subsection{Bolometric Light Curves}
\lSect{lbolo}

The peak bolometric luminosities and times for the standard models are
given in Appendix B. All used the new prescription for mixing
(\Sect{mix}). Also given in the tables are rise times ($t_{-1/2}$),
decline times ($t_{+1/2}$), and decline rates ($\Delta m_{15}$) for
the bolometric luminosity and similar information for the filtered
luminosities. From this information one can approximately reconstruct
any model light curve near its peak.

In \Fig{taddia}, the light curves of typical models, He4.10 (based on
a Salpeter IMF average of supernovae for models He2.70 to He8.00) and
He4.60 (based on average between He3.30 and He8.00), are compared with
the Carnegie sample of stripped-envelope supernovae
\citep{Tad18}. Also given are the light curves of the faintest
(He2.80) and the most luminous models (He5.38). Other models would
fill in the space between these two curves. The agreement with most of
the observations, including the faintest, is encouraging. There is a
substantial set though, perhaps as many as one-third of the observed
events, for which the brightest models are just too faint. This is
essentially the same dilemma noted by \citet{Ert20} restated with
better calculations of radiation transport and new observations. Is a
substantial fraction of the stripped supernovae called ``normal'' by
the observers outside of the reach of standard models based on a
neutrino-powered explosion and a radioactively illuminated light
curve?  For this model set, it seems so.

We also calculated the ``quasi-bolometric'' light curves of the same
models for direct comparison with the tabulations of
\citet{Pre19}. This entailed integrating the SEDONA spectral histories
assuming a complete measurement within the 4000 to 10000
\AA\ wavelength range and no detection outside. \Fig{prentice} shows
the comparison.  When peak luminosity is plotted against rise time,
the models rise too late and are too faint to explain the brightest
events. Some of these bright events are Type Ic-BL supernovae and
almost certainly involve other physics not included in this study. Two
other events are unusual. SN 2013bb (off scale in \Fig{prentice})
ejected a very large mass and is slowly moving. Its broad, faint light
curve is better explained by a more massive model than the common
events. The discussion of such events is deferred to \Sect{heavy}. At
the other other extreme, SN 2018ie is a faint, rapidly
evolving, high velocity, high temperature event. This might have been
a low mass explosion related to the fast blue transients discussed in
\Sect{fastblue}.

A comparison that excludes these anomalous cases is given in the
second panel of \Fig{prentice}. Since more supernovae had their decay
time measured than their rise time, there is slightly more data in this
figure. The agreement is significantly improved. One outlier, SN 2013F, 
is a fast rising, very luminous supernova. The host galaxy extinction, 
$E(B-V)_{\rm host} = 1.4 \pm 0.2$ mag, is much larger than for
any other supernova in the sample though and its colors are
uncertain. Perhaps its luminosity might be correspondingly uncertain.

\begin{deluxetable*}{lllcllcllccc}
\tablecaption{IMF Integrated Characteristics}
\tablehead{ \colhead{}  &
			\multicolumn{2}{c}{This work} &
			\colhead{}  &
			\multicolumn{2}{c}{This work} &
			\colhead{}  &
			\multicolumn{2}{c}{Ertl et al. (2020)} &
			\colhead{}  &
			\multicolumn{2}{c}{Observations}
			\\
			\colhead{}  &
			\multicolumn{2}{c}{$2.70\leq M_{\rm He} \leq 8.00$} &
			\colhead{}  &
			\multicolumn{2}{c}{$3.30\leq M_{\rm He} \leq 8.0$} &
			\colhead{}  &
			\multicolumn{2}{c}{$3.30\leq M_{\rm He} \leq 8.00$} &
			\colhead{}  &
			\multicolumn{2}{c}{Prentice et al. (2019)}
			\\
            \colhead{}  &
            \multicolumn{1}{l}{median}  &
            \multicolumn{1}{l}{mean}  &
            \colhead{} &
            \multicolumn{1}{l}{median}  &
            \multicolumn{1}{l}{mean}  &
            \colhead{} &
            \multicolumn{1}{l}{median}  &
            \multicolumn{1}{l}{mean} &
            \colhead{} &
            \colhead{median Ib}  &
            \colhead{median Ic}
            }\\
\startdata
$M_{\rm preSN}$ & 3.22 & 3.41 && 3.57 & 3.77 && 3.57 & 3.77 && --- & ---\\
$E_{\rm exp}$   & 0.77 & 0.89 && 1.26 & 1.06 && 1.26 & 1.06 && --- & ---\\
$M_{\rm ej}$    & 1.70 & 1.87 && 1.98 & 2.19 && 1.98 & 2.19 && 2.0$^{+1.2}_{-0.9}$  & 2.2$^{+3.1}_{-0.9}$ \\
$M_{\rm Ni}$    & 0.07 & 0.07 && 0.09 & 0.09 && 0.09 & 0.09 && 0.07$^{+.10}_{-.02}$ & 0.09$^{+.06}_{-.03}$ \\
\\
\multicolumn{12}{c}{Full Bolometric Luminosity}\\
$L_{\rm peak}$  & 42.21 & 42.20 && 42.32 & 42.27 && 42.30 & 42.25 && --- & --- \\
$t_{-1/2}$      & 11.9  & 12.3  && 11.8  & 11.9  && 8.57  & 9.00  && --- & --- \\
$t_{\rm peak}$  & 19.7  & 20.0  && 20.8  & 21.4  && 18.7  & 19.5  && --- & --- \\
$t_{+1/2}$      & 15.1  & 15.9  && 15.6  & 17.6  && 15.3  & 17.3  && --- & --- \\
\\
\multicolumn{12}{c}{4000 -- 10000 \AA}\\
$L_{\rm peak}$  & 42.11 & 42.10 && 42.21 & 42.16 && --- & --- && 42.2$^{+0.4}_{-0.1}$ & 42.3$^{+0.3}_{-0.2}$ \\
$t_{-1/2}$      & 11.0  & 11.1  && 11.9  & 11.9  && --- & --- && 10.4$^{+2.8}_{-1.7}$ & 9.8$^{+3.3}_{-3.1}$ \\
$t_{\rm peak}$  & 20.0  & 20.3  && 21.2  & 21.6  && --- & --- && --- & --- \\
$t_{+1/2}$      & 13.6  & 14.8  && 13.9  & 16.2  && --- & --- && 17.0$^{+4.7}_{-3.4}$ & 17.5$^{+7.1}_{-4.5}$
\enddata
\tablecomments{Observations are from \citet{Pre19}. Type Ib is their
  ``Ib +Ib(II)'' sample and Ic is their ``Ic-5/6/7'' sample in their
  Tables 4--7. Note that observationally inferred $^{56}$Ni masses are
  all based on Arnett's rule. Broad line supernovae are
  excluded. $E_{\rm exp}$ is in $10^{51}$ erg, masses are in \Msun,
  and times are all in days. Luminosities are in log base 10, and all
  models had mass loss with $f_{\rm WR}=1$.}  \lTab{imf_lum}
\end{deluxetable*}

\citet{Pre19} also give median luminosities, time scales, and ejected
masses for a larger set of several dozen supernovae with data
extracted from several sources. All luminosities are only for the
4000 - 10000 \AA\ range. \Tab{imf_lum} compares these median values to
the IMF-averaged models of this paper for the two different assumptions
about the progenitor mass range.

Now the agreement for all measured quantities is quite good.  The
median rise time is a day or so longer in the models and the decline
time two days shorter. The sum, i.e., the light curve width agrees
quite well. Even in the models, it is difficult to determine the exact
time of peak to better than a day (more so in heavier models) because
the luminosity varies so slowly there. Though the data is for a
  limited number of supernovae, the agreement is somewhat better
    than that of the same models with the Carnegie data set
    \citep[\Fig{taddia},][see also \Tab{imf_lum}]{Tad18}. Perhaps it
    is only the statistics of small numbers and the definition of what
    constitutes a ``normal'' supernova. We see no obvious explanation
    for a discrepancy.

Many others have extracted average ejected masses, $^{56}$Ni masses,
and kinetic energies from surveys of Type Ib and Ic supernovae
\citep{Ric06,Dro11,Can13,Tad15,Lym16,Pre16,Tad18}. For a summary table
of these other results see \citet{Tad18}. Most are consistent with
\citet{Pre19}, though often with larger error bars. Caution should be
exercised when using the $^{56}$Ni masses derived using Arnett's Rule,
as some of these surveys did. These are usually overestimates
(\Sect{arnett}). Using numerical modeling, \citet{Tad18}
  concluded, for Type Ib and Ic respectively, that $M_{\rm ej}$ = $3.8
  \pm 2.1$ and $2.1 \pm 1.0$ \Msun; $E_{\rm exp}$ = $1.4 \pm 0.9$ and
  $1.2 \pm 0.7 \times 10^{51}$ erg; and $M_{\rm Ni}$ = $0.14 \pm 0.09$
  and $0.13 \pm 0.04$ \Msun.

\begin{deluxetable}{lllcll}
\tablecaption{Dependence of Explosion Properties on Mass Loss}
\tablehead{ \colhead{}  &
			\multicolumn{2}{c}{$f_{\rm WR=1}$} &
			\colhead{}  &
			\multicolumn{2}{c}{$f_{\rm WR=1.5}$}
			\\
			\colhead{}  &
			\multicolumn{2}{c}{$4.40\leq M_{\rm He} \leq 8.00$} &
			\colhead{}  &
			\multicolumn{2}{c}{$5.00\leq M_{\rm He} \leq 13.0$}
			\\
			\colhead{}  &
			\multicolumn{2}{c}{$3.42\leq M_{\rm preSN} \leq 5.83$} &
			\colhead{}  &
			\multicolumn{2}{c}{$3.41\leq M_{\rm preSN} \leq 5.84$}
			\\
            \colhead{}  &
            \multicolumn{1}{l}{median}  &
            \multicolumn{1}{l}{mean}  &
            \colhead{} &
            \multicolumn{1}{l}{median}  &
            \multicolumn{1}{l}{mean}
            }\\
\startdata
$M_{\rm preSN}$ & 4.21 & 4.31 && 4.45 & 4.42 \\
$E_{\rm exp}$   & 1.42 & 1.38 && 1.41 & 1.37 \\
$M_{\rm ej}$    & 2.55 & 2.69 && 2.84 & 2.76 \\
$M_{\rm Ni}$    & 0.11 & 0.11 && 0.11 & 0.11 \\
\\
\multicolumn{6}{c}{Bolometric Luminosity}\\
$L_{\rm peak}$  & 42.39 & 42.36 && 42.35 & 42.36\\
$t_{-1/2}$      & 12.6  & 12.7  && 11.9  & 12.1 \\
$t_{\rm peak}$  & 22.0  & 22.7  && 21.0  & 21.8 \\
$t_{+1/2}$      & 18.5  & 19.7  && 21.8  & 20.9
\enddata
\tablecomments{This table is only to illustrate the effect of
    using different mass loss rates to evolve the same limited
    range of presupernova masses. It is not to be compared with
    observations that generally will include lower mass supernovae.
    See \Tab{imf_lum} for averages over the full range. Masses are in
  \Msun; explosion energy is in units of 10$^{51}$ erg; luminosity is
  log base 10 of L in erg s$^{-1}$; and time is in days. }
\lTab{imf_ml}
\end{deluxetable}

\Tab{imf_ml} compares the IMF-averaged characteristics of the two
model sets that used different mass loss rates ($f_{\rm WR}$ = 1 and
$f_{\rm WR}$ = 1.5; \Tab{main}). Unfortunately, no low mass explosions
with $M_{\rm preSN}$ less than 3.43 \Msun \ were computed for $f_{\rm
  WR} = 1.5$ by \citet{Ert20}, so one can only compare the averages
for a limited range of masses whose boundaries do not reflect the full
range of Type Ib and Ic progenitors. \Tab{imf_ml} should thus not be
compared with surveys that do not select against low mass (typically
low luminosity) events.

The use of a greater mass loss rate does not cause any major
differences in the bulk characteristics of the supernovae. Even though
the models to be averaged in \Tab{imf_ml} were selected on the basis
of a common presupernova mass, 3.42 \Msun \ to 5.63 \Msun \ for
$f_{\rm WR}$ = 1 and 3.43 \Msun \ to 5.64 \Msun \ for $f_{\rm WR}$ =
1.5, the average presupernova mass still differs by about 0.2 \Msun
\ because the weighting factors in the averaging process depend on the
estimated main sequence masses of the stars. The larger range of
higher masses for $f_{\rm WR}$ = 1.5 result in a more massive
average. Even so, the $^{56}$Ni masses and explosion energies are
virtually identical. Because of the greater emphasis on more massive
explosions, the ejected masses are a bit larger and that results in a
slight lengthening of the time scales and decrease in the peak
luminosity. From this limited study, one may conclude that the bulk
characteristics of the supernovae are insensitive to the mass loss
rate used to produce a given range of presupernova masses. This
insensitivity does not carry over to the colors and spectra, however
(\Sect{colors}).

\subsection{Photometry}
\lSect{colors}

\begin{figure}
\includegraphics[width=0.48\textwidth]{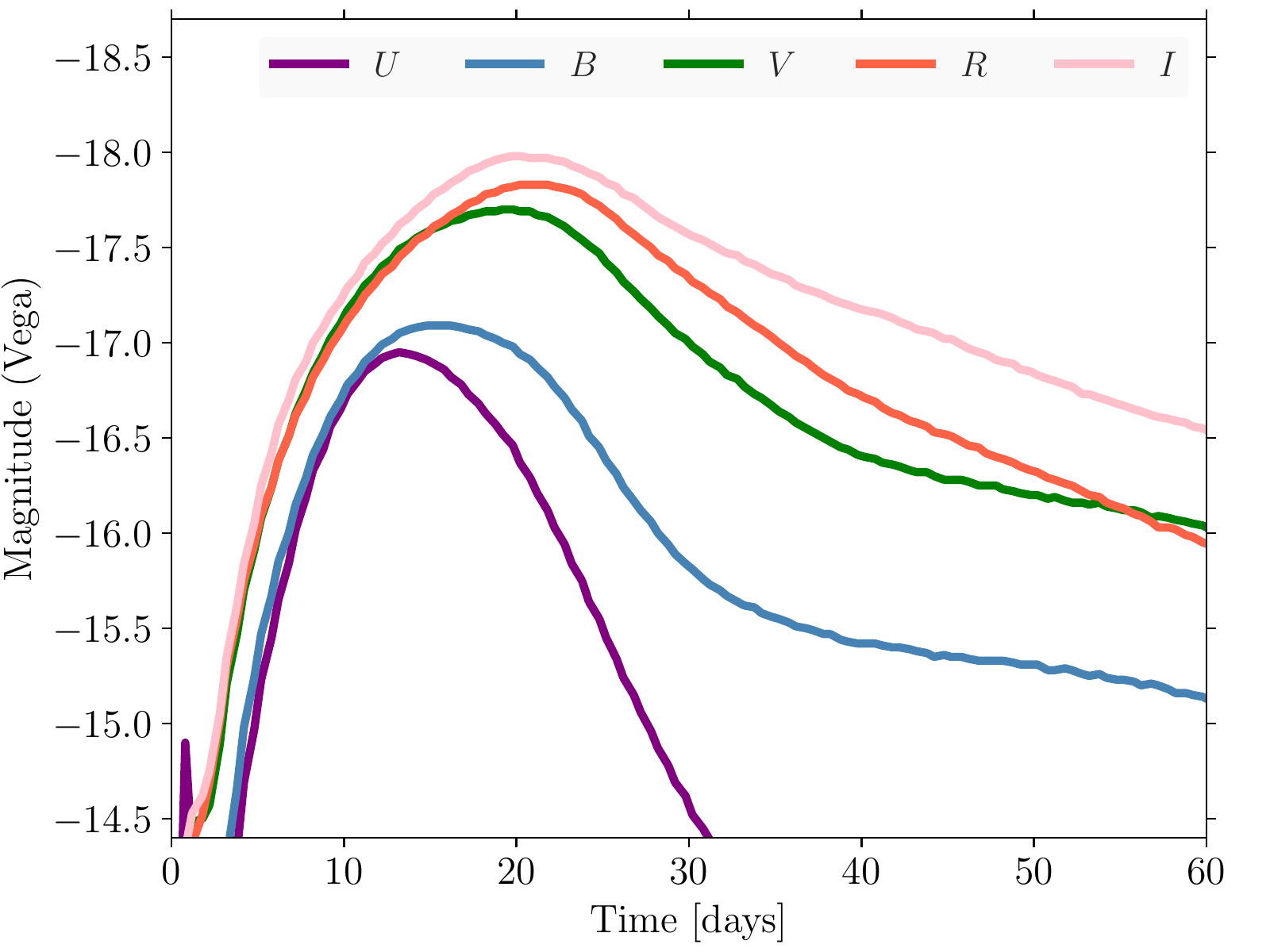}
\includegraphics[width=0.48\textwidth]{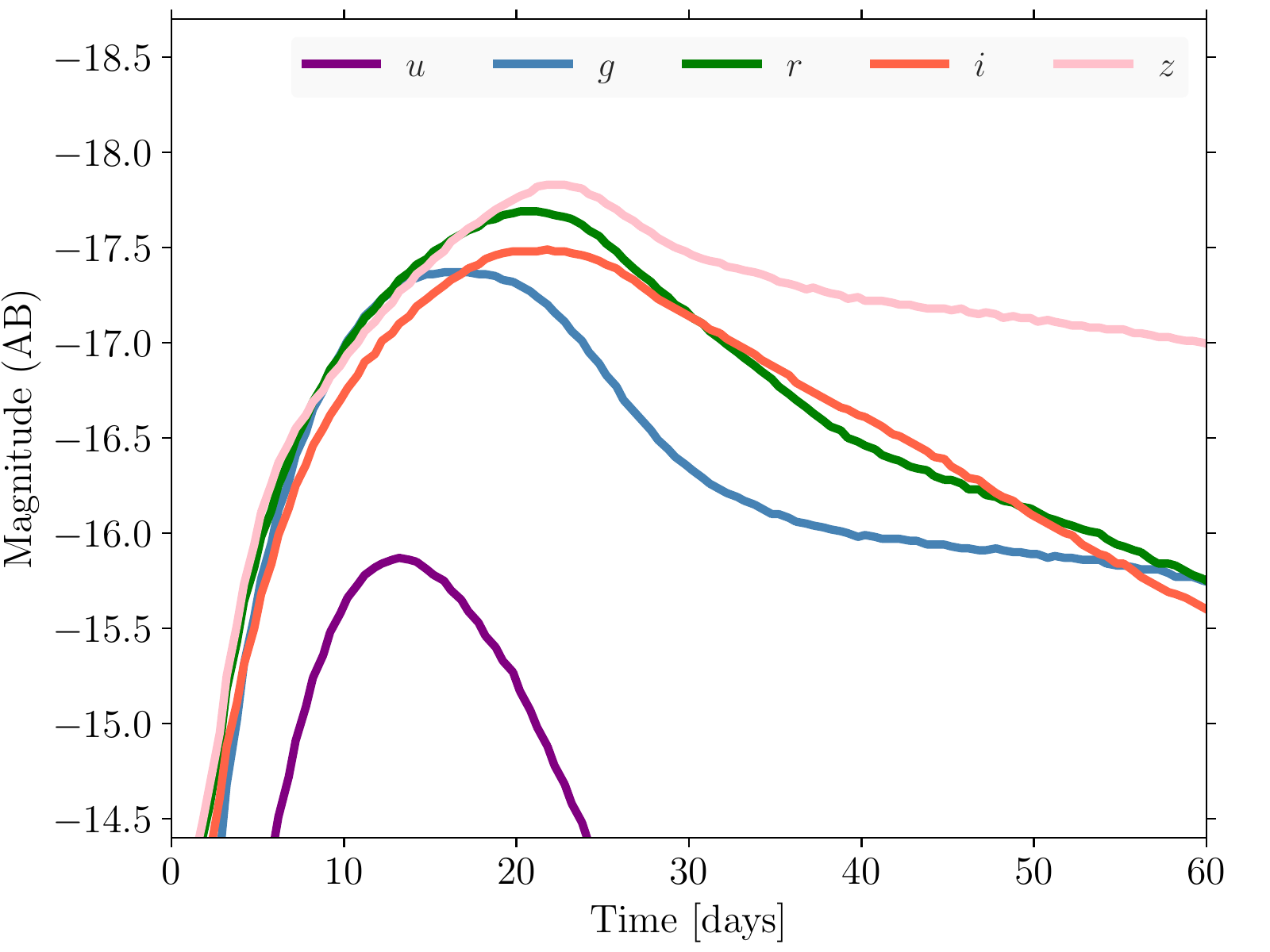}
\includegraphics[width=0.48\textwidth]{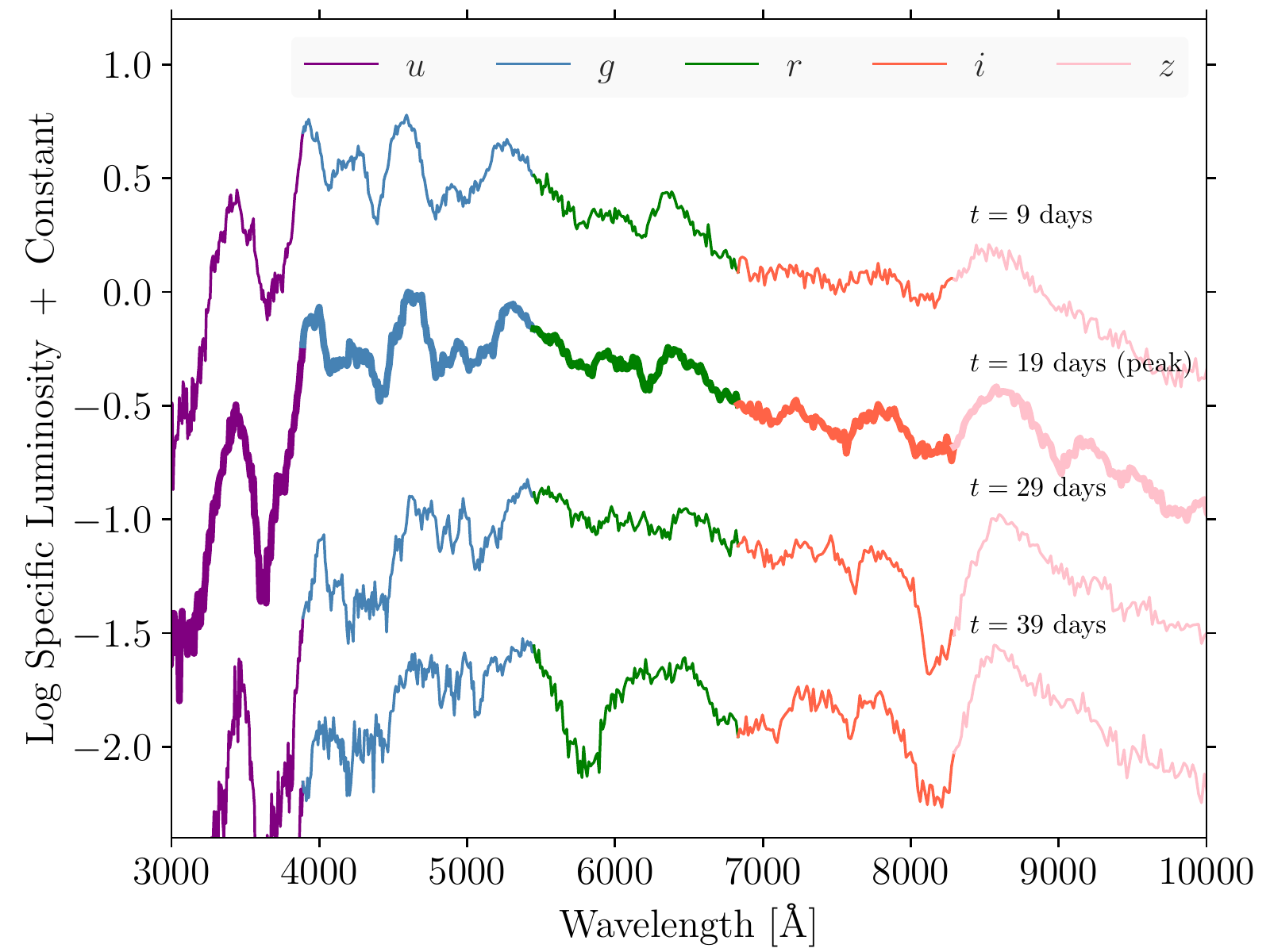}
\caption{Multiband Light curves and spectra for Model He4.50
  (\Tab{main}), a typical Type Ib model in our study. The model
  ejected 1.90 \Msun \ with a kinetic energy of $1.28 \times 10^{51}$
  erg and synthesized 0.099 \Msun \ of $^{56}$Ni. The peak bolometric
  luminosity was 10$^{42.39}$ erg s$^{-1}$. (top:) $UBVRI$
  light curves are given as indicated for Vega-based zero
  points. (Middle:) Light curves in the $ugriz$ wavebands based on an AB
  magnitude zero point. (Bottom:) The spectrum calculated using SEDONA
  at bolometric peak (19 d) and 10 days before and 10 and 20 days
  after. To facilitate comparison with the light curve, the spectrum
  has been color coded with the filters indicated corresponding to the
  wavelength range for AB-based magnitudes. \lFig{he4.5}}
\end{figure}

\begin{figure}
\includegraphics[width=0.48\textwidth]{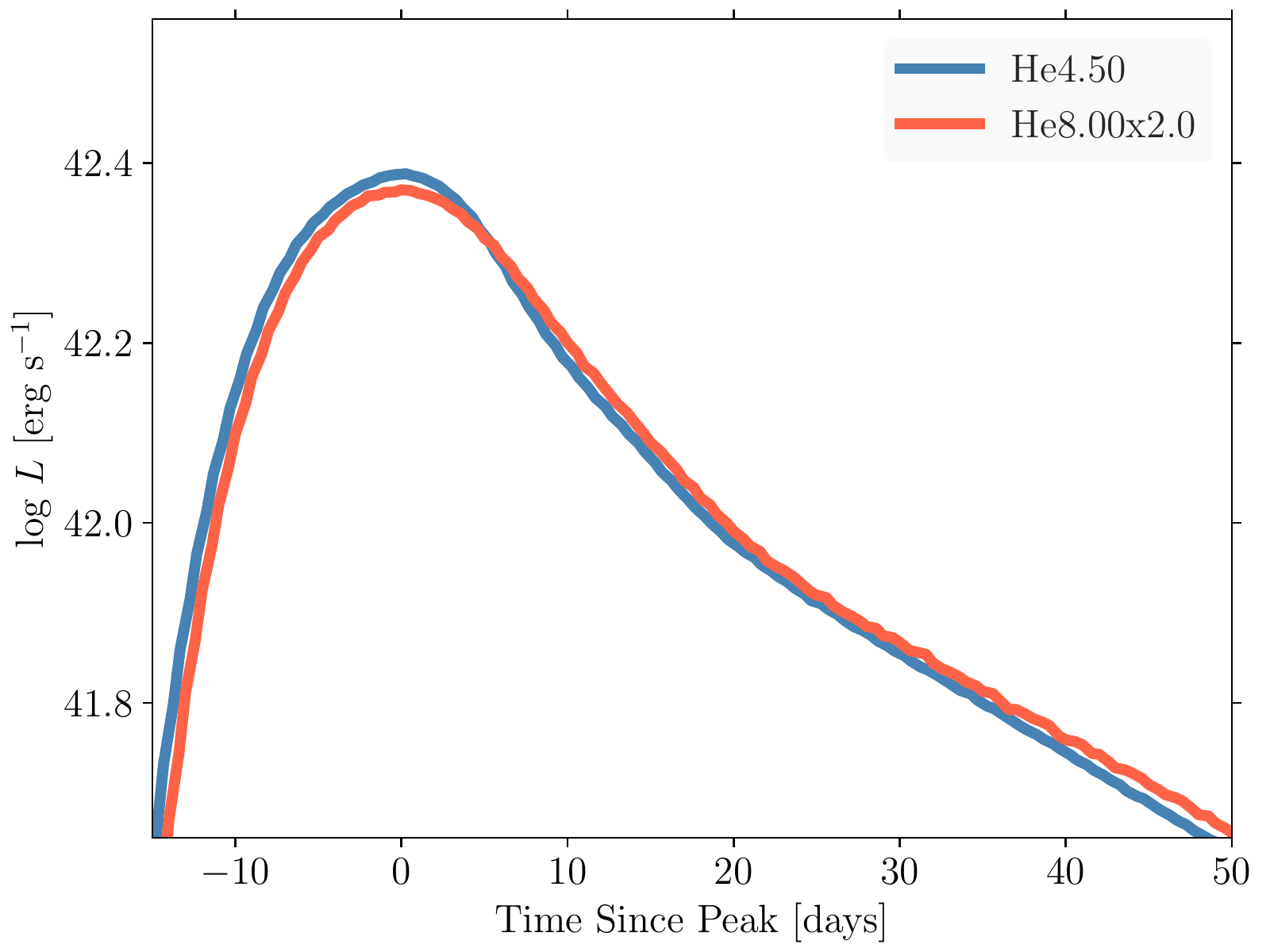}
\includegraphics[width=0.48\textwidth]{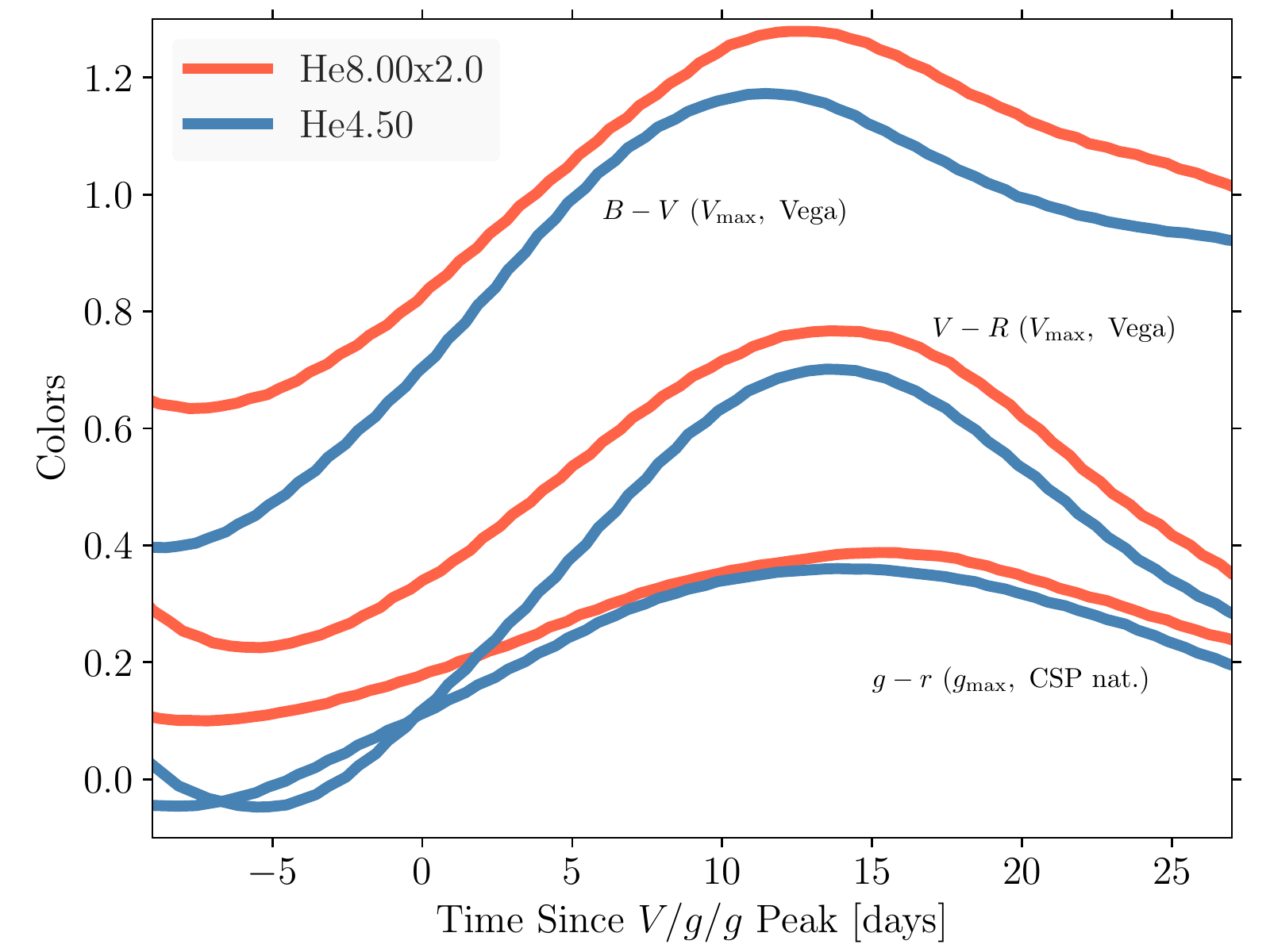}
\includegraphics[width=0.48\textwidth]{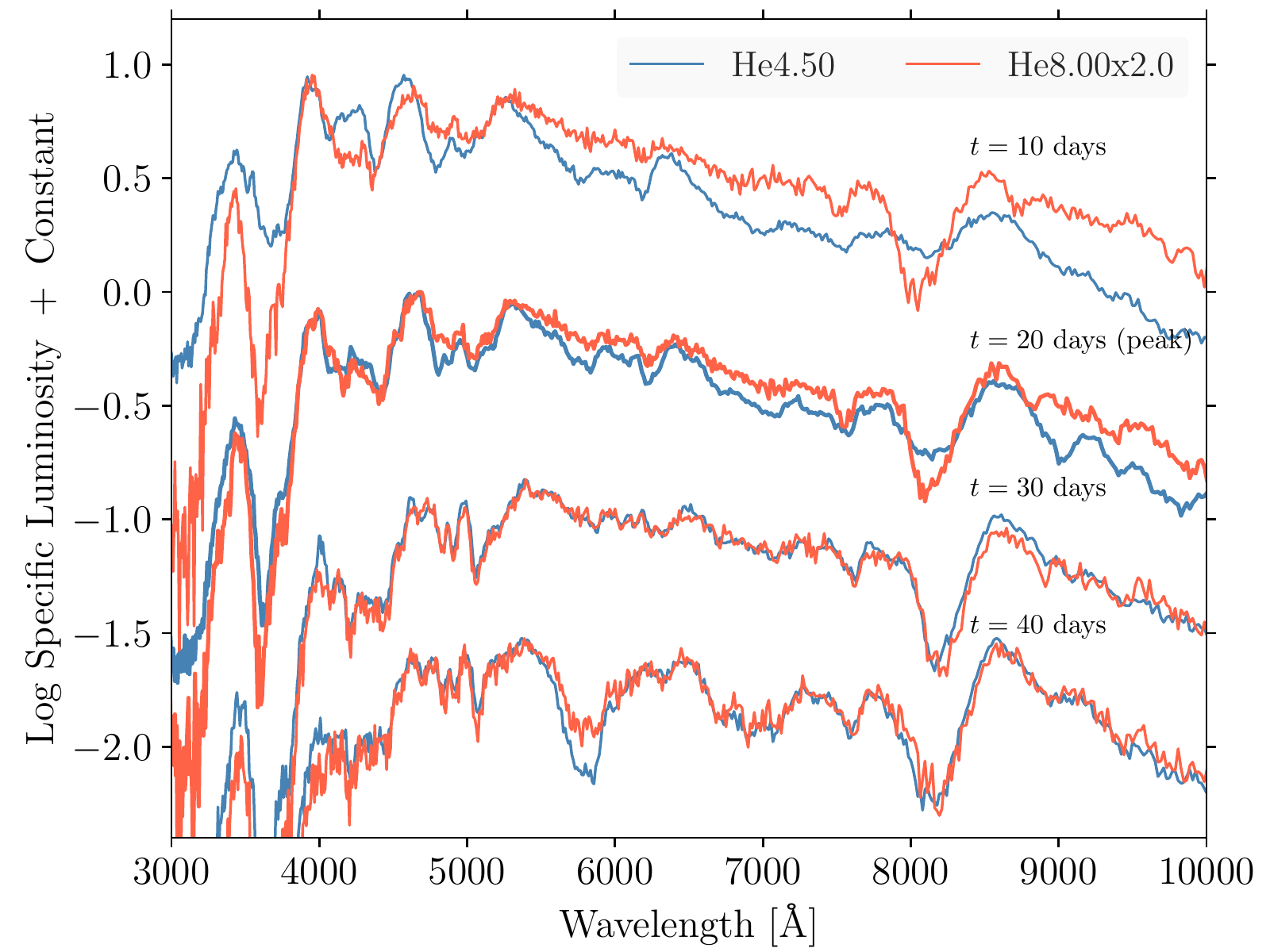}
\caption{Bolometric light curve (top), color indices (middle) and
  spectrum (bottom) for Model He8.00x2 compared with Model
  He4.50. Both models had identical explosion energies and $^{56}$Ni
  masses and similar ejecta masses. Model He8.00x2 was evolved from a
  larger mass star with a greater mass loss rate ($f_{\rm WR} = 2$)
  and thus lost almost all its helium before exploding. Model He4.50
  on the other hand (\Fig{he4.5}) was helium-rich with about half of
  its ejected mass being $^4$He. The carbon-rich Model He8.00x2 has a
  very similar bolometric light curve, but its colors are redder,
  especially at early times. The calcium infrared triplet is also
  stronger at earlier times in the carbon-rich model.\lFig{he8.0}}
\end{figure}

\begin{deluxetable}{lrrcrr}
\tablecaption{IMF Averaged Peak Magnitudes, Timescales, and Colors}
\tablehead{ \colhead{}  &
            \colhead{median}  &
            \colhead{mean}  &
            \colhead{} &
            \colhead{median}  &
            \colhead{mean}
            \\
            \colhead{} &
            \multicolumn{2}{c}{$f_{\rm WR}=1$} &
            \colhead{} &
            \multicolumn{2}{c}{$f_{\rm WR}=1.5$}
            }\\
\startdata
$U_{\rm peak}$  &  -16.50  &  -16.42  &&  -16.34  &  -16.47 \\
$B_{\rm peak}$  &  -16.74  &  -16.60  &&  -16.65  &  -16.54 \\
$V_{\rm peak}$  &  -17.30  &  -17.22  &&  -17.25  &  -17.13 \\
$R_{\rm peak}$  &  -17.39  &  -17.37  &&  -17.43  &  -17.29 \\
$I_{\rm peak}$  &  -17.54  &  -17.52  &&  -17.58  &  -17.45 \\
$u_{\rm peak}$  &  -15.40  &  -15.33  &&  -15.33  &  -15.39 \\
$g_{\rm peak}$  &  -17.05  &  -16.89  &&  -16.92  &  -16.82 \\
$r_{\rm peak}$  &  -17.25  &  -17.22  &&  -17.27  &  -17.14 \\
$i_{\rm peak}$  &  -17.06  &  -17.06  &&  -17.15  &  -16.99 \\
$z_{\rm peak}$  &  -17.40  &  -17.36  &&  -17.44  &  -17.28 \\
\\
$t_{-1/2,U}$  &  5.82  &  6.05  &&  5.68  &  5.78 \\
$t_{-1/2,B}$  &  8.03  &  8.47  &&  8.23  &  9.10 \\
$t_{-1/2,V}$  &  10.50  &  10.74  &&  10.68  &  10.68 \\
$t_{-1/2,R}$  &  11.64  &  11.76  &&  11.66  &  11.68 \\
$t_{-1/2,I}$  &  10.92  &  11.45  &&  11.02  &  11.36 \\
$t_{-1/2,u}$  &  5.64  &  6.09  &&  5.55  &  5.48 \\
$t_{-1/2,g}$  &  8.39  &  9.01  &&  8.94  &  8.97 \\
$t_{-1/2,r}$  &  11.36  &  11.57  &&  11.51  &  11.45 \\
$t_{-1/2,i}$  &  12.56  &  12.64  &&  12.07  &  12.35 \\
$t_{-1/2,z}$  &  11.22  &  11.40  &&  11.64  &  11.37 \\
\\
$t_{{\rm peak},U}$  &  13.82  &  13.52  &&  13.43  &  12.65 \\
$t_{{\rm peak},B}$  &  16.03  &  16.22  &&  15.82  &  16.15 \\
$t_{{\rm peak},V}$  &  19.34  &  19.82  &&  19.28  &  19.82 \\
$t_{{\rm peak},R}$  &  21.38  &  21.74  &&  21.33  &  21.57 \\
$t_{{\rm peak},I}$  &  20.35  &  21.27  &&  20.41  &  21.38 \\
$t_{{\rm peak},u}$  &  13.49  &  13.25  &&  13.29  &  12.43 \\
$t_{{\rm peak},g}$  &  16.83  &  17.20  &&  16.34  &  17.43 \\
$t_{{\rm peak},r}$  &  21.20  &  21.45  &&  20.84  &  21.28 \\
$t_{{\rm peak},i}$  &  22.83  &  22.88  &&  22.02  &  22.51 \\
$t_{{\rm peak},z}$  &  22.87  &  22.37  &&  22.97  &  22.38 \\
\\
$t_{+1/2,U}$  &  8.16  &  8.69  &&  8.94  &  9.29 \\
$t_{+1/2,B}$  &  10.38  &  10.92  &&  10.94  &  11.13 \\
$t_{+1/2,V}$  &  11.64  &  12.79  &&  11.77  &  13.48 \\
$t_{+1/2,R}$  &  13.21  &  14.09  &&  13.66  &  15.06 \\
$t_{+1/2,I}$  &  17.96  &  20.01  &&  17.93  &  21.63 \\
$t_{+1/2,u}$  &  7.56  &  8.13  &&  8.12  &  8.64 \\
$t_{+1/2,g}$  &  11.35  &  11.64  &&  11.96  &  11.74 \\
$t_{+1/2,r}$  &  12.33  &  13.32  &&  12.75  &  14.34 \\
$t_{+1/2,i}$  &  15.34  &  16.30  &&  17.23  &  17.84 \\
$t_{+1/2,z}$  &  37.06  &  40.45  &&  37.69  &  45.21 \\
\\
$(V-R)_{t_{\rm peak,V+10d}}$  &  0.340  &  0.336  &&  0.330  &  0.323 \\
$(V-R)_{t_{\rm peak,R+10d}}$  &  0.360  &  0.349  &&  0.361  &  0.345
\enddata
\tablecomments{$f_{\rm WR}=1$ averages were integrated between He2.70
  and He8.00, and $f_{\rm WR}=1.5$ averages between He3.00 and
  He13.0. See text for details.}  \lTab{imf}
\end{deluxetable}

\begin{figure}
\includegraphics[width=0.48\textwidth]{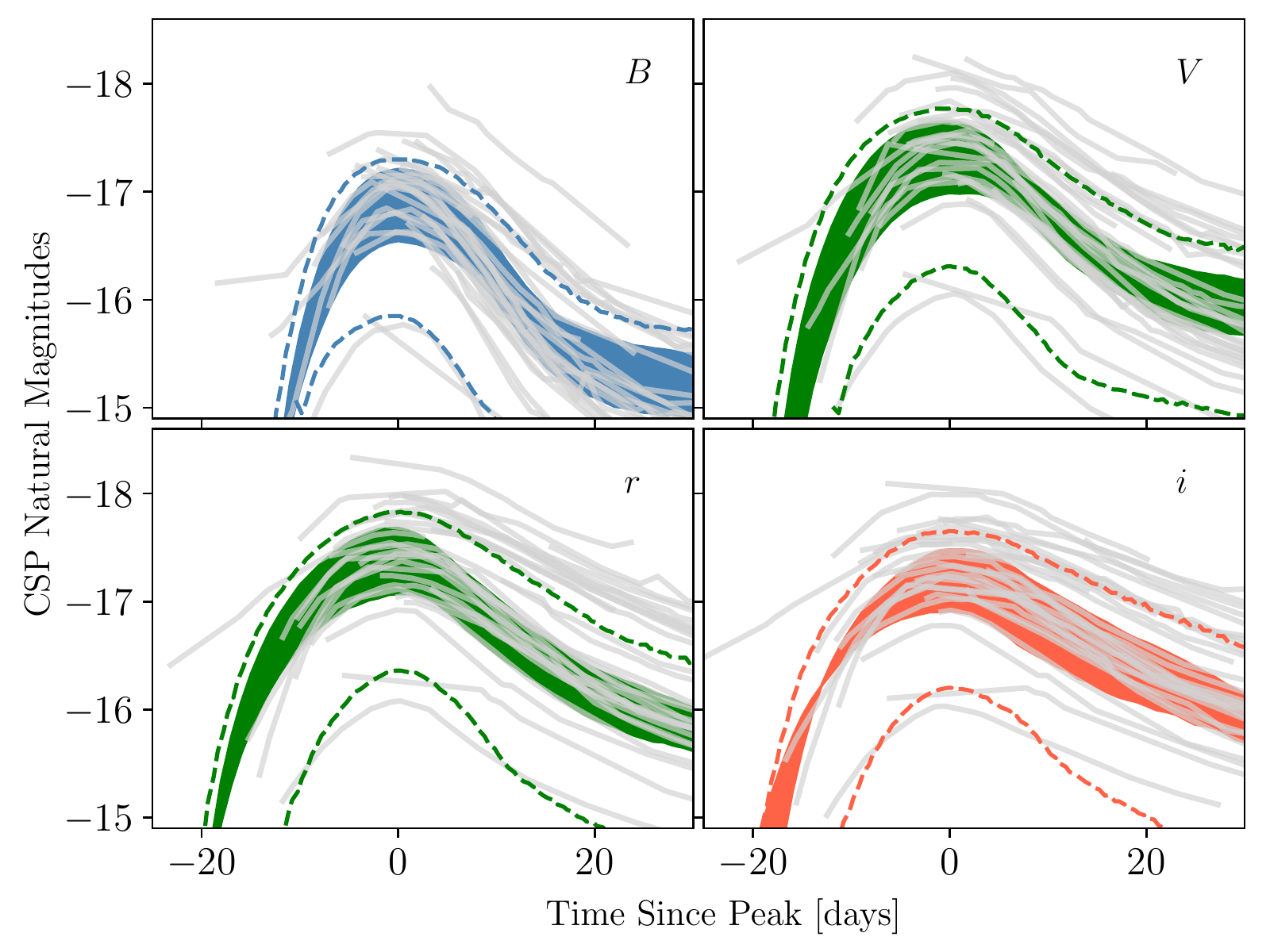}
\caption{Photometry for the Carnegie sample of light curves for Type
  Ib, Ic and IIb light curves \citep[gray
    curves][]{Tad18,Str18a,Str20} compared with models (colored
  curves). The dark band is bounded by Models He4.10 and He4.50, our
  typical explosions, and the dashed lines are for our faintest
  (He2.80) and brightest (He5.38) models using the standard W18 engine
  and mass loss ($f_{\rm WR} = 1$). Both the data and model are given
  in CSP ``natural'' magnitudes. No correction has been made in the
  data set for host galaxy extinction. \lFig{stritz}}
\end{figure}

Absolute magnitudes were computed by passing the SEDONA spectral
histories through appropriate filters and using the prescribed zero
points for various magnitude systems. In order to accommodate both 
convention and the data of several sets of observers, three different
magnitude systems were used: Vega for Johnson--Cousins $UBVRI$, AB for 
SDSS $ugriz$, and ``natural'' for CSP $uBgVri$. All filters, and also
zero points for Vega magnitudes, are given at
\url{http://svo2.cab.inta-csic.es/svo/theory/fps3}. The zero points 
for the AB system was taken to be 3631 Jy. For natural magnitudes, they 
were calculated according to \citet{Kri17}. Offsets at peak for a 
typical model are given in \Tab{offset} in the Appendix. Unless 
otherwise stated, $UBVRI$ magnitudes in this paper are on the 
Vega scale and $ugriz$ magnitudes are on the AB scale. Exceptions are 
when models are compared with data exclusively from the Carnegie group. 
Then natural magnitudes are used.

The photometry and spectrum of a typical model calculated with SEDONA
is shown in \Fig{he4.5}.  As noted previously, He4.50 is close to the
median presupernova mass, 4.6 \Msun, that one calculates for an
IMF-weighted distribution of helium cores in the mass range 3.3 to 8
using the equivalent ZAMS masses.  The mean ejected mass, kinetic
energy, and $^{56}$Ni mass for the same model set (\Tab{imf_lum}) are
also close to those of Model He4.50 (\Tab{main}). The spectrum in
\Fig{he4.5} has been color coded to show the wavelengths to which the
$ugriz$ filters are most sensitive.

\Fig{he8.0} shows that the bolometric light curve depends chiefly on
the bulk properties of the supernova - ejected mass, kinetic energy,
$^{56}$Ni mass - and not much on the composition of the presupernova
star. Models He4.50 with $f_{\rm WR} = 1$ and He8.00x2 with $f_{\rm
  WR} = 2$ end up ejecting similar masses with the kinetic energies
(\Tab{main} and \Tab{spcl}). They also eject, by design, the same
masses of $^{56}$Ni. It is then perhaps not too surprising that their
luminosities at peak are very similar, 10$^{42.39}$ erg s$^{-1}$ at
day 19.5 for He4.50 and 10$^{42.37}$ erg s$^{-1}$ at day 20.2 for
He8.00x2. The spectrum and colors are different though, especially at
early times \citep[see also similar results by][]{Yoo19}. The
carbon-rich star (He8.00x2) is redder than the helium-rich star
(He4.50). Even with large mass loss rates the surface is never devoid
of helium \citep[e.g.][]{Woo19} due to the recession of the helium
convective core in the presence of mass loss, but Model He8.00x2 has a
lot less total helium than He4.50. The helium, carbon, and oxygen
masses in the two explosions are 0.13 \Msun, 0.47 \Msun, and 0.84
\Msun \ for He8.00x2, but 0.96 \Msun, 0.13 \Msun, and 0.41 \Msun \ for
He4.50. The little bit of helium in He8.00x2 is in the outer layers
and is ejected with high velocity. Removing an electron from carbon or
oxygen is easier than taking one from helium, so the matter remains
ionized farther out for a longer time in Model He8.00x2. This sustains
a photosphere with a larger radius and may be why He8.00x2 is redder
at a given luminosity, especially at times so early that something
like a photosphere still exists. Another factor that could potentially
affect the color is the abundance of iron group elements at the
photosphere, but the abundance of iron in the unmixed outer layers was
the same (solar), and the mixing and iron-group synthesis was also the
same in the two models. This is therefore not an important effect.

While the rest of the paper will focus on the results using the
standard mass loss rate, $f_{\rm WR}$ = 1, one should keep in mind
these possible mass-loss-rate dependent variations in color indices.
The mass loss rate used in He8.00x2, $f_{\rm WR}$ = 2, is probably an
upper bound to the actual mass loss ate \citep{Yoo17,Woo20} so the
variations shown in \Fig{he8.0} may also be upper bounds.

Not all wavebands are treated equally well in SEDONA. $U$ (and $u$)
are probably the least accurate and will not be emphasized in this
study. The $U$-band is more sensitive to iron-line blanketing compared
to other bands and to the strength of the calcium H and K lines. That
means the $U$-band is quite sensitive to the distribution of iron-group
elements with velocity, hence to mixing. It is also sensitive to
temperature and luminosity, since small changes in temperature can
change the ionization state (e.g., the ratio of Fe III to Fe II) in
some zones, which has a big effect on the line blanketing.
Uncertainties due to non-LTE effects can also play a role. A small
error in, say, the LTE computation of the ionization state may have a
significant effect on the $U$-band while only a modest effect on on
the $R$-band.

Uncertainties in the input atomic line data can also play a role. The
calculations for the main models here used a large line list
(\Sect{sedona}). Calculations with a reduced set of 370,000 lines
showed a modest increase in the ultraviolet brightness and blue
magnitudes, but not much change for other filters.

IMF-averaged color characteristics are given in \Tab{imf} for the mass
range considered as typical for normal Type Ib and Ic
supernovae. Compared with observations, the absolute magnitudes of our
models in various filters agree quite well with the Carnegie
tabulation for 34 Type Ib, Ic and IIb supernovae (\Fig{stritz}). Since
the data in \citet{Str18a} and \citet{Str20} are given in CSP natural
magnitudes, model light curves were computed accordingly.  The
faintest observed curve in \Fig{stritz} is SN 2007Y, a brief Type Ib
supernova \citep{Str09,Str18a,Tad18}. This supernova is treated in
greater detail in \Sect{SN2007Y}. Explaining faint events generally
poses no problems since the $^{56}$Ni masses used here are upper
bounds (\Tab{main}).

\subsection{Peak Magnitude Distributions}
\lSect{distribution}

\begin{figure}
\includegraphics[width=0.48\textwidth]{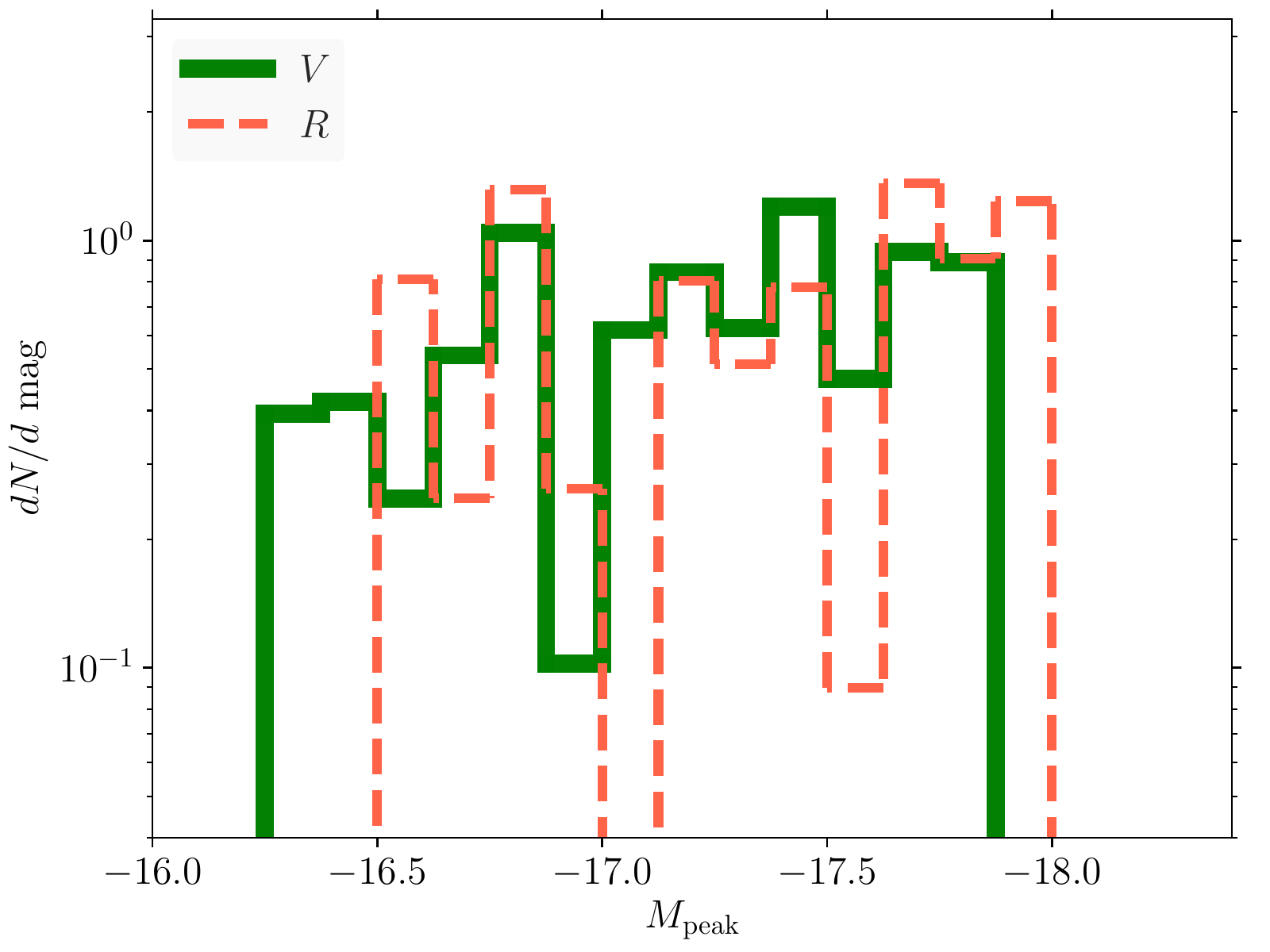}
\caption{IMF averaged peak magnitude in the $V$ and $R$ bands for the
  standard model set with starting helium star masses from 2.7 to 8.0 \Msun.
  \lFig{imf}}
\end{figure}

The distributions of peak magnitudes in $V$ and $R$ magnitudes are
given for our model set in \Fig{imf}. These are averages calculated
using a Salpeter IMF \citep{Sal55} applied on the zero age main
sequence mass distribution as described in \citet{Ert20}. The time of
maximum is defined for each color, i.e., the brightest $V$ magnitude
is evaluated at the time of $V$-band peak, not bolometric maximum. The
brightest $V$ magnitude for any of the standard models is
$-17.83$. The brightest $R$ magnitude is $-17.99$. This is for Model
He5.38. Slightly brighter models exist for other choices of explosion
energy and mass loss. For Model He5.25S, exploded with slightly more
powerful S19.8 engine, the magnitudes at peak are $B = -17.48$, $V =
-17.96$, and $R = -18.03$. For He5.00 with 1.5 times the standard mass
loss (\Tab{main}) and with W18 engine the peak magnitudes were $B =
-17.52$ , $V = -17.97$, and $R = -18.00$. Interestingly, the peak
luminosities reach a maximum value around He5.00 to He6.00 and decline
for heavier masses, especially above He8.00 (\Tab{peak_main}). This is
because the $^{56}$Ni masses cease rising (\Tab{main}), but the slower
expansion results in later peak times, more $^{56}$Ni decay before
peak, and thus lower luminosities. Our distribution of peak
brightnesses is not being influenced on the upper end by the neglect
of more massive models, at least for the neutrino-transport models
considered here.

These distributions agree reasonably well with observations
  for about 2/3 of the ordinary Type Ib and Ic supernovae given in Fig
  19 of \citet{Dro11} which have been corrected for host galaxy
  extinction. Fainter, briefer events could come from lower mass
stars (\Sect{fastblue}) and fainter broader events from higher mass
ones (\Sect{heavy}). Fainter values for peak magnitude could
  also be obtained by acceptable downwards adjustments of the
$^{56}$Ni yield in any of our models. The distributions in
\Fig{imf} lack, however, the brighter events in the observed
  distribution of \citet{Dro11}, even after broad-lined Type Ic
  supernovae are removed and we see no obvious reason for the
discrepancy.  A comparison with the more recent data of \citet[][see
  their Fig 8 and \Fig{taddia} here]{Tad18} shows somewhat better
agreement and some support for a cut off around $V = -18$.

\begin{figure*}
\includegraphics[width=\textwidth]{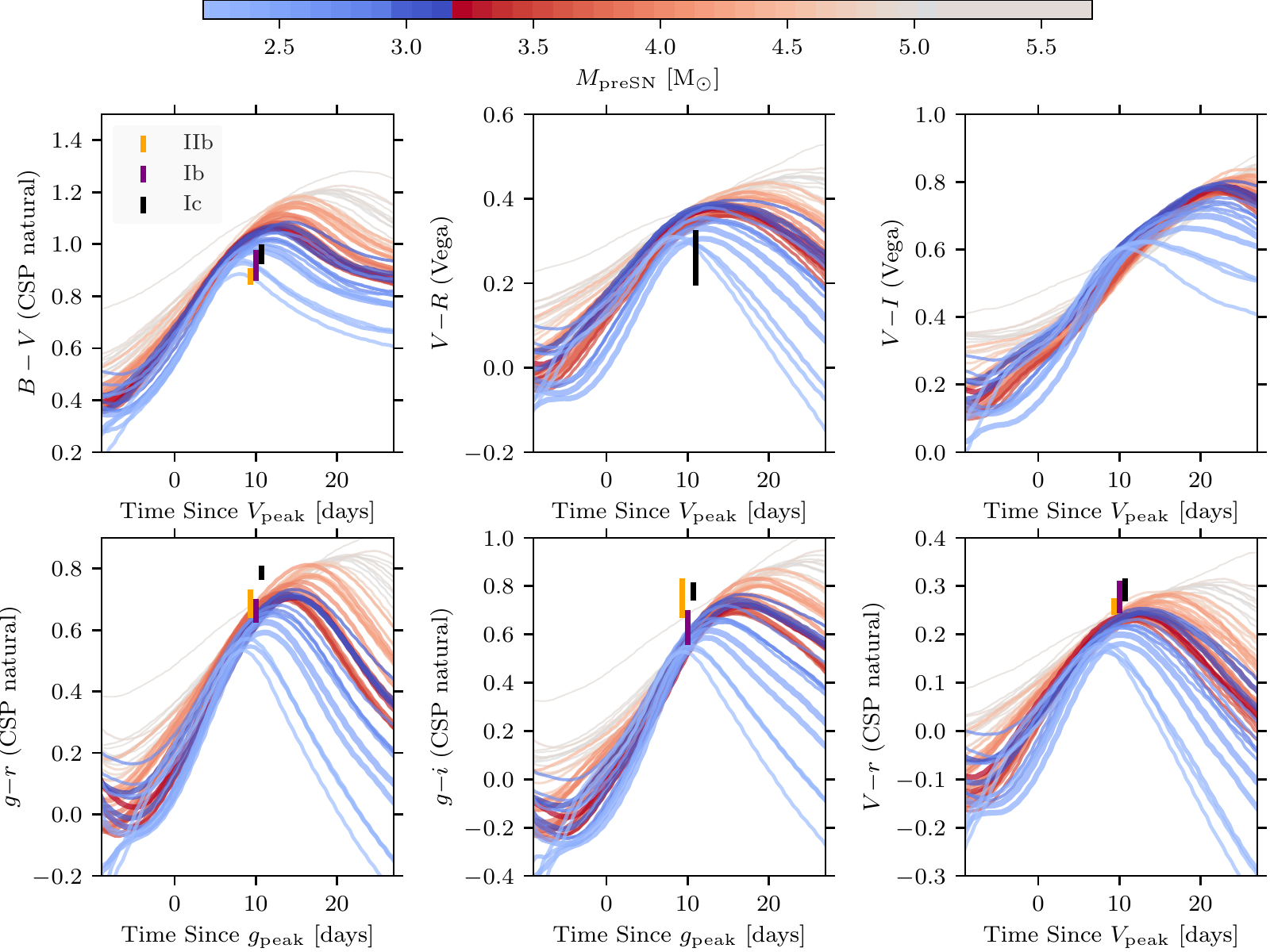}
\caption{Color indices as a function of time for the standard
    model set. The model lines, color coded according to the
    presupernova mass, are for Models He2.70 through He8.00. The most
    common model has a presupernova mass of 3.22 \Msun
    (\Tab{imf_lum}). More massive models are colored red, and less
    massive ones are blue, with a line weight that declines farther
    away from this median. Data points at 10 days post-peak are from
  \citet{Str18b}, with separate lines for Type IIb, Ib and Ic (left to
  right) for ($B-V$), ($g-r$), ($g-i$), and ($V-r$). These data points
  are all in CSP natural magnitudes. Some of the data at 10
  days have been offset slightly for clarity of display. The single
  data point for ($V-R$) is from \citet{Dro11}, given in Vega
  magnitudes. The models themselves are computed in CSP natural
    magnitudes except for (B-V) and (V-I) which are based on
    Johnson-Cousins filters in Vega zeropoint. See \Tab{offset} for
    approximate offsets. Note the ``pinches'' that occur in many
  color indices at about 10 days post maximum. For the purpose of
  presentation the model curves have been smoothed through a second
  order Savitzky--Golay filter.  \lFig{color}}
\end{figure*}

\begin{figure}
  \includegraphics[width=0.48\textwidth]{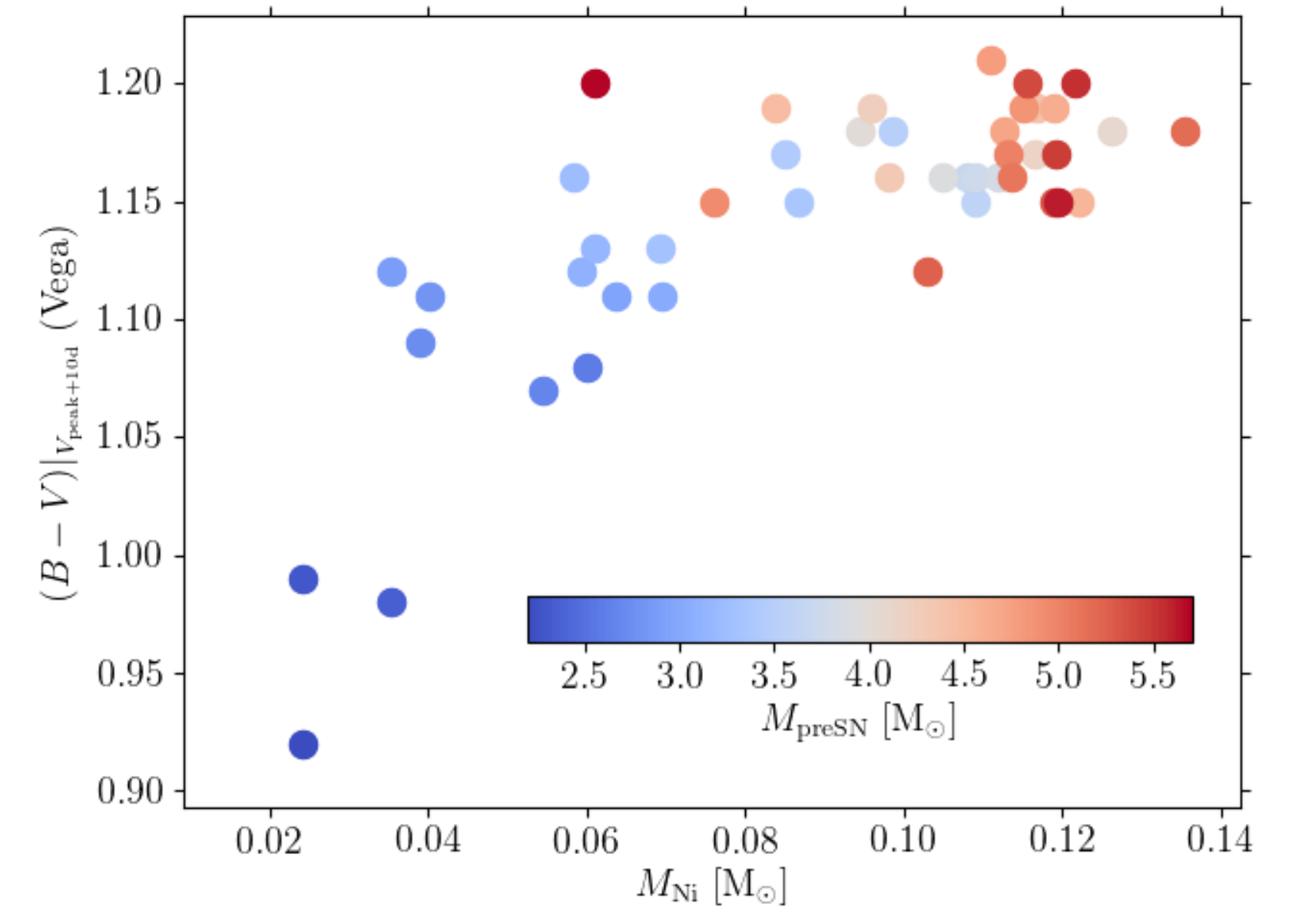}
  \includegraphics[width=0.48\textwidth]{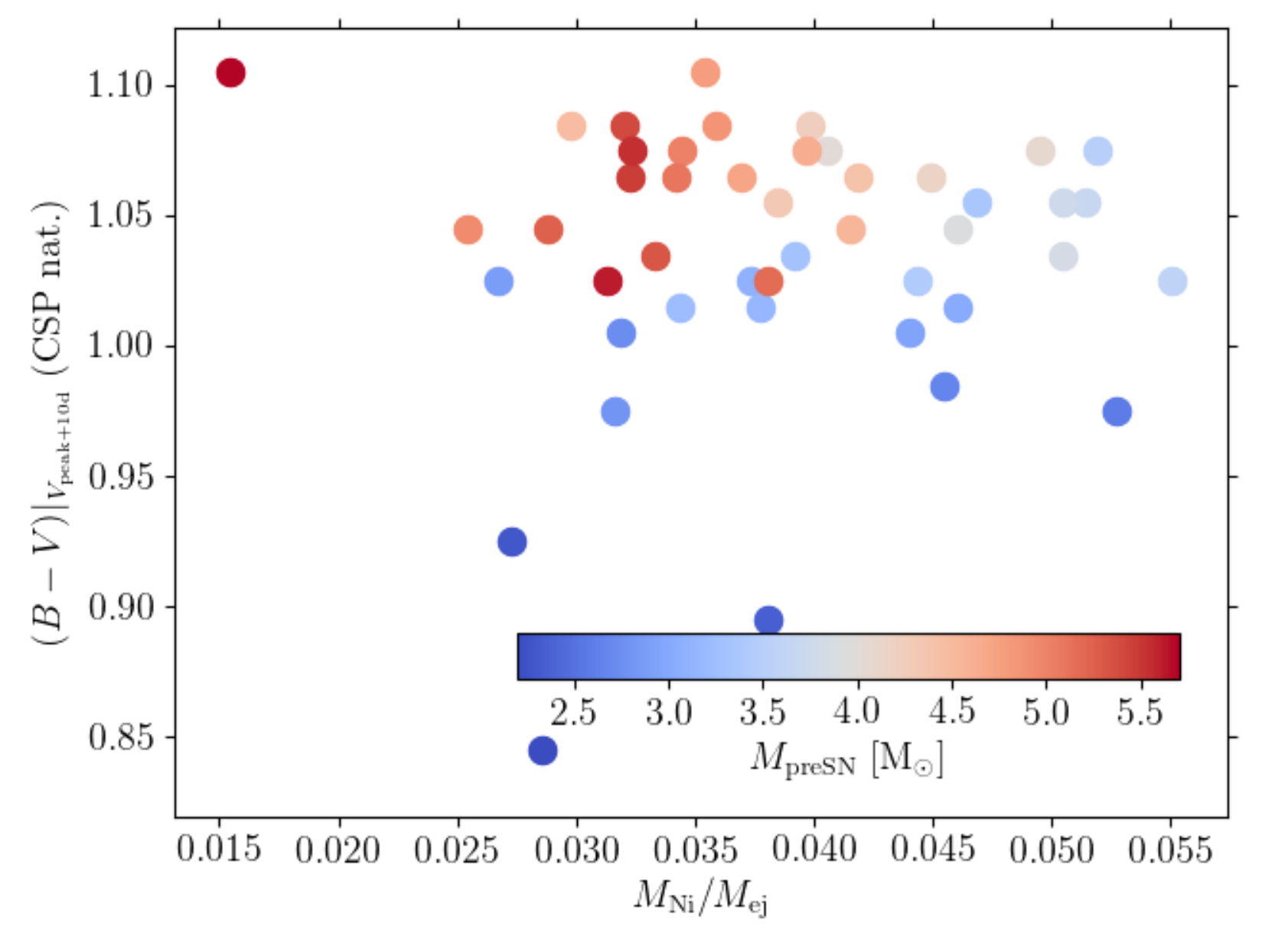}
  \caption{($B - V$) at 10 days post $V_{\rm max}$ vs $^{56}$Ni
      mass ejected (top) and $M_{\rm Ni}$/$M_{\rm ej}$ (bottom). More
      $^{56}$Ni makes the supernova brighter which tends to make it
      bluer at a given epoch, but more $^{56}$Ni mixed to large radius
      also increases line blanketing and maintains a larger photospheric
      radius. The overall effect at 10 days after peak is to redden
      the supernova. The bottom panel compares favorably with Fig. 8b
      of \citet{Str18b}, especially if the three very low mass models
      are ignored, although the range of $M_{\rm Ni}$/$M_{\rm ej}$
      sampled here is much smaller. \lFig{bmvni}}
\end{figure}

\begin{figure}
\includegraphics[width=0.48\textwidth]{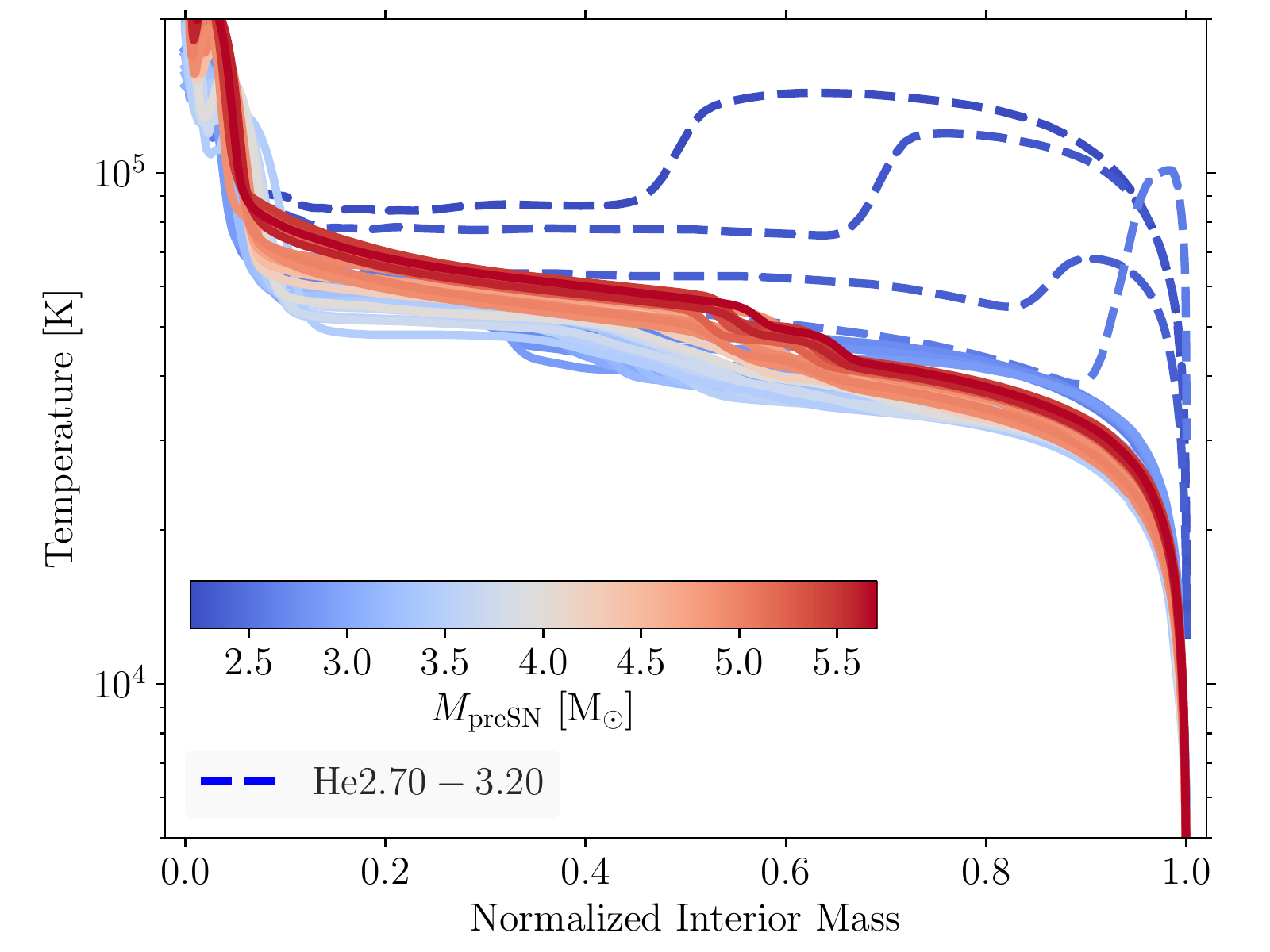}
\caption{Interior temperature for the initial models calculated by
  KEPLER one day after core collapse. The shock has exited the
  supernova and the ejected matter has commenced a stage of homologous
  expansion, except for the lightest inflated models (dashed). 
  For nearly the entire range of masses considered in the survey
  the temperature profile, normalized to the mass of the ejecta, is
  nearly a constant. Subsequent deposition of energy from radioactive
  decay will cause large deviations from the $T \propto 1/t$ behavior
  expected for adiabatic expansion in a gas dominated by radiation
  entropy.
  \lFig{tcent}}
\end{figure}

\subsection{Color Indices}
\lSect{indices}

\Fig{color} shows the evolution of several frequently used color
indices for the standard model set.  The first three days of the
explosion are not plotted because the KEPLER model is linked to SEDONA
on day one, after homologous expansion has become a good
approximation, and it takes another day or so for the radiation
transport to adjust to the new code. The evolution near and
  shortly after peak has been best studied observationally and there
  is considerable interest in ``pinches'' in the various color indices
  about 10 days post-maximum \citep{Dro11,Tad15,Str18b} since they
  might be used to estimate reddening.

Model results for ($B - V$), ($g - r$), ($g - i$), and ($V - r$)
  are plotted in CSP natural magnitudes and compared, at 10 days
  post-peak of $V$ or $g$, with measurements by \citet{Str18b}.  Model
  results for ($V - R$) and ($V - I$) are given in Johnson-Cousins
  Vega magnitudes and are compared, for ($V - R$), with data from
  \citet{Dro11}. Approximate offsets for other magnitude systems are
  given in \Tab{offset} and are less than 0.1 magnitudes.  All data
  has been de-reddened and reflects the intrinsic values at the
  host. Model results for ($V-I$) have previously been published by
  \citet{Des15}. Overall the agreement between observations and the
  models at 10 days post-peak in \Fig{color} is reasonably good,
  though there is a tendency for the models to be a little too red at
  early times and far too blue at very late times. Compare our model
  curves with, for example, the intrinsic color templates in Fig. 9 of
  \citet{Str18b}.

There are several reasons for the disparity. One is an inherent
  uncertainty in the time of maximum light in the various bands. The
  visual brightness is slowly varying in the models near maximum with
  a typical full width at 97\% maximum of about 6 days. Small changes
  in the spectrum would affect our estimated time of visual maximum
  appreciably and the observed dates of maximum brightness might also
  not be precise. At the same time, the color indices are rapidly
  evolving. An offset of 3 days would significantly affect the
  comparison between the models and observations. The model colors are
  also intrinsically sensitive to mixing, $^{56}$Ni production
  (\Fig{bmvni}), the LTE approximation used in our studies, and the
  geometry of the explosion. Less line blanketing in the outer layers
  would make the model bluer near peak. Greater ionization in the
  outer layers, possibly due to non-LTE effects at late times, would
  make them redder.

Larger values for the color indices in \Fig{color} are indicative of
cooler emission, so a positive slope indicates a time when the
supernova is becoming redder. Abrupt changes of slope in a color index
reflect evolving physical circumstances in the supernova.  During the
first several days (not shown), the color evolves sharply to the red
as shock deposited energy diffuses out of the expanding outer
layers. The luminosity declines or is nearly constant while the
photospheric radius increases. Helium recombines at about 10000 K and
a recombination front starts to move inwards in mass while still
moving outwards in radius. This brief phase was not accurately
captured in the SEDONA calculations and is not plotted, but is clearly
seen in the KEPLER light curves and effective temperatures.

Once the outward moving diffusion front carrying energy from
radioactive decay reaches the photosphere, the luminosity rises
rapidly while the radius continues a slow increase.  The combination
causes the color to become blue and this is the cause of the first
downward sloping part seen near the origin in some of the color
index plots.

Shortly before peak, the luminosity rises more slowly while the radius
rapidly increases. The color thus begins a long gradual ascent into
the red that is prominent in all the panels of \Fig{color}. By maximum
luminosity roughly half the supernova has recombined. The photospheric
radius is still increasing though and, as the luminosity declines, the
reddening of the color indices continues.  Other effects also come
into play and the idea of a well-defined photosphere with a color
governed by recombination like in a Type IIp supernova is increasingly
overly simplistic. The opacity is not just due to electrons, but to
lines that are sensitive to the ionization state and generally more
important at lower temperature. The deeper the photosphere, the more
heavy elements lie outside it, especially in mixed models, and this
line blanketing also drives the color to the red.

About 10 to 20 days after maximum luminosity, with the longer time
scale being appropriate to the more massive cases with longer
diffusion times and larger fractions of carbon and oxygen,
recombination reaches the center of the explosion and the photospheric
phase ends. The concept of a photosphere is only an approximation
anyway since different wavelengths originate from different depths and
recombination is never complete, but from this point on, the supernova
is increasingly just a cloud of ashes pumped by radiative decay. The
power deposited by this decay (that which is not escaping) balances
the luminosity. This is the onset of the nebular stage, though the
full transition will take some time.  Colors after this point vary
according to the lines present in the filter, the mass of $^{56}$Ni,
and the density. The LTE model becomes an increasingly poor
approximation. Among other things, it underestimates the ionization
in a radioactively energized plasma. Underestimating the ionization
means one sees to a deeper depth and hotter matter.  For these
reasons, the colors of our models are increasingly unreliable more
than 20 days after peak luminosity.

As previously noted \citep{Dro11,Des15,Str18b}, a ``pinch'' is seen in
the distribution of colors measured in many wavebands about 10 days
after peak. This reflects the regularity of the model set selected as
Type Ib and Ic candidates and basic physics. The ratio of explosion
energy to ejected mass and the $^{56}$Ni mass do not vary greatly in
the standard set (\Tab{main}). One day after the explosion, all the
models, except a few low mass ones with presupernova radius expansion,
have a similar distribution of internal temperature (\Fig{tcent}). The
entropy in most of the ejecta at this time is about half due to
radiation; the other half is electrons and ions. For adiabatic
expansion, temperature thus declines more slowly than (radius)$^{-1}$,
and hence slower that 1/t. This common thermodynamic configuration
means that the light curves for the models without radius expansion
will be quite similar during the first week.  Nevertheless, the high
density cores of all models would be fully recombined by 10 days after
the explosion, well before peak, were it not for radioactive decay.

Helium recombines at $\sim$10,000 K. Carbon and oxygen combine with
their last electrons at a lower temperature, typically $\sim$5000
K. For a blackbody, these two temperatures correspond to ($B-V$) = 0
and 1 which is very roughly the range of ($B-V$) in \Fig{color}. These
are approximate values because of the variable density, the
non-uniformity of the temperature, and gradients in the composition
after mixing, but for temperatures cooler than about 5000 K, the
supernova is effectively recombined.  Prior to recombination, the
photosphere continues to move out in radius, forcing the color to the
red. This is the epoch corresponding to the well-defined upward slopes
in \Fig{color}. The lower mass models have faster expansion speeds in
their deep interiors and greater helium fractions. They experience
recombination earlier, so their color-index curves (e.g., for $B-V$)
are steeper. These differences in slope result in a convergence, a
``pinching'' of the distribution as time passes. The convergence is
truncated when the supernova recombines. After that the sudden
collapse of the photosphere sends things to the blue, though what
happens later is not well determined in the present study.

As \Fig{color} shows, there is observational evidence for these
pinches.  \citet{Dro11} have found a convergence of ($V - R$) evaluated
10 days after $V$-band maximum. For their ``gold'' set of well-studied
Type Ib and Ic supernovae, ($V-R$)$_{\rm 10 \ days} = 0.26 \pm 0.06$ mag.
In a study similar to ours, \citet{Des16} found a
similar convergence, but with a value $0.33 \pm 0.035$ magnitudes, if
they removed two high energy outliers. \citet{Str18b} also observed
similar pinches in a number of different filters, especially
($V-r$). These are shown as vertical error bars on the plot and agree 
reasonably well with the models.

It is encouraging that four studies, two observational and two
theoretical agree on ($V-R$) at 10 days.  As \citet{Des16} note, weaker
mixing favors higher temperatures in the inner ejecta, causing redder
colors early on, but bluer colors around maximum \citep[see
  also][]{Des12}, so this color index might be useful for constraining
mixing. As discussed by \citet{Dro11} and \citet{Str18b}, the narrow
range of several color indices in \Fig{color} might also be useful for
better estimating the host galaxy extinction.

\Fig{carncolor} shows the time-dependent color indices for the
Carnegie collection of of 34 stripped supernovae \citep{Str18a}. Here
both data and models have been displayed in the ``natural'' magnitudes
in which they were measured. The agreement for ($B-V$) is quite good.
Both the average supernova models (between He4.10 and He4.50) and the
band defined by the brightest and faintest models show near congruence
with the observations. Interestingly the narrowing in ($B-V$) for the
models at 10 days post-peak is more pronounced than in the observations.
Typical error bars on the observations are 0.03 magnitudes.

The agreement is not as good for ($V-r$) and ($V-i$) showing
  again that our models emit inadequate radiation above about 5500
  \AA\ at late times. See also the spectroscopic comparisons to
  individual supernovae in \Sect{spectra}. This could be due to the LTE
  approximation used in the models. Inadequate blue radiation is being
  processed into the red. The photosphere is collapsing and becoming
  quasi-nebular prematurely.

\begin{figure}
\includegraphics[width=0.48\textwidth]{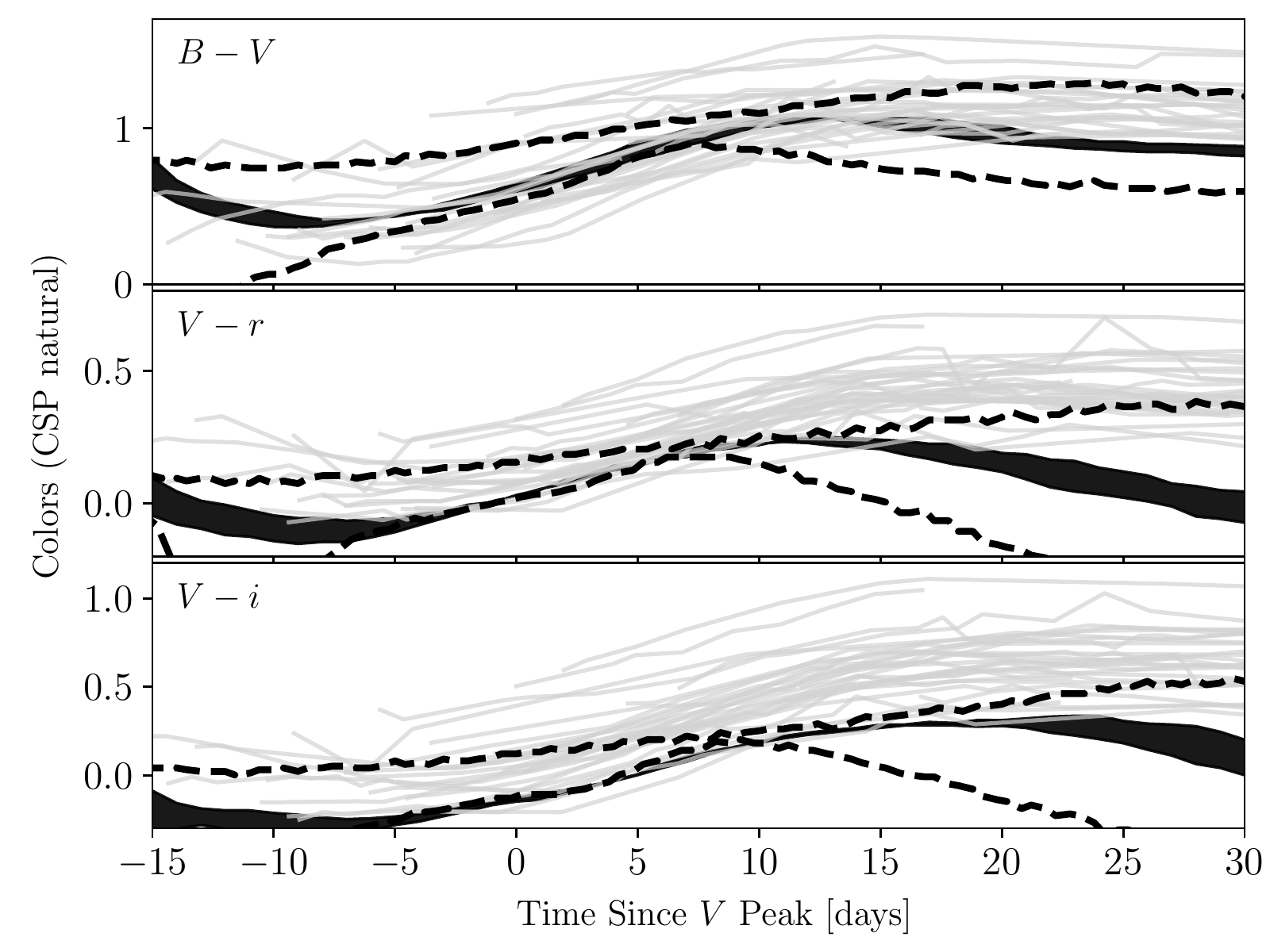}
\caption{Color indices from the Carnegie sample compared with the
  models of this paper. \citet{Str18a} recommend the interval
    from 0 to 20 days as most reliable. While the agreement with
  ($B-V$) is good, the models are too faint in the $i$-band by about
  0.3 magnitudes. As in \Fig{stritz}, our model colors have been
  computed in CSP natural magnitudes to directly compare with the
  data. No host galaxy extinction is included. The dark solid curve is
  the range between our standard models He4.10 and He4.50. The
  brightest and faintest events have similar color histories to these
  typical models and are not plotted here. Instead, we show as dashed
  lines the lowest (He2.70) and highest ejecta mass (He8.00)
  models. The higher mass models, though rarer in nature, resemble the
  data for ($V-r$) and ($V-i$) better.  \lFig{carncolor}}
\end{figure}

\subsection{Spectra}
\lSect{spectra}

\Fig{spectra} shows the spectra for our models. Similar to
\citet{Shi19}, the spectra within 10 days of peak are rather uniform,
similar to the uniformity in colors in \Fig{color}. This is especially
true if the very light and very heavy models, those with
initial masses over 7 \Msun \ and below 2.9 \Msun \ are excluded from
the sample. The lighter models have experienced radius expansion and
are unusual because of it. The heavier models are fainter, rarer, and
may be unusually red. The differences that exist in the models are
mostly in the ultraviolet short of 4000 \AA\ where there is a
deficiency in emission, perhaps due to line blanketing. This same
deficiency is also seen in the similar (but non-LTE) models of
\citet{Des15}. The 5880 \AA\ line of He I here is not particularly
strong, even 10 days post-maximum, suggesting that all of these
supernovae could be identified as Type Ic. Helium is certainly present
in the models, ranging from 83\% of the ejected mass in He2.90 to 21\%
in He8.00.  Probably the weak line reflects the use of the LTE
approximation which does a poor job of reproducing the level
populations in a mixture of $^{4}$He irradiated by the radiation from
$^{56}$Ni decay \citep{Luc91,Des12,Woo97}. Other line identifications
can be made by comparison with Fig. 9 of \citet{Shi19} and Fig. 14 of
\citet{Des15}. The O I 7740 \AA\ line is apparent, but not strong, and
the 6200 \AA\ line may be due to Si II. No hydrogen was included in
the present study.

\begin{figure}
\includegraphics[width=0.48\textwidth]{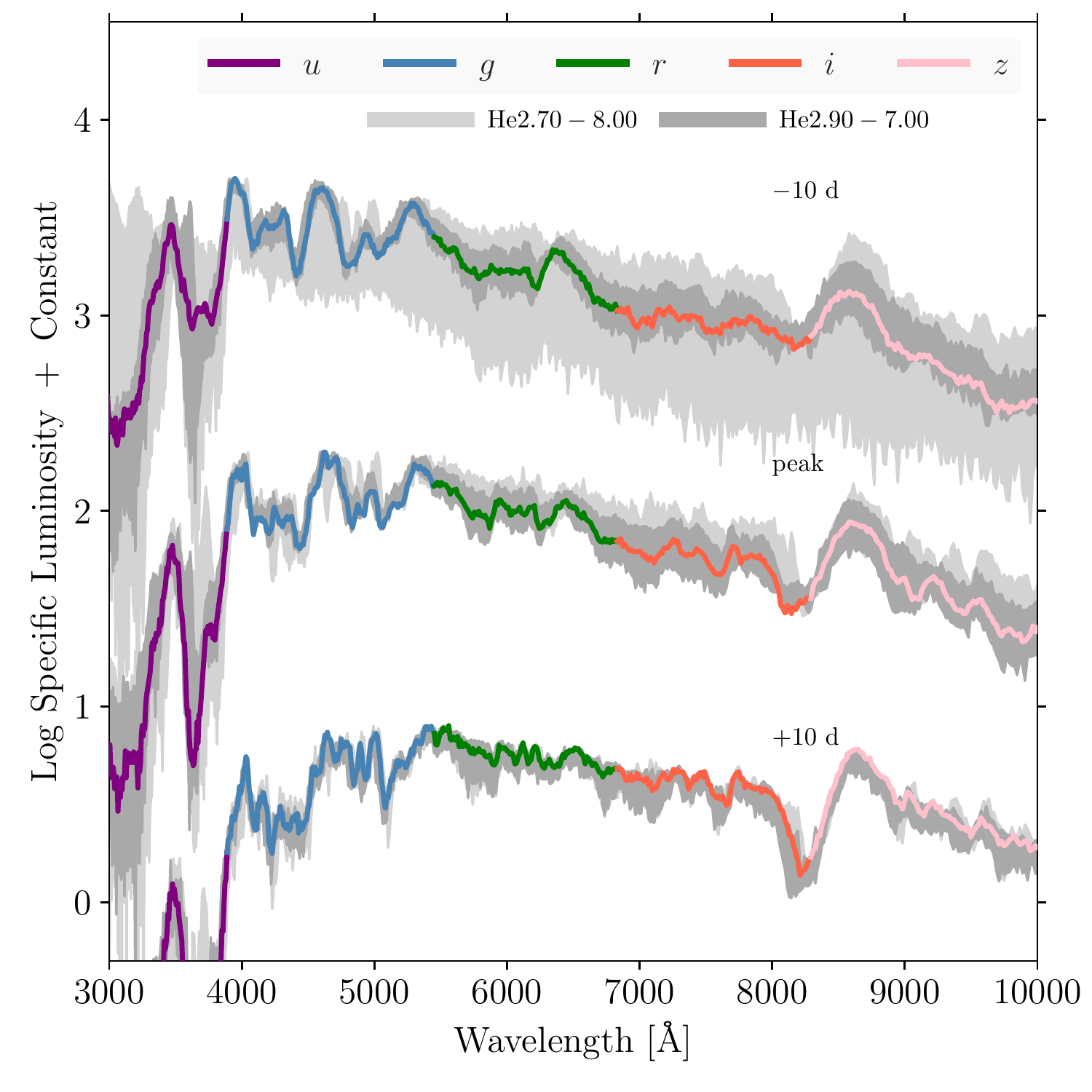}
\caption{Spectra for our models. For a subset of models spanning the
  lowest and highest masses, spectra are presented at the time of peak
  luminosity and 10 days before and after. The light gray range is
  bounded by Models He2.70 and He8.00, while the darker gray range
  spans between Models He2.90 and He7.00. A typical model, He4.10, is
  highlighted as a colored line. Except at low and high mass,
    the the models display little diversity, especially 10 days after
    peak.\lFig{spectra}}
\end{figure}

\subsection{Comparison with Observations}
\lSect{observations}

To illustrate the strengths and shortcomings of our models and to gain
a better perspective on their differences in peak luminosity, five
models were compared with five well-analyzed Type Ib and Ic
supernovae.  The events chosen were selected by observers (see
``Acknowledgments'') to be representative cases. No effort was made to
seek out the observations with which we agreed best. The five cases
span a range of luminosities from very faint (SN 2007Y) to quite
bright (SN 2009jf) and low mass to high mass. They include two events
with double peaks (LSQ13abf and SN 2008D) and one low mass Type Ic
event (SN 2007gr).

Spectra from the UV to the near IR for the 100 days following
explosion were calculated for all models using SEDONA and photometry was
obtained by numerically passing those spectra through various filters.
As discussed in \Sect{indices}, the supernova undergoes a major
adjustment about 20 to 30 days after the explosion (\Fig{color}) when
it ceases to be substantially ionized. In reality, gamma-rays from
radioactive decay may keep the gas at least partly ionized longer than
calculated in SEDONA, which assumes thermal equilibrium. This may
have the effect of maintaining something resembling a photosphere for
a longer time and make the supernova generally redder than calculated
here. We thus trust the SEDONA calculations of spectrum and photometry
only to about 20 - 30 days past peak, or about 50 days after the
explosion.

\begin{deluxetable*}{lrllll}
\tablecaption{Supernova Observations and Models}
\tablehead{ \colhead{}                     &
            \colhead{$t_{\rm peak}$}       &
            \colhead{$\log\ L_{\rm peak}$} &
            \colhead{$M_{\rm ej}$}         &
            \colhead{$E_{\rm exp}$}        &
            \colhead{$M_{\rm Ni}$}
            \\
            \colhead{}                     &
            \colhead{[day]}                &
            \colhead{[erg s$^{-1}$]}       &
            \colhead{[\Msun]}              &
            \colhead{[10$^{51}$ erg]}      &
            \colhead{[\Msun]}
            }\\
\startdata
\multicolumn{6}{c}{SN 2007Y}\\
\multicolumn{6}{c}{}\\
Str09    & 18       & 42.1                    &    $\cdots$    & $\sim0.3$           & 0.06                 \\    
Lym16    & $>13.6$  & 42.00                   & $1.4^{+1.3}_{-0.4}$ & $0.7^{+0.7}_{-0.3}$ & 0.04$^{+0.01}_{-0}$            \\
Pre16    & 18.8     & 42.01$^{+0.10}_{-0.12}$ &       $\cdots$     &  $\cdots$            & 0.051$^{+0.015}_{-0.013}$    \\
Tad18    & 18       & 41.90                   & 1.9                 & 0.6                 & 0.03                 \\
He3.40   & 19.5     & 42.01                   & 1.20                & 0.40                & 0.039                \\
He3.40Ni & 17.5     & 41.85                   & 1.20                & 0.40                & 0.025                \\  
\multicolumn{6}{c}{}\\
\multicolumn{6}{c}{SN 2007gr}\\
\multicolumn{6}{c}{}\\
Hun09    & $\sim14$ & 42.24                   & 2.0 - 3.5           & 1 -- 4              & 0.076$\pm$0.02        \\
Lym16    & $>13.5$  & 42.28                   & $1.8^{+0.6}_{-0.4}$ & $1.1^{+0.5}_{-0.4}$ & 0.08$^{+0.01}_{-0.01}$          \\
Pre16    & 13.1     & 42.30$^{+0.10}_{-0.08}$ &    $\cdots$             &   $\cdots$        & 0.073$^{+0.020}_{-0.013}$     \\
He3.90   & 22       & 42.15                   & 1.59                & 0.59                &  0.059               \\
\multicolumn{6}{c}{}\\
\multicolumn{6}{c}{SN 2009jf}\\
\multicolumn{6}{c}{}\\
Val11    & $\sim22$ & 42.62                   &  5 -- 7             & 5 -- 10            & 0.23$\pm$0.02         \\
Sah11    & 19$\pm$1 & 42.48$^{+0.08}_{-0.05}$     &  4 -- 9             & 3 -- 8             &  0.17$\pm$0.03        \\
Lym16    & $>20.5$  & 42.60                   & $4.7^{+1.7}_{-1.1}$    & $2.5^{+2.2}_{-0.9}$    & 0.24$^{+0.03}_{-0.02}$    \\
Pre16    & 21.3     & 42.68$^{+0.07}_{-0.06}$     &   $\cdots$          &  $\cdots$         & 0.271$^{+0.051}_{-0.035}$  \\
He5.38   & 21.5     & 42.48                   & 2.55                & 1.65                & 0.126                 \\
He5.38Ni & 23       & 42.61                   & 2.55                & 1.65                & 0.20                  \\
\multicolumn{6}{c}{}\\
\multicolumn{6}{c}{LSQ13abf}\\
\multicolumn{6}{c}{}\\
Str20    & 23       & 42.41                   & 5.94$\pm$1.1        & 1.27$\pm$.23        & 0.16$\pm$.02          \\
He6.0    & 24.5     & 42.23                   & 2.82                & 1.07                & 0.084
\enddata
\tablecomments{Hun09 = \citet{Hun09};Str09 = \citet{Str09}; Val11 =
  \citet{Val11}; Sah11 = \citet{Sah11}; Lym16 = \citet{Lym16}; Pre16 =
  \citet{Pre16}; Tad18 = \citet{Tad18}; Str20 = \citet{Str20}. The
  hydrodynamical model of \citet{Tad18} was used to get $M_{\rm ej}$,
  $E_{\rm exp}$, and $M_{\rm Ni}$.}  \lTab{snobs}
\end{deluxetable*}

\subsubsection{SN 2007Y}
\lSect{SN2007Y}

\Tab{snobs} and \Fig{2007Y} show the comparison between Model He3.40
and Type Ib SN 2007Y \citep{Str09}. The bolometric light curve 
for \Fig{2007Y} is taken from \citet{Tad18}.  Apparent magnitudes
given at \url{https://csp.obs.carnegiescience.edu/data/} were
converted to absolute magnitudes using the offsets at peak provided in
Tables 4 and 5 of \citet{Tad18}. Typical error bars at peak in the
absolute magnitudes are 0.37 magnitudes, dominated by the uncertain
distance. SN 2007Y is one of the faintest Type Ib supernovae ever
studied (\Fig{taddia}). 

Model He3.40 had a presupernova mass of 2.74 \Msun \ and ejected 1.2
\Msun \ with a kinetic energy of $4.0 \times 10^{50}$ erg. These
values are consistent with previous estimates based on model light
curves, and spectral analysis, all of which suggest that SN 2007Y was
a low-mass, low-energy event that produced less than typical $^{56}$Ni
(\Tab{snobs}). Based on its late-time spectrum, \citet{Str09}
estimated 0.44 \Msun \ of ejecta moving slower than 4500 km s$^{-1}$,
of which $\sim0.20$ \Msun \ was oxygen. Model He3.40 has 0.55 \Msun
\ moving slower than 4500 km s$^{-1}$, of which 0.13 \Msun \ was
oxygen. Total oxygen in the ejecta is 0.15 \Msun. All in all, the
model is similar to the 3.30 \Msun \ helium star suggested as a
prototype for SN 2007 by \citet{Str09}, although the presupernova mass
here is more like the 5.0 \Msun \ model of \citet{Woo95}.

\Fig{2007Y} shows the light curve of Model He3.40 compared with
observations. The unmodified model is too bright. Since our $^{56}$Ni
yield is an upper bound \citep[\Tab{main};][]{Ert20}, we are at
liberty to use a  smaller value while maintaining the overall
composition and dynamical structure of the standard
case. Consequently, we also calculated for comparison a version of
Model He3.40 with 0.025 \Msun\ of $^{56}$Ni instead of 0.039
\Msun\ (\Tab{snobs}). The agreement with both the bolometric
luminosity and the color magnitudes is greatly improved.



The calcium infrared triplet at around 8500 -- 8600 \AA is poorly
modeled. By default, the Sedona calculations treat all lines as purely
absorbing. This is a reasonable approximation for modeling the
pseudo-continuum of weaker lines from complex iron group species;
after a photon excites an iron ion, fluorescence will redistribute the
energy in cascades across numerous lines.  For strong transitions in
simple ions like the Ca II triplet, however, there are few
fluorescence channels, and the interaction is better modeled as purely
scattering. Sedona has the option to adjust epsilon, the ratio of
absorptive to absorptive plus scattering opacity, for each element
individually \citep{Kas06b}.  In the third panel of \Fig{2007Y}, the
curve labeled ``$\epsilon_{\rm Ca}=0$'' is a case for which the lines
of calcium, and only calcium were treated as purely scattering. This
resulted in substantial improvement with the observed spectrum. The
overall photometry was not greatly affected by this change however.

Another major discrepancy in the +12 day spectrum of SN 2007Y, is the
lack of strong features at 4900, 5800, 6600, and 6900 \AA.  The latter
three are probably lines of He I not replicated correctly in our LTE
study \citep{Shi19}.

A degeneracy in the models should be noted. An almost identically good
fit, including good spectroscopic agreement, could be obtained using
the Model He3.60 or Mode He2.90. Getting the characteristics at peak
right largely depends on the use of an appropriate $^{56}$Ni mass, and
the mapping between $^{56}$Ni and presupernova mass is not monotonic
(\Tab{main}). Similarly, the time of bolometric peak which sets the
light curve width, only varies by 3 days for the He2.90, He3.40, and
He3.60 models.

\begin{figure}
  \includegraphics[width=0.48\textwidth]{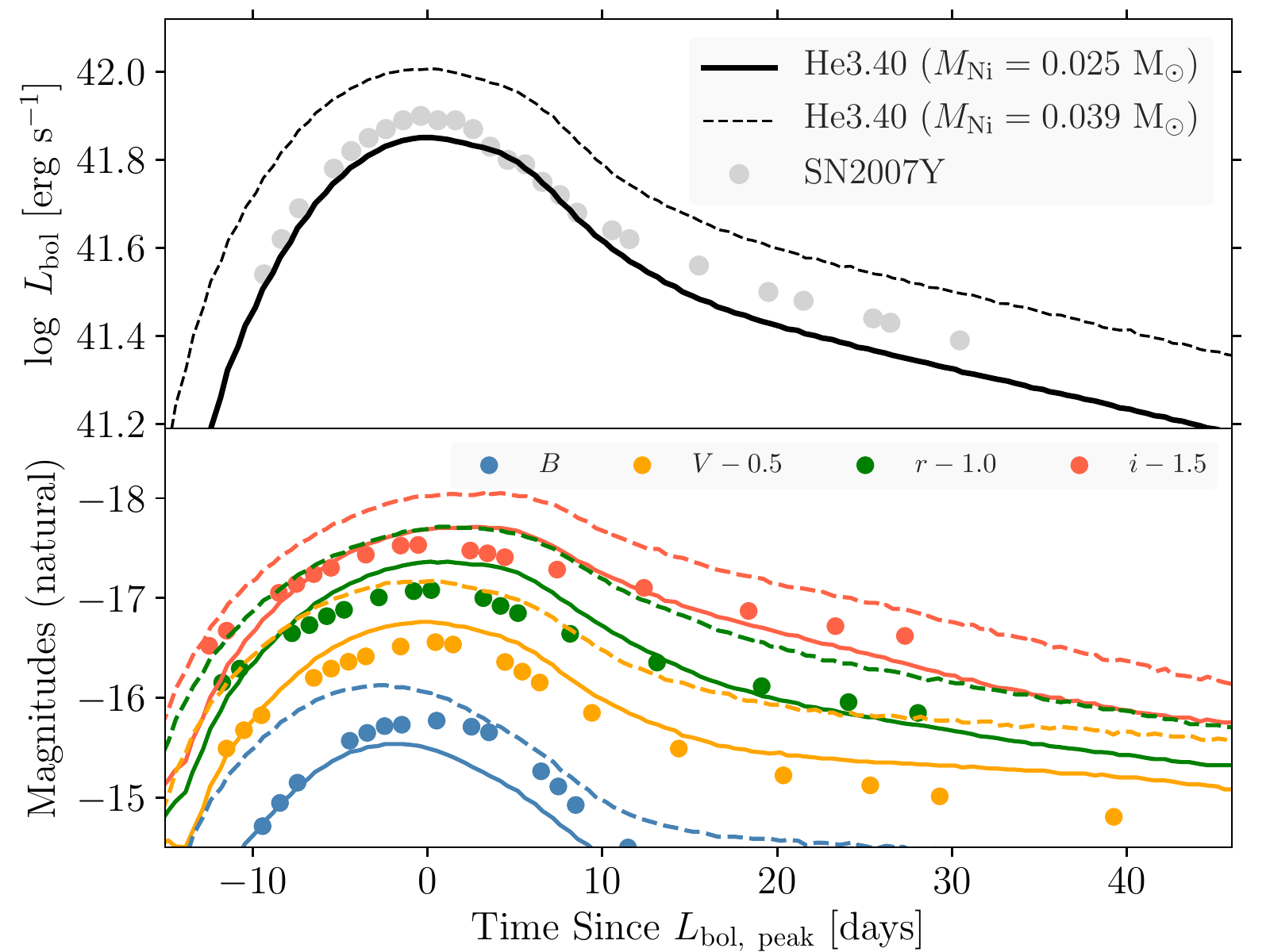}
  \includegraphics[width=0.48\textwidth]{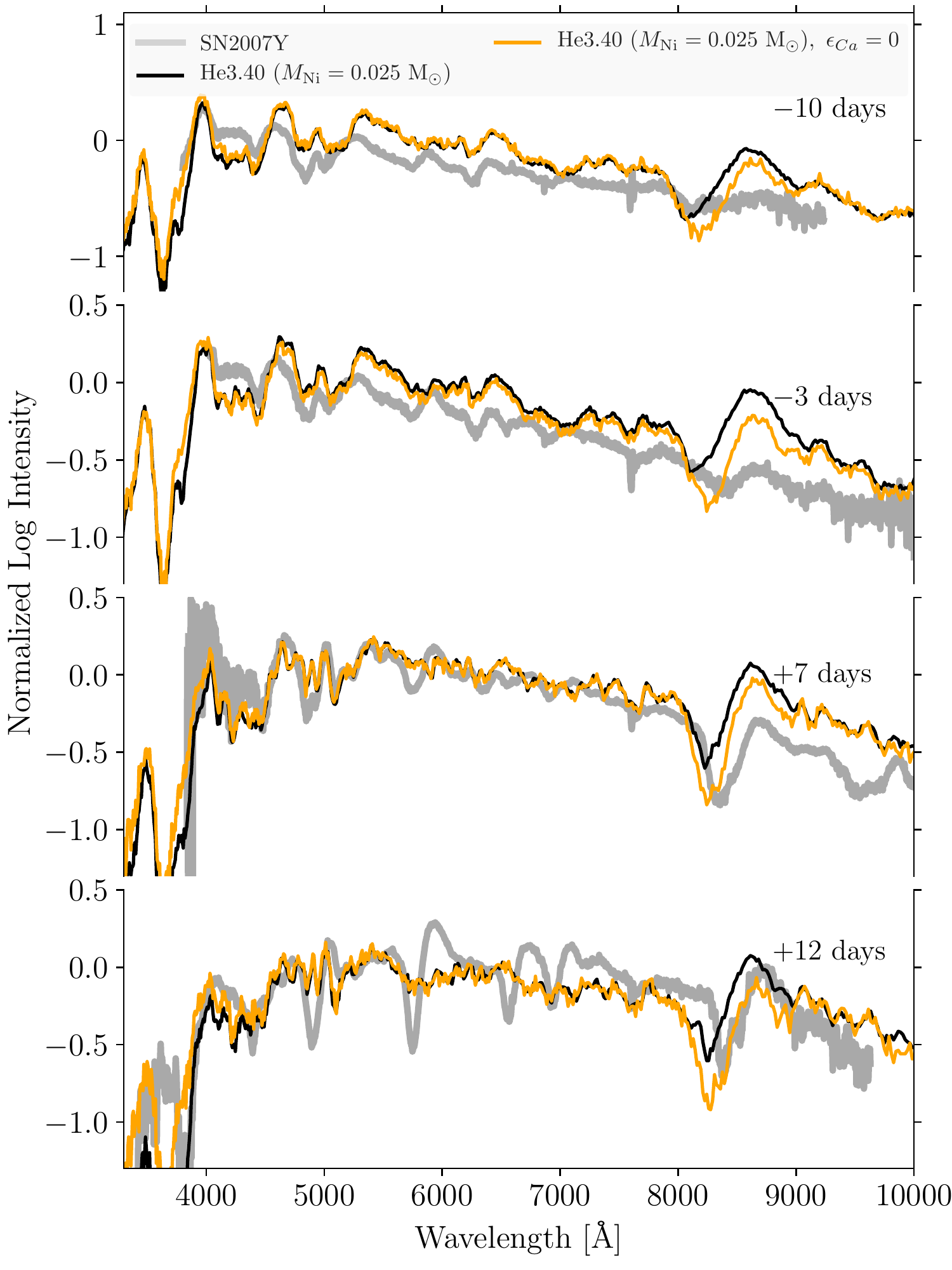}
  \caption{Bolometric luminosity, color magnitude evolution, and
    spectra for SN 2007Y \citep{Str09} compared with the standard
    Model He3.40 (0.039 \Msun; \Tab{main}; dashed line) and an
    equivalent model with with a reduced $^{56}$Ni mass (0.025 \Msun;
    solid lines in the top two panels). The model and observed
    spectra are normalized to give the same value at 5200 \AA.  The
    model and observations are in good agreement for the interval
    given for the smaller $^{56}$Ni mass. \lFig{2007Y}}
\end{figure}

\subsubsection{SN 2007gr}
\lSect{SN07gr}

SN 2007gr is one of the closest stripped-envelope supernovae ever
discovered. Measurements by \citet{Val08} gave an absolute
  $R$-band magnitude, at maximum, of $-17.3$, an estimated $^{56}$Ni
  mass of 0.07 - 0.1 \Msun, and showed compelling evidence for carbon
  in the spectrum. Later measurements (\Tab{snobs}) gave similar
characteristics. \citet{Mau16} estimate that the progenitor of SN
2007gr was $M_{\rm ZAMS} \sim 30$ \Msun.

Here we take for comparison the unmodified Model He3.90 with
characteristics given in \Tab{main} and \Tab{snobs}. For the standard
mass loss rate, $f_{\rm WR = 1}$, the main sequence progenitor of this
model would have been 18 \Msun, though larger progenitors could be easily
tolerated if the mass loss rate were greater. The important quantity
again is the presupernova mass, 3.09 \Msun. The comparison is given in
\Fig{2007gr}.  Digitized photometry and spectra were provided by Simon
Prentice and were corrected for redshift, reddening, and
extinction. Photometric data is from \citet{Hun09} and spectra are from
\citet{Val08} and \citet{Mod14}. Magnitudes for both the model and
observations are expressed in the Vega system. The luminosity plotted
from the model is just the portion in the 4000 - 10000 \AA\ range, which
is appropriate for the comparison.

\begin{figure}
  \includegraphics[width=0.48\textwidth]{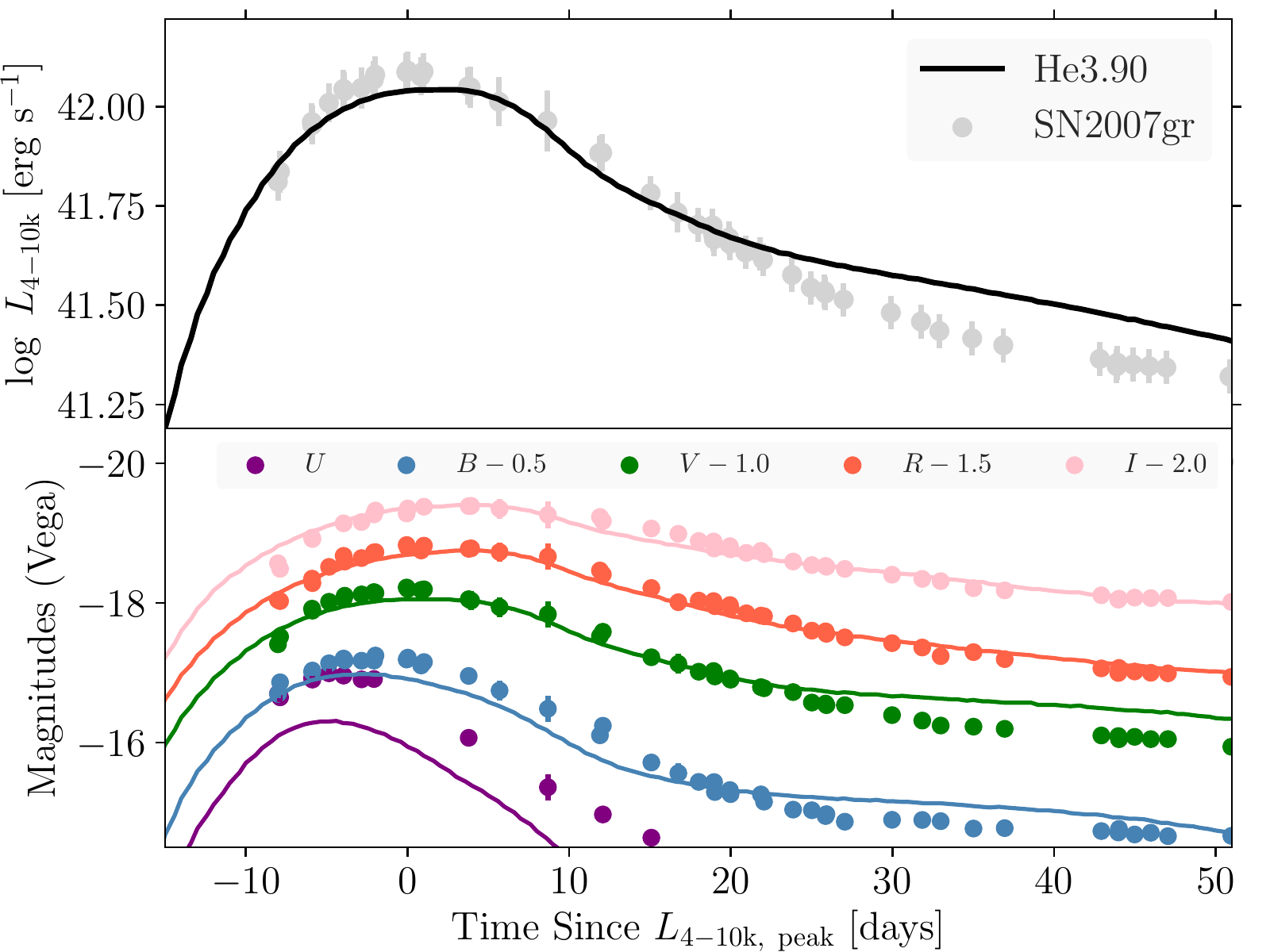}
  \includegraphics[width=0.48\textwidth]{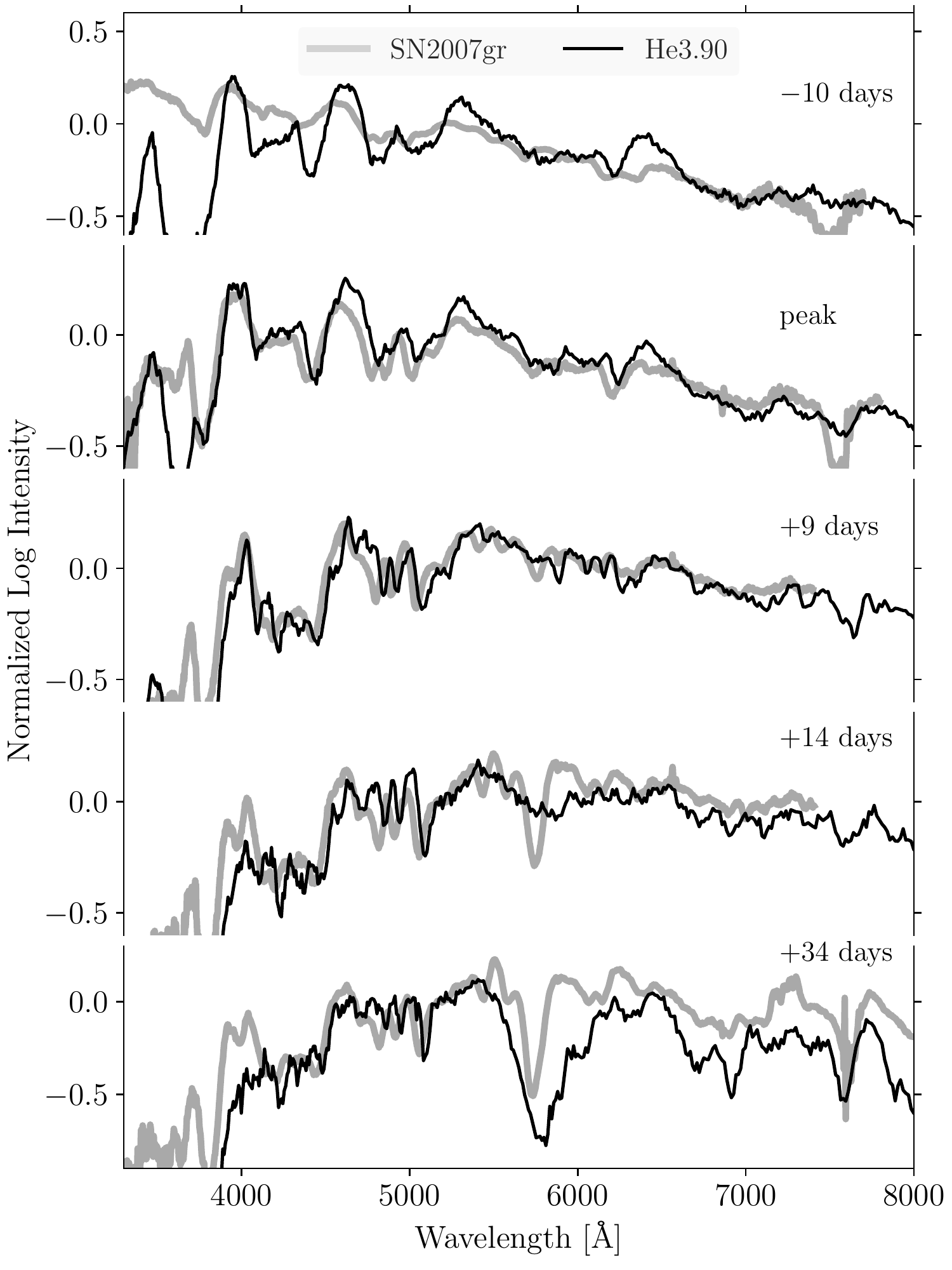}
  \caption{Bolometric luminosity \citep{Pre19}, color magnitude
    evolution \citep{Hun09}, and spectra \citep{Mod14,Shi19} for SN
    2007gr, a Type Ic supernova, compared with Model He3.90
    (\Tab{main}). The model and observations are in good agreement for
    the interval given. \lFig{2007gr}}
\end{figure}

Overall the agreement is excellent even though the model has a
significantly longer time to peak, lower ejected mass, lower explosion
energy, and smaller $^{56}$Ni mass than previously estimated by the
observers. Except for the time of peak, the discrepancy is not
substantial. Our peak time includes 5 days spent at low optical
luminosity ($\lesssim\ 2 \times 10^{41}$ erg s$^{-1}$) before the main
diffusion dominated display commenced. The time of peak luminosity is
also not very precisely defined. From 17 to 25 days after the star's
initial explosion, the bolometric luminosity varied only 3\% from its
peak value. The ratio $M_{\rm ej}^3/E$ (in units of \Msun \ and
10$^{51}$ erg) was 6.8 for the model, but varies from 1.39 to 0.35 for
an assumed constant opacity 0.05 to 0.1 cm$^2$ g$^{-1}$ for the
observations \citep{Pre16}. Our opacity is not constant and our time
to peak is quite different. Based on nebular spectroscopy
\citet{Maz10} estimates that $\sim1$ \Msun \ of heavy elements were
ejected. Model He3.90 ejected 0.62 \Msun \ of heavy elements (besides
neon) and 0.97 \Msun \ of helium and neon. A model with larger mass
loss and the same presupernova mass would probably have the same
ejecta mass, but more of its ejecta in the form of heavy
elements. Also consistent with \citet{Maz10}, the intermediate mass
elements in the model constitute 0.17 \Msun \ of the ejecta and the
carbon to oxygen ratio is about 1/3.  \citet{Maz10} says $\sim0.1$
\Msun\ and a ratio of $<1/5$. The latter ratio may be sensitive to
convection physics and the rate for $^{12}$C($\alpha,\gamma)^{16}$O.

The model spectrum again displays a striking deficiency in the
ultraviolet, though the $B$ magnitude appears well behaved. Unlike
models for SN 2007Y and LSQ13abf, the pseudo-bolometric luminosity
(4000 - 10000 \AA) and the absolute magnitudes seem to be
consistent. About 20 days after peak, the model spectrum begins to
exhibit prominent features not apparent in the observations and we no
longer trust the model. The bolometric luminosity should remain
accurate til later though.

\subsubsection{SN 2009jf}
\lSect{SN09jf}

We next consider a supernova on the brighter end of the observed
distribution, SN 2009jf. Observations in \Tab{snobs} suggest a
luminosity and $^{56}$Ni mass both greater than our fiducial limits of
10$^{42.5}$ erg s$^{-1}$ and 0.15 \Msun.  Our brightest model with
standard mass loss is Model He5.38, with a peak bolometric luminosity
of 10$^{42.45}$ ergs$^{-1}$. Indeed, this is the brightest model for
any central engine we considered \citep{Ert20}.

\citet{Val11} suggest that SN 2009jf arose from a main sequence star
of 25 - 30 \Msun. The main sequence mass that produced Model He5.38
was 22 \Msun, but the usual uncertainty regarding mass loss applies
More mass loss implies a bigger main sequence progenitor.

\Fig{2009jf} and \Tab{snobs} show the comparison between Model He5.38
and observations.  The photometry and spectra were again provided by
Simon Prentice and have been corrected for redshift, extinction and
reddening. The pseudo-bolometric luminosity is shown which should be
compared with our 4000 - 10000 \AA\ evaluation.  The original
photometry is from \citet{Val11} and the spectra are taken from
\citet{Val11} and \citet{Mod14}. Despite the inadequate luminosity,
the agreement is pretty good. It can be made better by artificially
invoking the synthesis of 0.20 \Msun \ of $^{56}$Ni. Consistent with
the estimates of what is needed in \Tab{snobs} this is a 60\% increase
for the standard model (0.126 \Msun). Interestingly, the spectrum is
not altered very much by this enhancement at times before and during
peak, but is appreciably better with enhanced $^{56}$Ni at later
times. The enhanced nickel maintains ionization to a later time and
keeps the photosphere from collapsing.

While the improvement resulted here from augmented nickel, probably
any source of centrally concentrated, energetic radiation would have a
similar effect.  It is an interesting question whether there might be
spectroscopic signatures of a point source (magnetar) in a mixed
supernova that also contains radioactivity. Whether that is all
meaningful in the context of a one-dimensional model could be
questioned. It does seem though that the spectrum and light curve are
better explained by an augmented central source than by,
e.g., circumstellar interaction.

\begin{figure}
  \includegraphics[width=0.48\textwidth]{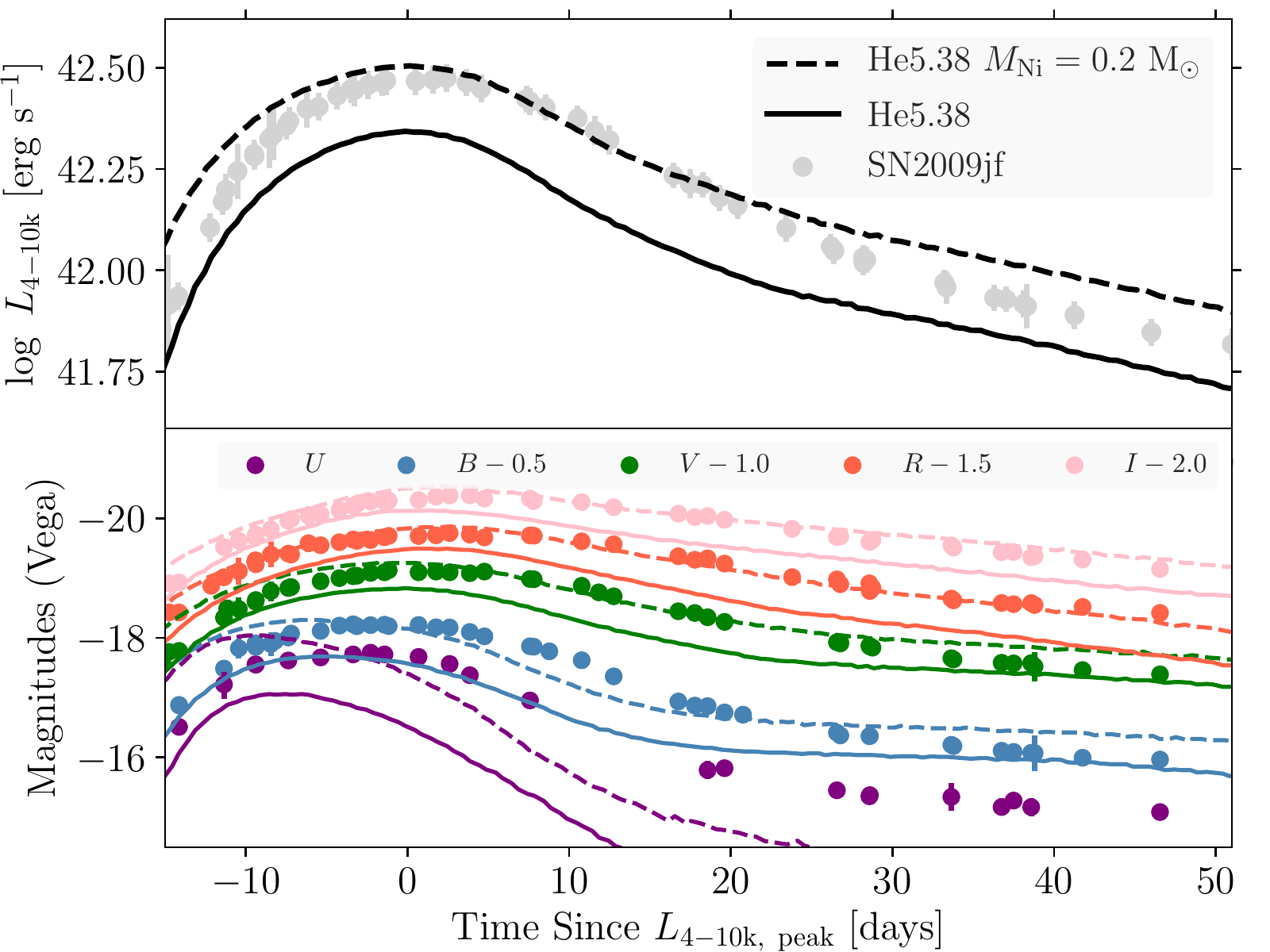}
  \includegraphics[width=0.48\textwidth]{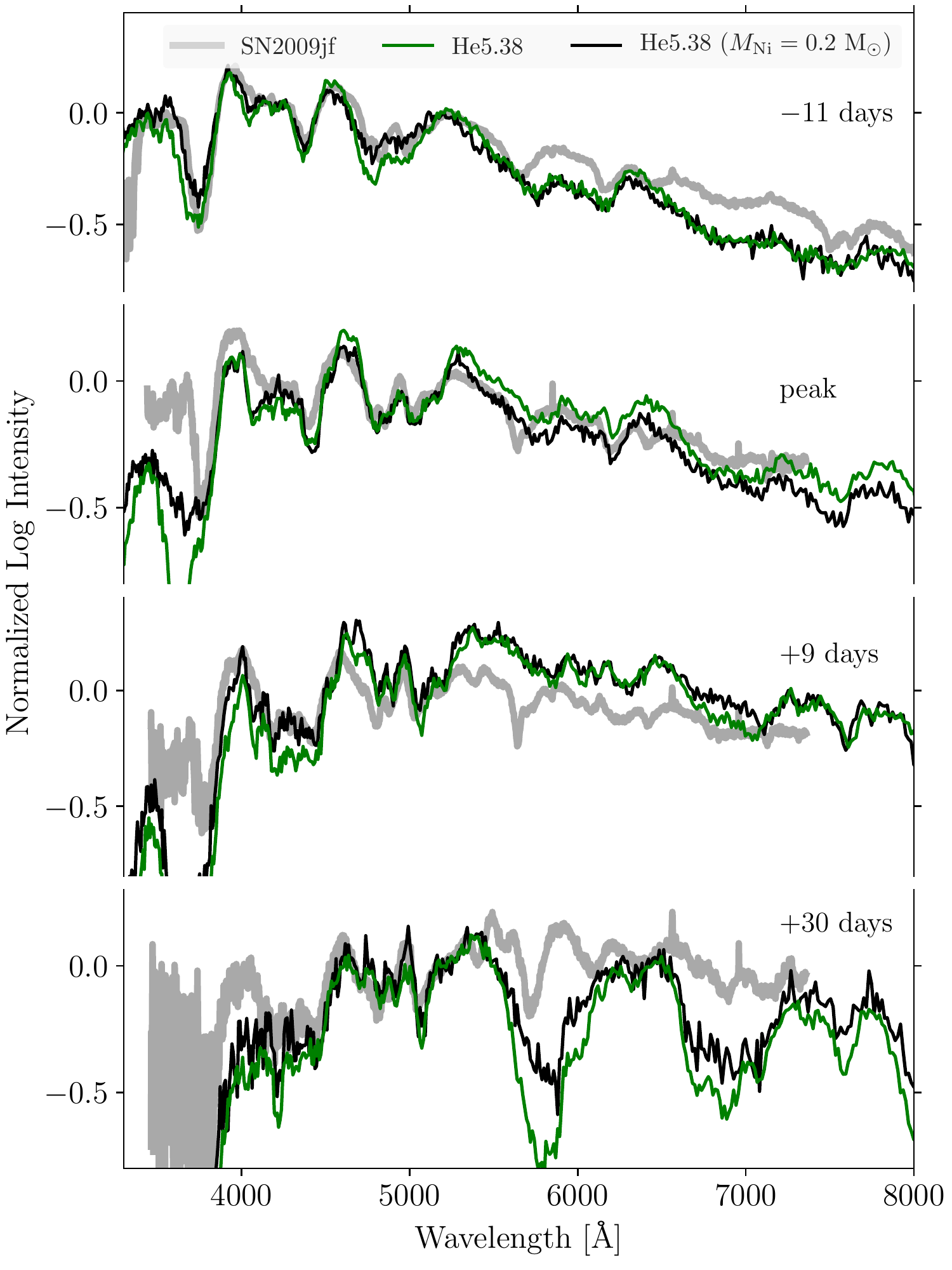}
  \caption{Bolometric luminosity \citep{Pre19}, color magnitude
    evolution, and spectra for SN 2009jf \citep{Val11}, a Type Ib
    supernova, compared with Model He5.38 (\Tab{main}). Solid lines in
    the upper two frames are the unmodified model which produced 0.126
    \Msun \ of $^{56}$Ni. Dashed lines show the effect of increasing
    the $^{56}$Ni yield to 0.20 \Msun \ to improve agreement. The
    model and observations are in good agreement for times up to 10
    days after the peak. By 30 days after peak, the model has broken down 
    and the spectrum cannot be trusted. \lFig{2009jf}}
\end{figure}

From an analysis of the nebular spectrum, \citet{Sah11} estimated a
lower limit to the oxygen ejected in SN 2009jf of $\sim1.34$
\Msun. Model He5.38 only ejected 0.72 \Msun. There are three
possibilities: a) their measurement could be an overestimate; b) SN
2009jf came from a more massive presupernova star; or c) our
presupernova mass is correct, but SN 2009jf was derived from a more
massive star that experienced greater mass loss so that much of what
was helium in Model He5.38 (0.97 \Msun) was actually $^{16}$O. With
regards to b), Model He7.13 (\Tab{main}) fits the observations near
peak reasonably well (with the same augmentation to $^{56}$Ni) and
ejects 1.35 \Msun \ of $^{16}$O, but its light curve is overly broad
and it declines too slowly compared with the observations. A larger
explosion energy would help, but that is not a free parameter in the
present study. With regards to c), Model He12.00x2, a 12 \Msun
\ helium core evolved with twice the mass loss rate ($f_{\rm WR} = 2$)
has almost the same presupernova mass (4.03 \Msun) as He5.38 (4.05
\Msun) and is thus expected to have similar luminosity, color, and
time scale. The presupernova model for He12.00x2 contained 1.09 \Msun
\ of $^{16}$O. This would change slightly in the explosion.

\subsubsection{LSQ13abf}
\lSect{LSQ13}
  
Heavier supernovae have broader, redder light curves.  \Fig{LSQ} shows
a comparison of Model He6.00 (presupernova mass 4.44 \Msun) with Type
Ib supernova LSQ13abf as reported by \citet{Str20}. Photometry for
LSQ13abf is taken from their Table 2, where magnitudes are given in
the CSP natural system. These apparent magnitudes are translated into
absolute magnitudes using a distance modulus for SDSS
J114906.64+191006.3 of 34.96 and a correction for extinction by the
Milky Way of $A_V = 0.087$ mag assuming reddening law characterized by
$R_V = 3$ \citep{Str20}. The resulting values agree with what is
plotted in Fig. 3 of \citet{Str20}. Assuming that LSQ13abf exploded on
JD 2456395.80, yields the values compared with Model He6.00 in
\Fig{LSQ}.

Model He6.00 agrees quite well with the observations except during the
first week when factors other than the expansion of a compact
Wolf-Rayet star are clearly at play. The similarity with SN 2008D
\citep{Str20} naturally leads to the suspicion of an extended envelope
or circumstellar interaction, as was invoked by many to explain the
very early observations of that event
\citep{Ber13,Che08,Kat10,Sap13,Pir15,Des18,Iok19}.  An extended
envelope is a natural solution \citep{Des18,Woo19,Ert20}, especially
given the possibility of degenerate silicon flashes that can
occasionally eject substantial amounts of matter during the weeks and
months prior to the final explosion. For mild flashes, as in Model
He3.20, the effects are difficult to distinguish from ordinary
envelope inflation during earlier burning stages. Even Model He3.20 is
too faint, especially in the red, to give the early light curve of
LSQ13abf (\Fig{LSQ}). The deep dips in the light curve when the
supernova is only a few days old, might be weaker in a
multidimensional model or one where mixing was treated more
realistically. Even then a greater mass at a larger radius seems to be
required.

The problem is that for our presupernova model set \citep{Woo19} both
radius inflation and silicon flashes are absent in models more massive
than He3.20 (presupernova mass 2.7 \Msun), yet the width of the light
curve at peak seems to demand a greater mass. We are thus unable to
fit the entirety of the LSQ13abf light curve. The situation might
change if the envelope of the presupernova star for LSQ13abf,
  and similar supernovae like SN 2008D contained a trace of hydrogen
  \citep{Des18}. This is well worth pursuing. Could a bit of high
velocity hydrogen be hiding in the spectrum of many of these
supernovae that show evidence for radius inflation? All our models
are, by construction, hydrogen free, so we do not see this
effect. \citet{Des18} also described a single helium star model,
He\_R173, with large presupernova radius (173 \Rsun) and moderate
presupernova mass (2.73 \Msun). Our equivalent model in terms of
presupernova mass, Model He3.40, had a much smaller radius, 7.7 \Rsun,
but this star is near the boundary between large expansion and
small. Different treatments of semiconvection \citep[][have
  less]{Des18} and opacities might explain the dichotomy. In any case,
their Model He\_R173, like our He3.20, was only capable of explaining
the early evolution of the light curve and not its broad width.

If LSQ13abf did not contain hydrogen then something else may be going
during that first week.  Perhaps a magnetar is providing extra energy
the first few days and becoming less important later on \citep{Kas16}?
Or perhaps there was circumstellar interaction. Our models have
sufficient surface resolution to estimate the energy in mildly
relativistic ejecta, $\beta > 0.1$.  For Model He6.00, the outer $\sim
5 \times 10^{-4}$ \Msun \ had velocity greater than 30,000 km
s$^{-1}$, corresponding to a kinetic energy in excess of $5 \times
10^{48}$ erg. The total energy in the early light curve sampled during
the first 10$^6$ seconds of the evolution of LSQ13abf is of order
10$^{48}$ erg. Provided the fast moving ejecta encounters its own mass
inside $\sim3 \times 10^{16}$ cm, the energetic needs of the light
curve could be satisfied. Assuming a wind or pre-explosive ejection
speed of 1000 km s$^{-1}$, this requires a mass loss rate of
$\sim10^{-4}$ \Msun \ y$^{-1}$. This is much more than our standard
Wolf-Rayet mass loss rate, but perhaps not too unusual for the last 10
years of the star's life when advanced burning stages and
semi-degenerate flashes are in progress.

In any case, \Fig{LSQ} shows the good agreement between the
observations of LSQ13abf, especially for $B$ and $V$ near peak, and
Model He6.00. The bolometric luminosity at peak is only
65\% of the observed value \citep[\Tab{snobs} here and][]{Str20}, but half
of that difference might reasonably be accommodated by a slight
increase in the $^{56}$Ni mass from 0.084 \Msun \ to say 0.11
\Msun. This would also brighten the photometry in the $B$ and $V$
bands, but perhaps not unacceptably.

\begin{figure}
\includegraphics[width=0.48\textwidth]{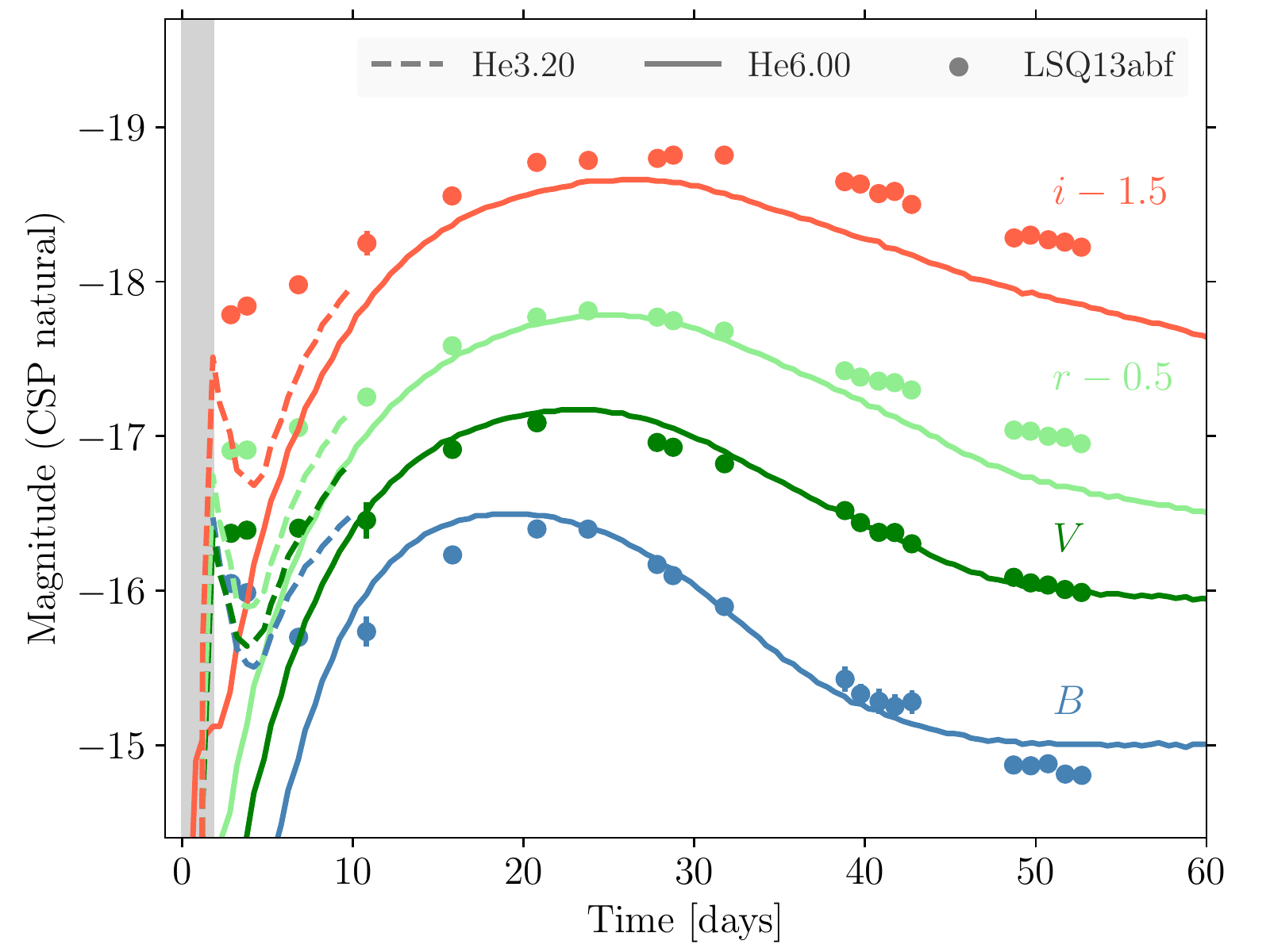}
\includegraphics[width=0.48\textwidth]{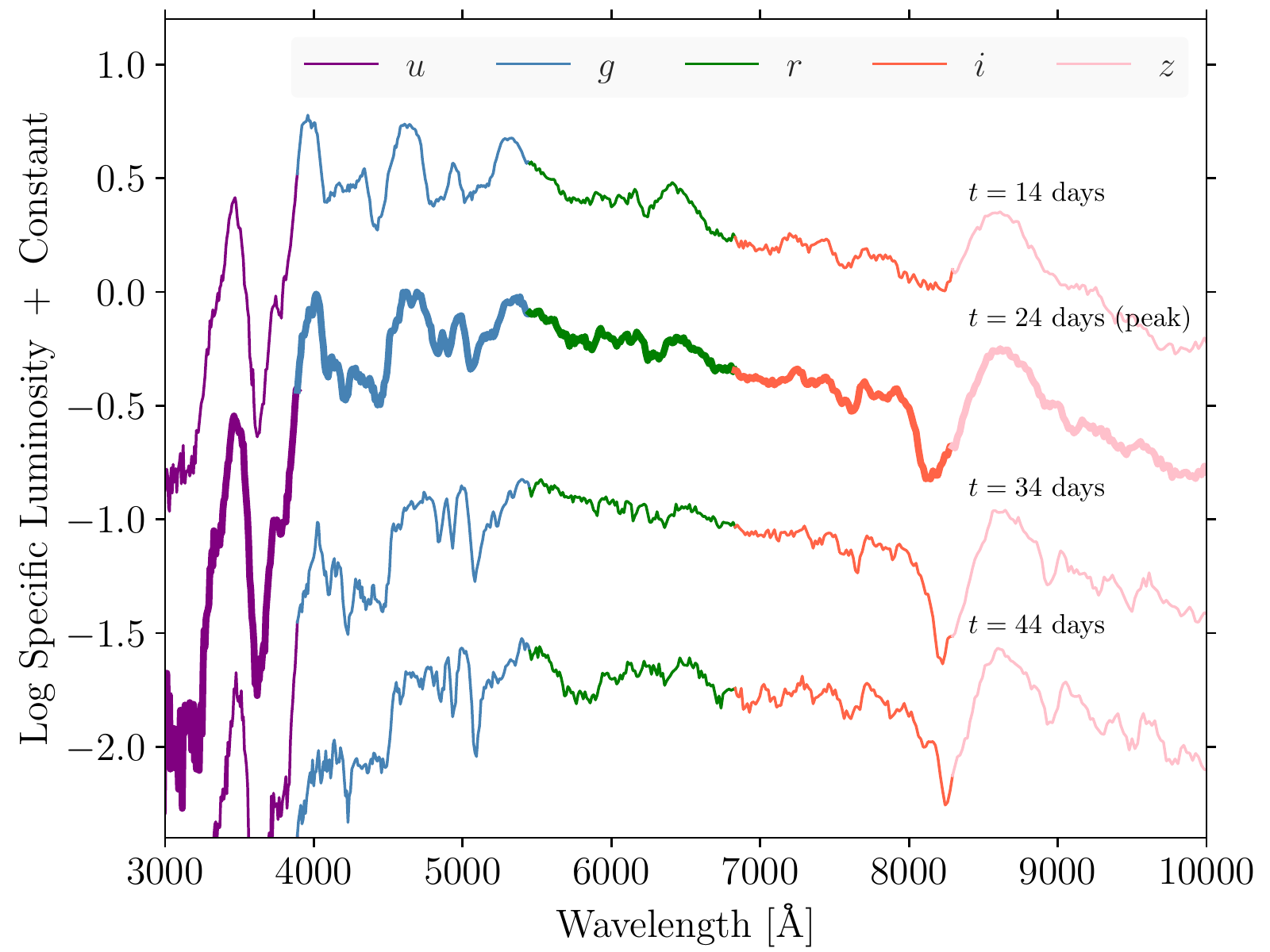}
\caption{Comparison of Model He6.00 with supernova LSQ13abf.  (top:)
  Photometry of the supernova LSQ13abf \citep{Str20} compared with
  Model He6.00 (solid line). The fit is remarkably good after the
  first week in all bands except $i$. The early light curve may be due
  to envelope expansion and the first 10 days of Model He3.20 are
  plotted for comparison (dashed line). (bottom:) The spectrum of
  Model He6.00 calculated using SEDONA at bolometric peak (24 days)
  and 10 days before and 10 and 20 days after. To facilitate
  comparison with the light curve, the spectrum has been color coded
  with the filters indicated corresponding to the approximate
  wavelength range for SDSS $ugriz$ filters.  \lFig{LSQ}}
 \end{figure}

\subsubsection{SN 2008D}
\lSect{sn08D}

Although similar to LSQ13abf in appearance and possibly in
explanation, SN 2008D is worth separate mention because it is so well
studied and has an accurately determined explosion time. Like
LSQ13abf, the light curve has two peaks and is very bright during the
first week possibly due to a large radius for the presupernova star
or a recently ejected shell. Here apparent magnitudes are taken from
\citet{Bia14} in standard (Vega) magnitudes. These are converted to
absolute magnitudes by normalizing the peak $V$ magnitude to
−17.0$\pm$0.3 and $B$ magnitude to −16.3$\pm$0.4. These values account
for a substantial host reddening ($E(B - V)$ = 0.6$\pm$0.1 mag). An
explosion date of MJD 54474.56 is assumed \citep{Sod08}.

\begin{figure}
\includegraphics[width=0.48\textwidth]{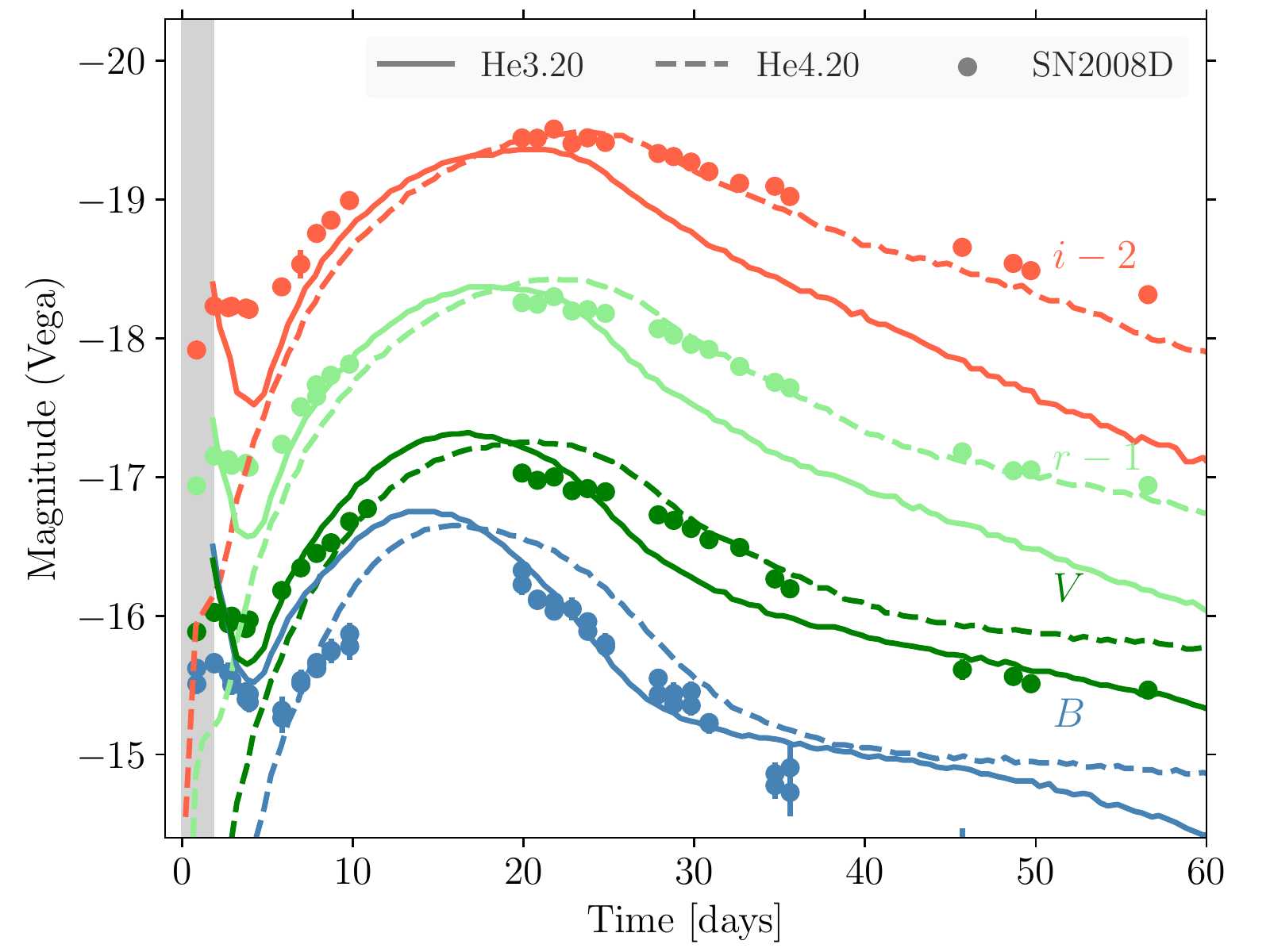}
\caption{Comparison of Model He3.20 (solid line) and He4.20 (dashed
  line) supernova 2008D. Model He3.20 experienced a weak silicon flash
  and substantial radius expansion before exploding and that accounts
  for its brilliance at early times. The model overall gives too
  narrow a light curve however, and Model He4.20 is a better fit to
  the peak. \lFig{sn2008d}}
 \end{figure}

\citet{Mod09} estimate that SN 2008D reached its bolometric peak at
19.2$\pm$0.3 days post-explosion with a luminosity of
$10^{42.2\pm0.1}$ erg s$^{−1}$.  \citet{Mal09} give 10$^{42.3}$ erg
s$^{-1}$.  \Fig{sn2008d} shows a comparison of the $BVri$ light curves
of SN 2008D and Models He3.20 and He4.20. Model He3.20 has a peak
luminosity of 10$^{42.22}$ erg s$^{-1}$ at 16.5 days. Model He4.20 has
a peak luminosity of 10$^{42.22}$ erg s$^{-1}$ at 21.0 days which
agrees better with observations. The display from He3.20 might be made
brighter and redder during the first week if it had a larger initial
radius or an envelope containing more mass. Both could come from an
earlier or more energetic silicon flash.

The problem with SN 2008D is the same as for LSQ13abf, though not
as severe. Models with significant radius expansion give light
curves that are too narrow to explain the peak. Models that explain
the peak fail to reproduce the first week's data. A more massive model
with a supergiant structure or shell ejection is needed \citep[see
  also \Sect{LSQ13} and][]{Des18}. The presence of residual hydrogen
near the surface might facilitate this expansion for the more massive
stars. Other similar ``double-peaked'' supernovae with ambiguous
evidence for high velocity hydrogen might be noted, e.g.  SN 2013ge
\citep{Dro16}.

\section{Low Mass Explosions}
\lSect{fastblue}

For presupernova masses less than about 2.6 \Msun \ (standard
  initial helium star masses less than 3.3 \Msun), advanced burning
stages are punctuated by off-center burning and, at the lower end, by
degenerate flashes \citep[Table 7 of][]{Woo19}. For presupernova
masses less than about 2.1 \Msun, electron-capture supernovae may
occur. While presupernova masses less than 2.6 \Msun \ should be
common in nature, their properties - low explosion energy, small
$^{56}$Ni mass, and small ejected masses - make them unlikely to
appear as common Type Ib and Ic supernovae
\citep{Tau15,Mor17,Woo19,Ert20}. But what would explosions in
presupernova stars with masses in this range look like?

Major uncertainties for those models that experience a strong silicon
flash: He2.50, He3.00, He3.10, and He3.20, are the energy and
timing of the flash \citep{Woo19}.  If the flash is powerful, it
ejects a lot of mass and, since the recovery time for the remaining
core is longer for a strong flash, this matter coasts to a large distance
before the iron core finally collapses. In extreme cases, the ensuing
supernova is powered by circumstellar interaction and can be very
bright. For the weaker flashes, the star ejects less mass at slow
speed. The effective radius is not greatly increased, and the
supernova is similar to what would have happened without a flash - a
faint, fast Type Ib supernova.

The version of SEDONA employed here is incapable of treating
supernovae powered by shock interaction. Homologous coasting is
assumed. For illustration, we thus consider only two cases. Model
He2.50 is characteristic of stars where the silicon flash has a
moderate influence, leading to the substantial expansion of the outer
layers of the presupernova star prior to iron core collapse, but not
to the extent ($\sim10^{15}$ cm) where circumstellar interaction
becomes a major contributor to the light curve. The other, Model
He2.60, is a case where the silicon flash is avoided, or very weak,
but the star already has an initial radius of over 100 \Rsun. As we
shall show, both have similarities to one another and to SN 2010X and
similar fast blue transients. These stars are similar to He3.20 which
we treated as an ordinary Type Ib supernova, but are treated
separately here because their lower explosion energies, small
$^{56}$Ni production, and large radii cause them to appear
qualitatively different.


Model 2.5B as defined in Table 7 and Fig 14a of \citet{Woo19},
experienced a silicon flash 19 days before core collapse and ejected
0.25 \Msun \ with a typical speed of just a few hundred km
s$^{-1}$. By the time the shock from core collapse arrived at the
surface, the outer edge of the ejected matter had only coasted to $\sim7
\times 10^{13}$ cm. The subsequent explosion thus resembled what would
have happened in a helium red supergiant with an unusually large
radius. The shock interaction in the outer layers was different (the
shock decelerated rather than accelerating, generating a strong reverse
shock), but the light curve was similar. The main shock took about a
day to reach the photosphere and, after expanding an additional two days,
was coasting approximately homologously. At that point the KEPLER
model was linked into the SEDONA code.  Except for a very thin high
velocity layer at the surface, the exploding star was all 
optically thick at this point

The results of the explosion \citep{Ert20} for this model and several
variations are shown in \Fig{he2.5}. For the standard case, the total
ejected mass was 0.737 \Msun\ including that already ejected in
  the silicon flash; the kinetic energy was $1.1 \times 10^{50}$ erg
and 0.0151 \Msun \ of $^{56}$Ni was produced. The light curves in the
$U$, $B$, $V$, and $r$ bands are shown. Shock break out and early
expansion were not followed. A plateau, as shock deposited
  energy is released by helium recombination and diffusion, is
followed by the abrupt fall off to a faint radioactive tail. On
the plateau typical photospheric temperatures, from KEPLER, ranged
from 16000 K on day 5 to 10000 K on day 15 and the photospheric
  velocity ranged from 12,000 to 4,000 km s$^{-1}$ in the same
  period.  A similar model was considered by \citet{Kle18a}, but here
we use different approach to model the presupernova evolution and
explosion. The radius is larger due to the silicon flash and the
kinetic energy and $^{56}$Ni masses were computed in a less
parametrized model for the explosion. Also shown on the plot are data
points for the $r$-band light curve of SN 2010X. Data is taken from
\citet{Kas10} with Julian date 55231.8 days in their table
corresponding to day 0 on the plot, and with a distance modulus of
$34.11$.  (The Open Supernova Catalog;
\url{https://sne.space/sne/SN2010X/}).

The unmodified model is similar to the $r$-band light curve of SN
2010X, but a bit too faint and expands too slowly at late times
\citep{Kas10}. Reasonable modifications can improve the
fit. \citet{Kle18a} obtained a similarly good fit to the $r$-band
light curve with a dialed-in explosion energy of $1.0 \times 10^{51}$
erg and 1.33 \Msun \ of ejecta (their Model M2.73-E1). The
neutrino-transport models of \citet{Ert20} that underlie the present
study suggest an explosion energy of, at most, a few $\times 10^{50}$
erg for this low mass range. Even with an ejected mass half as large,
we would be unable to approach the brilliance of SN 2010X using the
radii adopted by \citet{Kle18a}. The large radius from the silicon
flash helps substantially, and a still larger radius is well within
the range of possible silicon flash characteristics, but this does not
address the low velocity.

Raising the explosion energy from 0.11 to 0.3 $\times 10^{51}$ erg
helps with both issues and, while not specifically what \citet{Ert20}
calculated for this star, a higher energy is found in other cases
with low mass, e.g., Model He3.00 had an explosion energy of 0.3
$\times 10^{50}$ erg and also experienced a silicon flash. Using this
energy and a slight adjustment to the $^{56}$Ni synthesis yields a
better fit to the $r$-band light curve of SN 2010X and raises the
velocity by about 50\%.


One merit of fitting SN 2010X this way is that similar models might
fit other more luminous supernovae with rapid rise and decline rates.
Explaining supernovae like SN 2002bj \citep{Poz10,Kas10,Kle18b} is
difficult because the $^{56}$Ni abundance required to explain its peak
luminosity, if the supernova light curve is powered by radioactivity,
would also produce a bright tail. If the luminosity at peak is
determined by recombination releasing shock-deposited energy
  however, then one is at liberty to vary $^{56}$Ni to fit the
tail. SN 2002bj had a similar rapid rise and decline to SN 2010X, but
was 1.5 magnitudes brighter and seemingly lacked a radioactive tail.
A good first approximation to that event might be Model 2.5C of
\citet[][his Fig. 14]{Woo19}.

\begin{figure}
\includegraphics[width=0.48\textwidth]{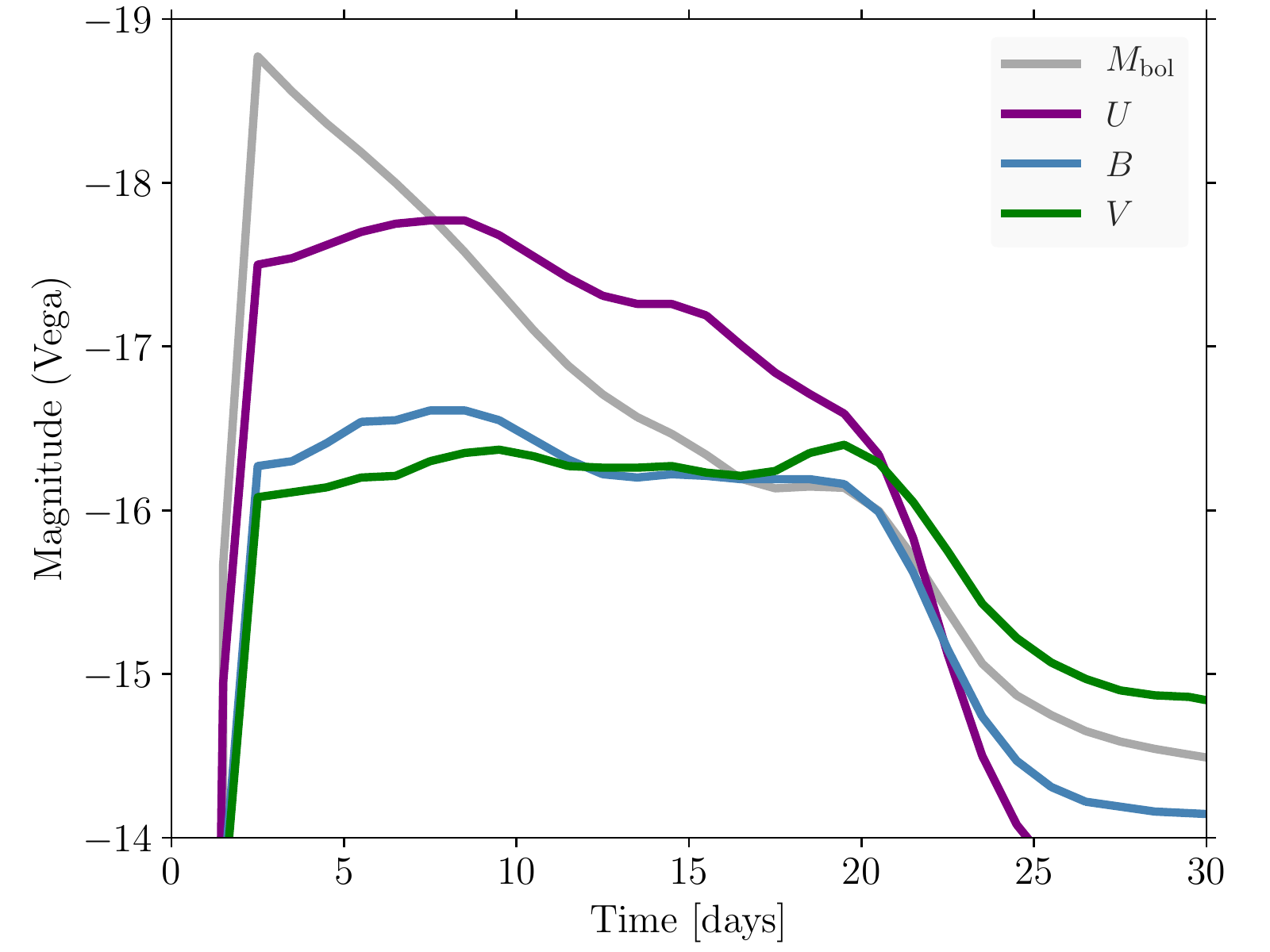}
\includegraphics[width=0.48\textwidth]{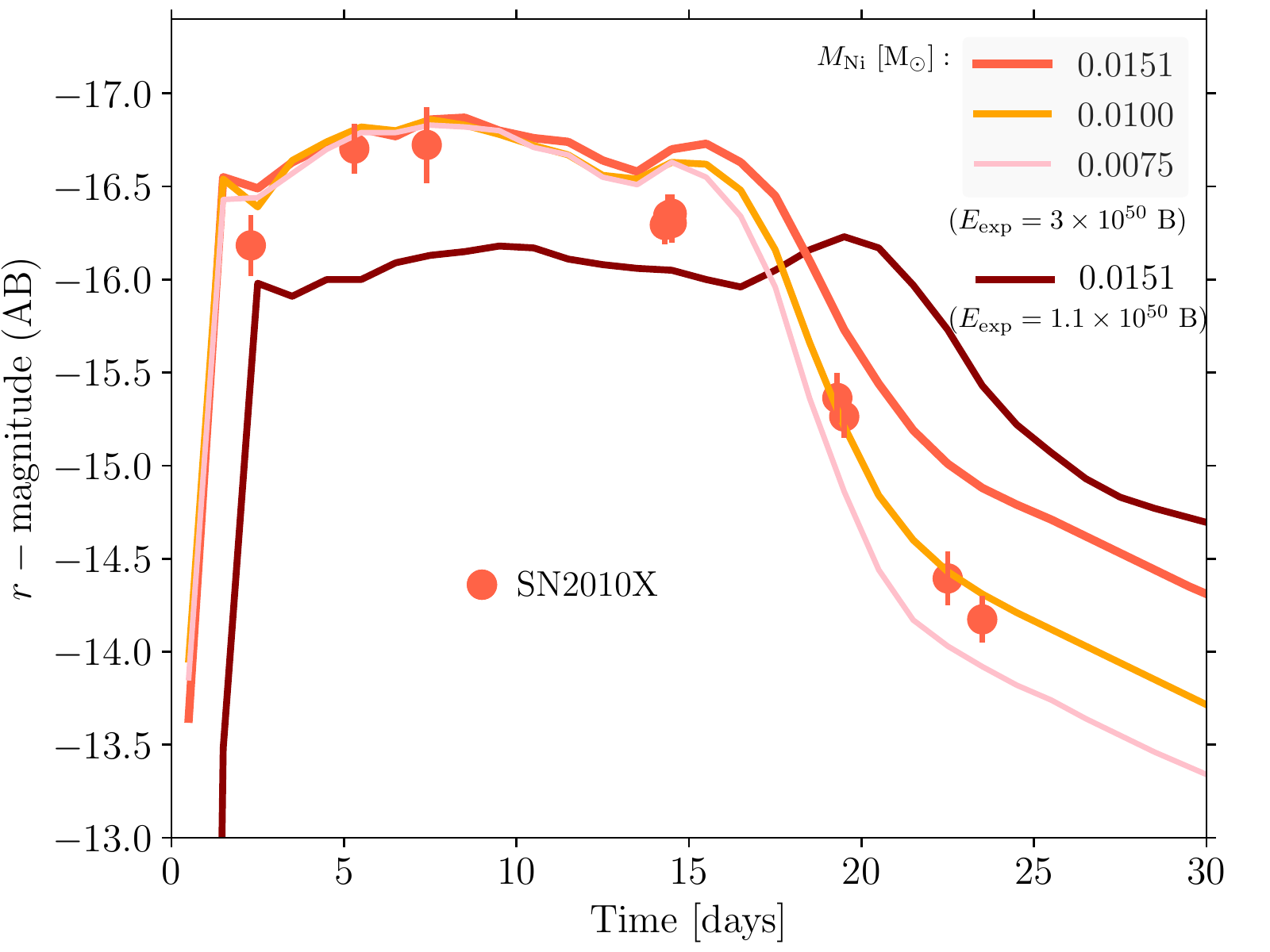}
\caption{Light curves for Model He2.50. This model had a large initial
  radius, $6.40 \times 10^{13}$ cm, owing to a degenerate silicon
  flash 19 days before final core collapse \citep[see Model 2.5B
    of][]{Woo19}.  The model ejected a mass of 0.737 \Msun\ of which
  0.0151 \Msun\ was $^{56}$Ni. The kinetic energy of the unmodified
  explosion was low, $1.1 \times 10^{50}$ erg. The top panel shows the
  bolometric and $UBV$ light curves for this explosion in the Vega
  magnitude system.  Results before 3 days are not reliable due to the
  initialization of the calculation in SEDONA.  The bottom panel shows
  the $r$-band light curves for the same model and some modifications.
  The solid dark purple curve shows the results for the unmodified
  model. The pink, orange, and red curves are artificial adjustments
  that used a larger explosion energy of $3 \times 10^{50}$ erg with
  varying masses of $^{56}$Ni, 0.0075 \Msun \ (pink); 0.01 \Msun
  \ (orange) and 0.0151 \Msun (red). The higher energy explosions with
  slightly lower $^{56}$Ni mass are a good fit to SN 2010X.
  \lFig{he2.5}}
\end{figure}

A similarly good fit to the $r$-band light curve of SN 2010X
and similar events is possible using a different sort of
model. Model He2.60 had a presupernova mass of 2.15 \Msun, a radius of
$7.8 \times 10^{12}$ cm. Using the standard W18 central engine
\citep{Ert20}, the star exploded with a final kinetic energy of $1.5
\times 10^{50}$ erg and produced 0.018 \Msun \ of $^{56}$Ni. The
explosion left a bound remnant of 1.36 \Msun \ and thus ejected 0.782
\Msun, mostly composed of helium.  The resulting light curves are
shown in \Fig{he2.6}. 

\begin{figure}
\includegraphics[width=0.48\textwidth]{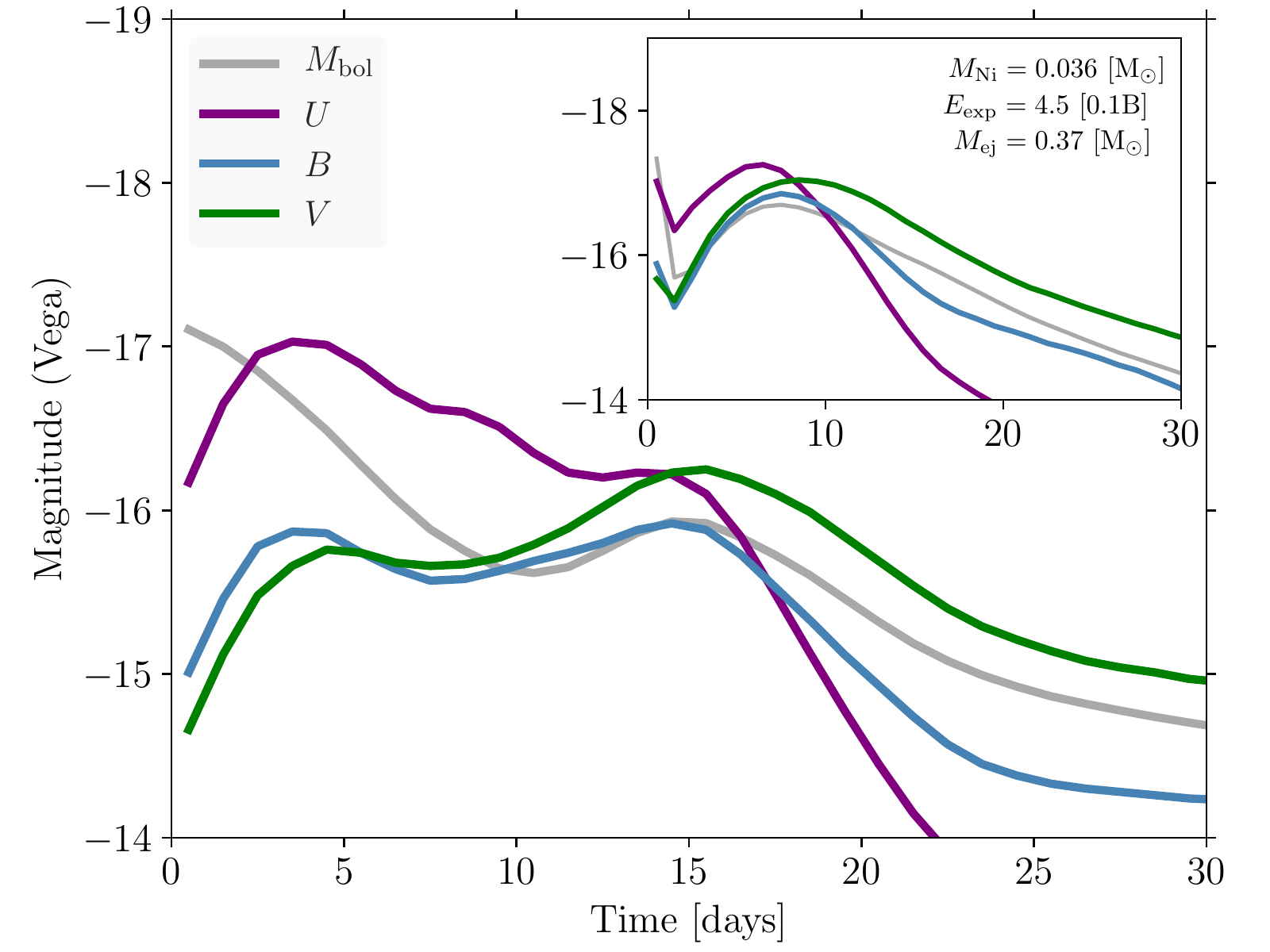}
\includegraphics[width=0.48\textwidth]{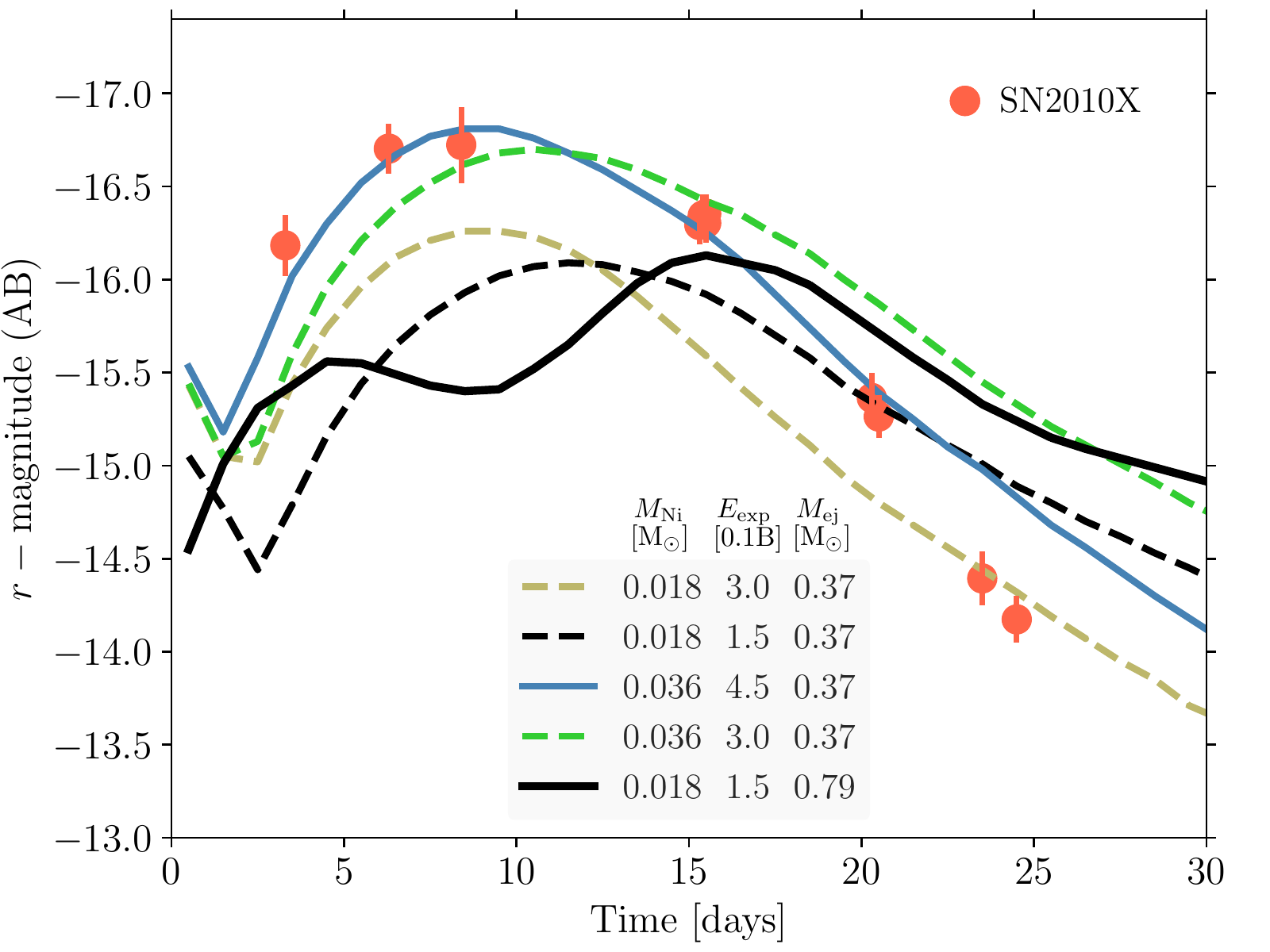}
\caption{Light curves for Model He2.60.  This model experienced no
  silicon flash and had an initial radius of $7.8 \times 10^{12}$
  cm. The explosion ejected 0.782 \Msun\ with a kinetic energy of $1.5
  \times 10^{50}$ erg and produced 0.018 \Msun \ of $^{56}$Ni. (top:)
         $U$, $B$, $V$, and bolometric light curves for the
         unmodified model with the best-fit modified model from
           the lower panel shown as an inset. (bottom:) the $r$-band
         light curve for the standard case (solid black line) and four
         variations that reduced the ejecta mass by about a factor of
         two to 0.374 \Msun\ and varied the amount of $^{56}$Ni.  The
         $^{56}$Ni mass and kinetic energies are indicated in the
         inset table.  Unlike \Fig{he2.5}, all light curves here are
         powered at peak by radioactive decay. Red data points are for
         SN 2010X \citep{Kas10} and are fit reasonably well by the
         blue curve.  $U$, $B$, $V$, and bolometric magnitudes for
           the blue curve are given as an inset in the top panel.
         \lFig{he2.6}}
\end{figure}


\begin{figure}
\includegraphics[width=0.48\textwidth]{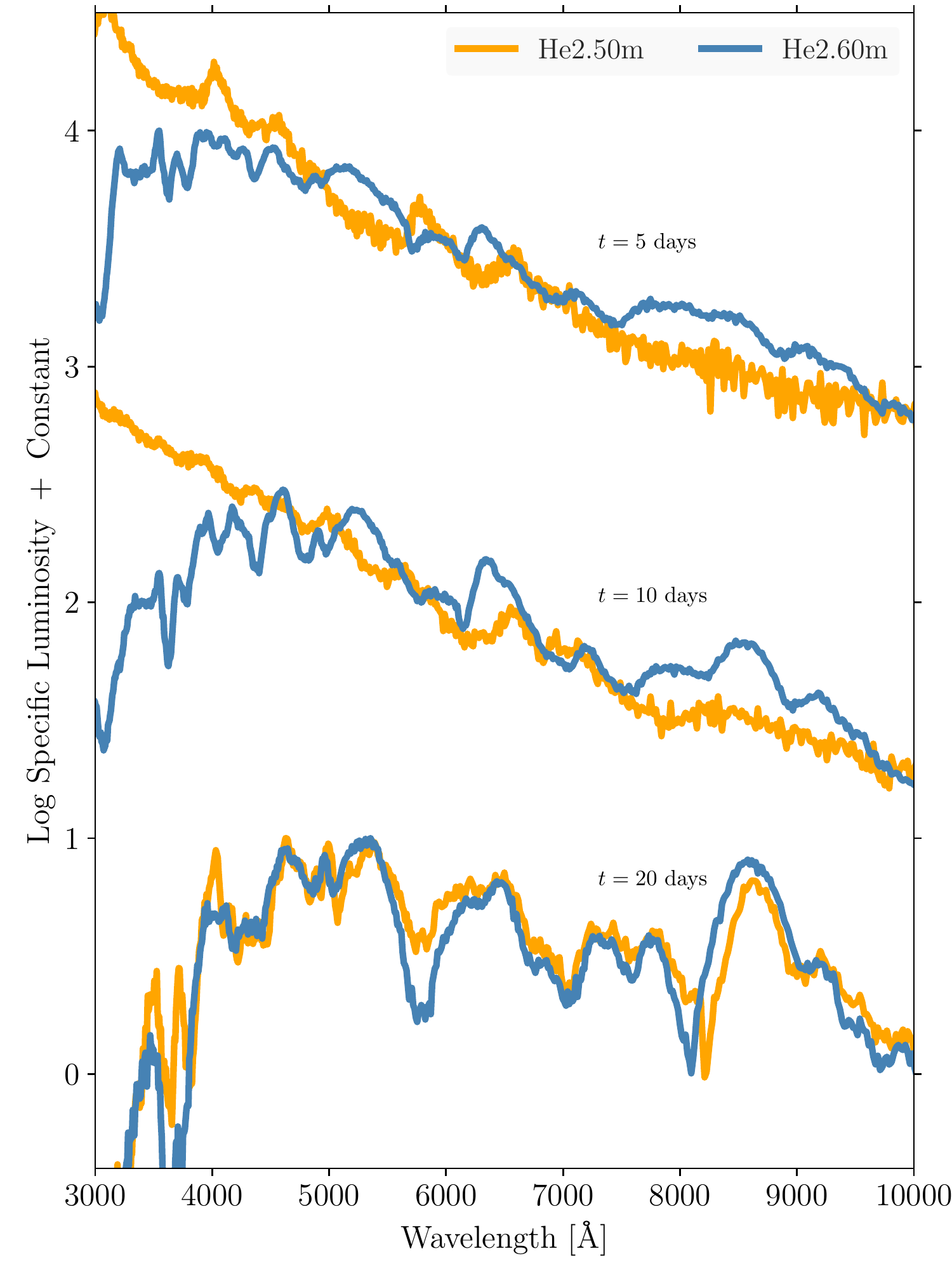}
\caption{Spectra for the modified (``m'') best fit models in the
    lower panels of \Fig{he2.5} and \Fig{he2.6}. Both models are
    evaluated (from top to bottom) at 5, 10, and 20 days post
    explosion. Model He2.50m declines rapidly bolometrically during
    its $r$-band plateau phase (5 and 10 days), and becomes similar to
    Model He2.60m as both approach their radioactive tails (20 days).
    Model He2.50m declined monotonically after shock break out and had
    bolometric magnitude of -18.19 and -17.10 at days 5 and 10
    respectively. This corresponds to luminosities of 5.7 and 2.1
    $\times 10^{42}$ erg s$^{-1}$. It reached an $r$-band maximum
    magnitude of -16.82 at 8.7 days.  Model He2.60m had bolometric
    maximum of $M_{\rm bol}$ = -16.7, $L_{\rm bol}$ = 1.4 $\times
  10^{42}$ erg s$^{-1}$ at 8.0 day and a $r$-band maximum magnitude of
  -16.89 at 10 days. t = 10 days in the figure is thus approximately
  $r$-band peak.  \lFig{2010xspec}}
\end{figure}

The bolometric light curve of the unmodified model shows a strong
first peak resulting from the expansion and recombination of the
helium, similar to but fainter than Model He2.6 because of the
  smaller radius. Because the ejected mass is smaller, the velocity
is higher, $\sim$7000 km s$^{-1}$ at day 10.  Following helium
recombination, the photospheric radius is small and the high
luminosity again gives a hot temperature and a blue color.  A
secondary maximum in bolometric emission occurs around 20 days due to
radioactivity. For a star with a smaller presupernova radius and
greater $^{56}$Ni production, this secondary peak dominates and one
would have a more typical Type Ib or Ic supernova. In the
unmodified model the two components are comparable and, depending on
the filter employed, one might see a double-peaked light curve.

The bottom frame of \Fig{he2.6} shows the $r$-band light curve with
several modifications again compared with observations of SN
  2010X. A model with a smaller presupernova mass was generated by
artificially increasing the mass loss rate so that the ejected mass
was about half as large, 0.37 \Msun. With this smaller mass the early
display from helium recombination is reduced and the radioactive peak
occurs earlier (black dashed line in \Fig{he2.6}). Additional
artificial increases in the $^{56}$Ni mass and explosion energy both
brighten and shorten the display. Increasing the energy by a factor of
three to $4.5 \times 10^{50}$ erg and doubling the $^{56}$Ni mass to
0.036 \Msun \ results in a Model He2.6m (solid blue line in \Fig{he2.6})
that agrees quite well, except at very late times, with SN 2010X.  As
with Model He2.50, these changes are substantial, but not
incredible. Indeed, this model is quite similar to what \citet{Kas10}
themselves proposed to explain their observations.  The outer third of
the ejecta now moves at over 10,000 km s$^{-1}$ and this is also in
better agreement with observations.

Model He2.6m is also similar to the ``ultra-stripped''
  supernovae discussed by \citet{Tau13}, \citet{Tau15}, and
  \citet{Mor17}. In their typical model for SN 2010X, there is no
  silicon flash and the entire light curve after breakout is powered
  by radioactive decay. In that model, core collapse ejects a small
  amount of mass ($\sim0.16$ \Msun) with a low explosion energy (1 to
  $2.5 \times 10^{50}$ erg) that contains a modest amount of $^{56}$Ni
  ($\sim0.02$\Msun). Our Model 2.60m (\Fig{he2.6}), on the other hand,
  ejects 0.37 \Msun, of which 0.036 \Msun \ is $^{56}$Ni, with an
  explosion energy of $4.5 \times 10^{50}$ erg. In terms of peak
  bolometric luminosity, $1.4 \times 10^{42}$ erg s$^{-1}$, peak time
  of about 8 days, and velocity $\sim(E/M)^{1/2}$, the two models are
  similar. Model He2.60m had a greater presupernova mass though, 1.70
  \Msun\ vs Moriya's 1.50 \Msun. More critically, the presupernova
  models of \citet{Tau13} had a small helium envelope (0.033 \Msun)
  and a much smaller radius (0.4 \Rsun) due to the assumption of
  evolution in a very compact binary. Our modified best fit model
  ejected 0.28 \Msun \ of helium and had a presupernova radius of
  about 100 \Rsun. Because of this large radius, it is much brighter
  early on than that of Moriya (\Fig{he2.6}).

The best model for a particular 2010X-like supernova, He2.50m or
  He2.60m, can be better determined from the time history of the
  colors and spectrum. \Fig{2010xspec} gives the spectrum for the two
  best-fit, modified models in \Fig{he2.5} and \Fig{he2.6} at three
  different times. He2.50m is derived from a helium supergiant and
  stays hotter longer with a temperature set, in part, by the partial
  recombination of helium. It is bluer near peak and much bluer at
  earlier times. Once the diffusion of decay energy dominates though,
  the spectrum of He2.5m resembles closely that of He2.60m. If the
  time of explosion were not well known, one model could be confused
  for the other, especially if data were lacking at wavelengths less
  than 4500 A. Comparing with Fig. 2 of \citet{Kas10}, Model He2.6m,
  the radioactive model, is a better fit to the observed spectrum of
  SN 2010X itself 3 or 4 days before visual peak (approximately +5
  days here).

  \section{Heavier Models - Long Red Transients}
\lSect{heavy}

For presupernova masses above 5.6 \Msun \ which, for $f_{\rm WR} = 1$
correspond to initial helium star masses above 8 \Msun\ (13 \Msun\ for 
$f_{\rm WR}=1.5$), the light curves become too broad and faint to be 
common Type Ib and Ic supernovae \citep{Ens88,Ert20,Pre19}. \Fig{he12.0} 
and \Fig{he19.0} show the multiband light curves and spectra for two 
cases. Compared with the events in \Sect{sn1bc}, these massive explosions 
are not only fainter and broader, but redder than common events and 
characterized by narrower spectral lines. The color index ($B-V$) for 
typical Type Ib supernova Model He4.50, at $V$-peak, is 0.70. For 
He12.0 and He19.0 it is 0.98 and 1.07. Based on KEPLER calculations 
the photospheric velocity for Models He6.00, 12.0 and 19.0 at peak are 
7900, 5400, and 4800 km s$^{-1}$ \citep{Ert20}. The explosion energies 
and $^{56}$Ni masses are comparable (\Tab{main} and \Tab{spcl}), but 
the ejecta masses are greater, so the velocity is lower and the time 
scales longer.

Because of their large mass, the frequency of such events is small.
\citet{Ert20} speculated that at least a fraction of them might
involve explosion mechanisms and light sources other than neutrinos
and radioactivity.  Some may be the progenitors Type Ic-BL supernovae
and gamma-ray burst supernovae, but probably not all.

It is interesting to compare Model He12.0 with SN 2013bb
\citep{Pre19}. Though classified as a Type IIb supernova, the
appearance of SN 2013bb after the first few weeks might not be so
different from a Type Ic event. It has the longest light curve yet
observed for a stripped envelope explosion \citep{Pre19}. SN 2013bb
had a peak luminosity (4000 - 10000 \AA) of $\log L_{\rm 4-10k}$ =
$42.0 \pm 0.1$ erg s$^{-1}$; an uncertain peak time of at least
$\sim25$ days; a decline time $t_{+1/2}$ of $56 \pm 10$ days; an
inferred ejected mass of 4.8 \Msun; a velocity at peak of 7000 km
s$^{-1}$, and an inferred $^{56}$Ni mass of $0.07 \pm 0.02$ \Msun.
Model He12.00 had a peak luminosity (4000 - 10000 \AA) of $\log L_{\rm
  4-10k}$ = 41.92 erg s$^{-1}$; a peak time of 39 days; a decline time
$t_{+1/2}$ of 41 days, an ejected mass of 5.33 \Msun; a velocity at
peak of about 5400 km s$^{-1}$; and an inferred $^{56}$Ni mass of
0.079 \Msun.  The decline time for the model is perhaps a bit short,
but the time of peak luminosity is poorly defined in the model and
presumably in the observations as well. The luminosity only varies 1\% in an
interval of 6 days around peak and 5\% in 10 days. Shifting our
definition of peak back 5 days would improve the agreement
considerably. Other models with initial helium star masses in the range 9 to 13
\Msun\ had similar properties \citep[Table 5 of][]{Ert20}. Minor
modifications of the uncertain $^{56}$Ni mass and explosion energy and
including a small residual hydrogen envelope could account for the
small differences with SN 2013bb.

Apparently SN 2013bb was of the same family as common Type Ib and Ic
supernovae, but happened in a more massive star with a small residual
hydrogen envelope.  Observers should be on the look out for more of
these and similar cases where there is less hydrogen.

\begin{figure}
\includegraphics[width=0.48\textwidth]{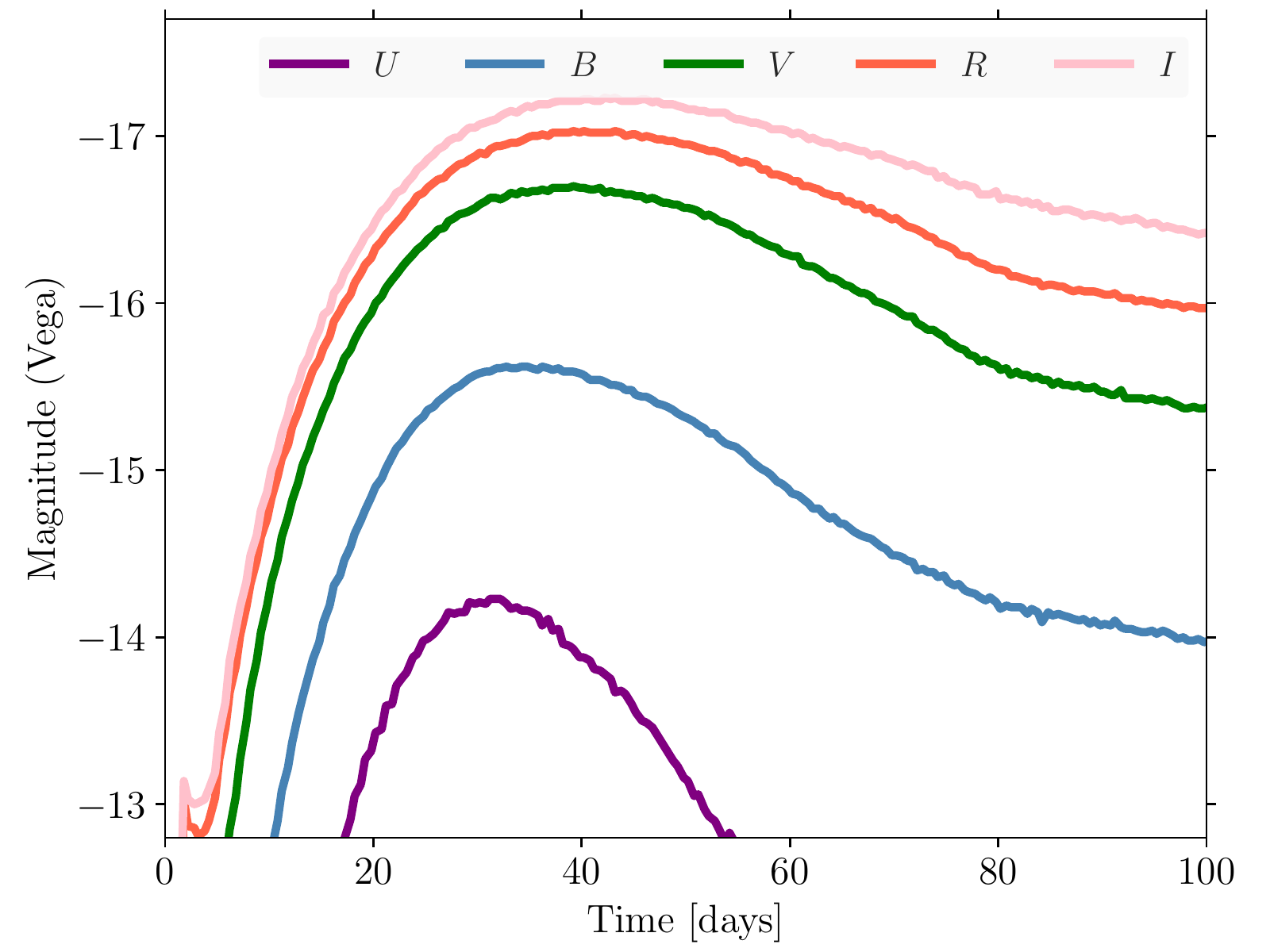}
\includegraphics[width=0.48\textwidth]{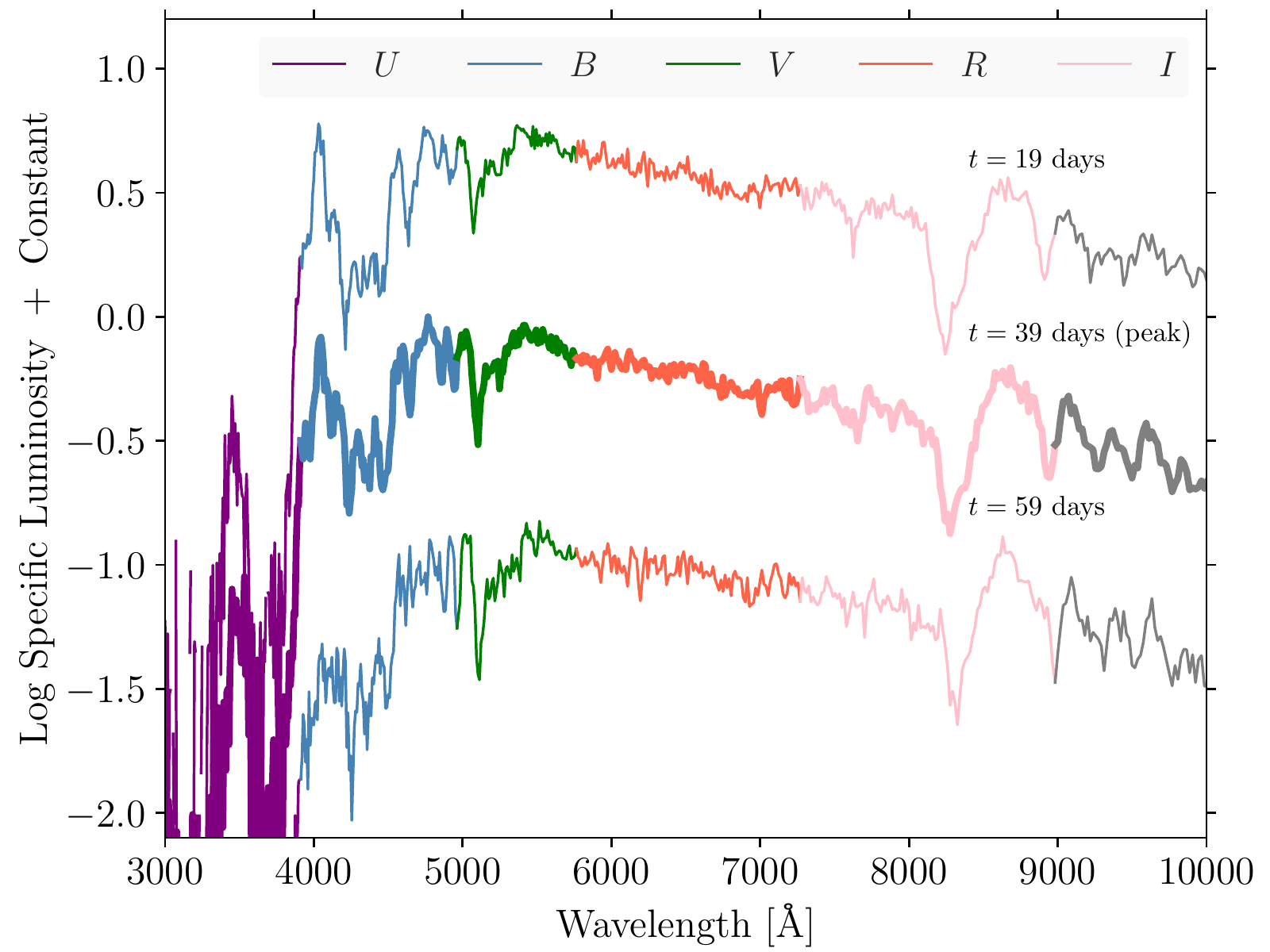}
\caption{(Top:) Light curve for Model 12.0. The explosion ejected 5.33
  \Msun \ with a kinetic energy of $0.81 \times 10^{51}$ erg and
  synthesized 0.0792 \Msun \ of $^{56}$Ni. (top:) Vega-based $UBVRI$ 
  light curves until 100 days. (bottom:) Spectra at peak (39 days) and at 
  minus and plus 20 days.
  \lFig{he12.0}}
\end{figure}

\begin{figure}
\includegraphics[width=0.48\textwidth]{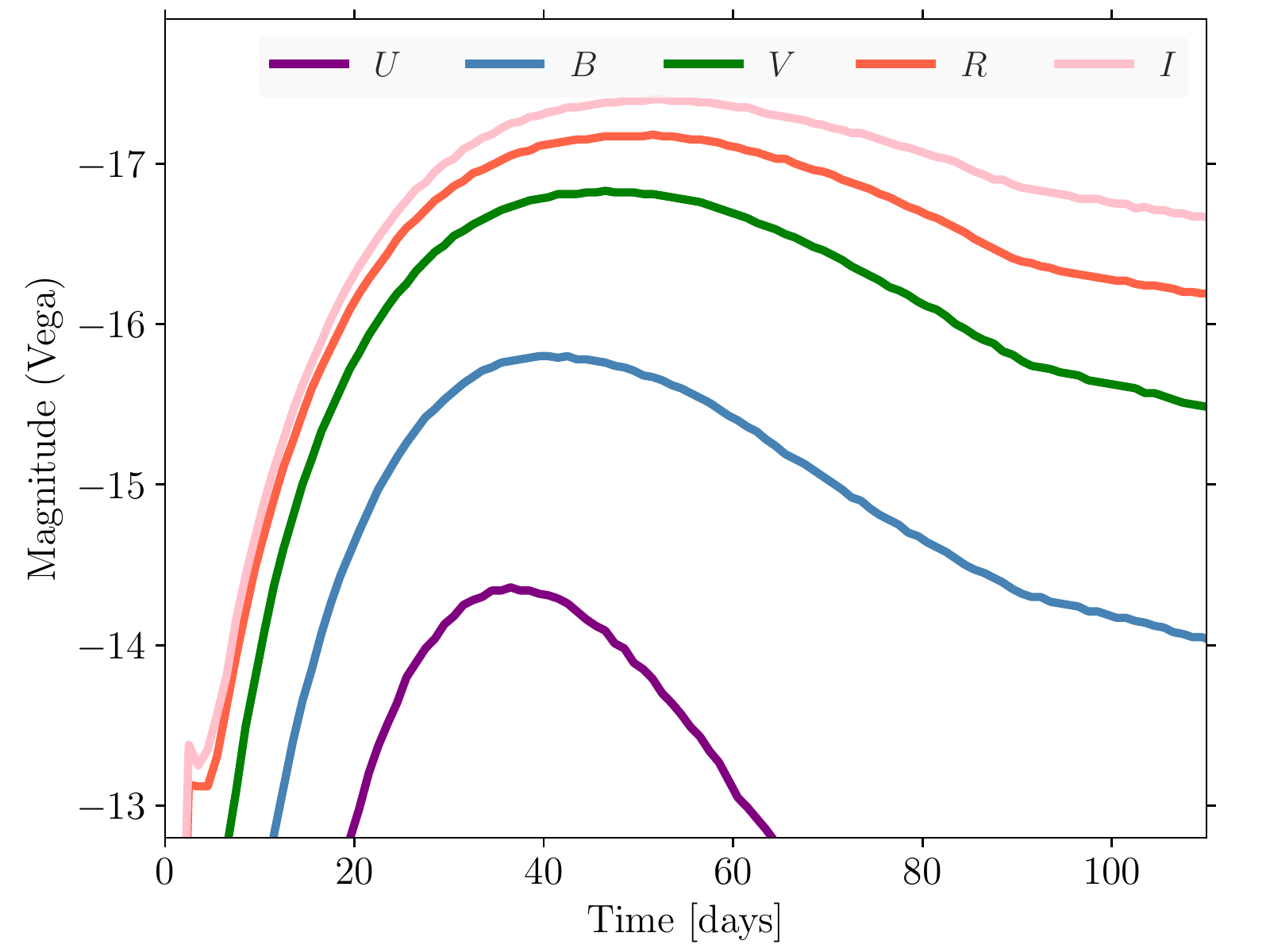}
 \includegraphics[width=0.48\textwidth]{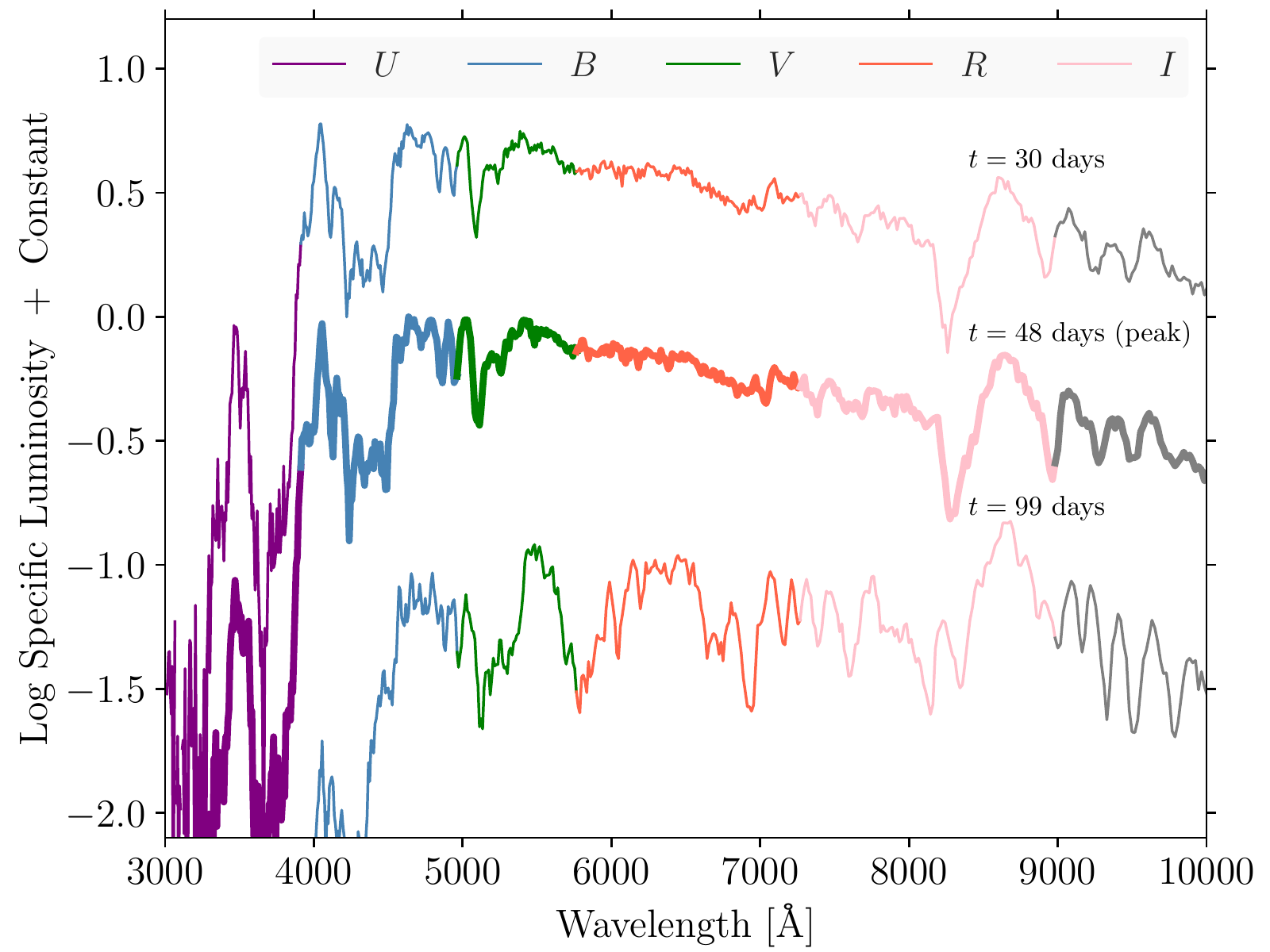}
\caption{Multiband light curves and spectra for Model 19.0. (top:)
  $UBVRI$ light curves as in \Fig{he12.0}. Due to the kinetic
  energy per unit mass the light curve is very broad, faint, and
  red. (bottom:) Spectra at peak light (48 days) and at 30 days and 99
  days. Note the presence of many narrows lines reflecting the
  slow expansion speed.
  \lFig{he19.0}}
\end{figure}

\section{Conclusions}
\lSect{conclude}

The spectra and light curves resulting from the explosion of a broad
range of mass-losing helium stars have been calculated and compared
with observed ``stripped envelope'' supernovae.  The models are
representative of the endpoints of evolution in close, mass-exchanging
binary systems. The presupernova evolution of these stars, their
neutrino-powered explosions, their nucleosynthesis, and the
distribution of remnant masses they leave behind has been published
previously \citep{Woo19,Ert20,Woo20}. Here emphasis is on a careful
treatment of the radiation transport using the SEDONA code and a
comparison with modern observations.

For a range of presupernova masses from 2.2 to 5.6 \Msun, the models
are in good qualitative agreement with observations of common Type Ib
and Ic supernovae. The median values of some major observables are
given in \Tab{imf_lum} and the comparison depends on how well observers
have sampled this broad mass range. Light curves from lower mass
explosions are brief and faint and some may have been
overlooked. Similarly, the light curves of very massive models are
broad, faint, and red and could also be under-sampled. Depending upon
the assumed mass loss rate, the presupernova mass range 2.2 to 5.6
\Msun \ corresponds to initial helium star masses of 2.7 to 8 \Msun \
for $f_{\rm WR}$ = 1 and 3.0 to 13 \Msun \ for $f_{\rm WR} =
1.5$. These helium stars are, in turn, derived from main sequence mass
stars of approximately 14 - 28 \Msun \ for $f_{\rm WR} = 1$ and 15 -
38 \Msun \ for $f_{\rm WR} = 1.5$. Still larger initial masses could
be accommodated using larger values for $f_{\rm WR}$, but probably not
smaller ones.

A typical ejected mass is 1.7 to 2.0 \Msun \ with kinetic energy 0.8
to 1.3 $\times 10^{51}$ erg (\Tab{imf_lum}), though a range of both is
allowed in individual events (\Tab{main} and \Tab{spcl}).  These
energies and masses agree well with published values
\citep[\Tab{imf_lum};][]{Dro11,Lym16,Tad18,Pre19}.  Similarly the rise
times, peak times, decline times, spectra, and colors are in good
agreement with observations (e.g., \Tab{imf_lum}, \Tab{snobs},
 \Fig{stritz}, \Fig{color}). The paradigm that most Type IIb, Ib
  and Ic supernovae originate in mass exchanging binaries with light
  curves powered by the decay of radioactive $^{56}$Ni and $^{56}$Co
  is strongly supported. Depending on the mass loss rates assumed, a
  subset of Type Ib and Ic supernovae could still come from single
  stars, but the presupernova mass would need to be less than 5.6
  \Msun \ \citep[Model He8.0;][]{Ens88} in order not to have a light
  curve that was broader and fainter than common events. Presupernova stars
  with much greater mass might also be difficult to explode using only
  neutrinos. Producing such a low mass presupernova star from a
  massive star that only lost its envelope to radiative winds would
  require a main sequence mass well in excess of 30 \Msun \ and mass
  loss rates substantially larger than given by \citet{Yoo17} once the helium
  core was revealed. While we have attempted no studies of population
  synthesis, the number of lower mass stars (12 to 30 \Msun) in mass
  exchanging binaries is probably much greater.

The situation with peak luminosities remains less clear. Our models
agree with the bolometric luminosities and absolute magnitudes of most
normal Type Ib and Ic supernovae, but fail to explain the brighter
events (\Fig{taddia} and \Fig{prentice}). This is true even when
anomalies such as broad-lined supernovae, superluminous supernovae,
and gamma-ray burst supernovae are removed from the observational
sample and the theoretical maximum $^{56}$Ni production is employed in
the models. The disagreement exists both for bolometric light curves
\citep{Ert20,Lym16,Pre16}, as well as peak magnitude distributions
(\Sect{distribution}). Interestingly, the disagreement might not be as
great when the comparison is made with specific colors (\Fig{stritz})
or for limited wavelengths \citep[\Tab{imf_lum},\Fig{prentice}][]{Pre19},
though even then quite a few bright events remain unexplained. This
suggests that part of the discrepancy with peak bolometric values
could be in the way the observers have assembled their bolometric
light curves by extrapolations into the infrared and ultraviolet, or
by using overly simple models for the light curve.

On the other hand, our treatment of the ultraviolet and infrared light
curves in SEDONA is not definitive and 1D models of neutrino transport
could be missing important aspects of a putative ultimate 3D
treatment. This includes e.g., convection in the neutrino-heated
``hot bubble''. The models of \citet{Ert20} seek to compensate for
this by normalizing key parameters to the observed properties of SN
1987A and the Crab, but if, for example, the explosions were to
develop earlier and the mass separation was deeper, $^{56}$Ni
production would be modestly increased. No existing 3D models of
which we are aware show such large $^{56}$Ni production (greater than
0.15 \Msun) and the tendency of current 3D models is towards explosion
at late times. Nucleosynthetic constraints and the observed masses of
neutron stars limit increases in $^{56}$Ni production for all but a
few rare cases \citep{Ert20}.  Further work is needed.

We are now confident though that we have calculated the bolometric
luminosity for this particular set of models correctly. The results of
two very different codes, KEPLER and SEDONA, give essentially the same
answer and mixing, though important for the colors and spectrum does
not affect the peak bolometric luminosity greatly. A substantial
fraction of bright supernovae categorized as normal Type Ib and Ic
continue to be mysterious. We encourage observers to undertake new
surveys and reanalyze past ones. It may be convenient to avoid
constructing ``bolometric light curves'' and compare instead with
theoretical models adjusted to individual passbands now that such
predictions are becoming available. We have provided many (\Tab{imf},
\Tab{peak_main}, \Tab{tscales_main}), and we and others could provide
more. Host-galaxy extinction and bolometric corrections seem to have
played an important role in some of the observational estimates, and
may need to be carefully examined.  Major exceptions might be studied
for other signs of peculiarity.

In any case, it is risky to arbitrarily vary $^{56}$Ni
masses and explosion energies for bright supernovae that may not be
exploded by neutrinos and hence, whose source of illumination --
rotation or radioactive decay -- is uncertain. Our predicted maximum
bolometric luminosity remains 10$^{42.5}$ erg s$^{-1}$. Our maximum
explosion energy is $2 \times 10^{51}$ erg and maximum $^{56}$Ni mass
is 0.15 \Msun. The brightest $B$, $V$, $R$, and $I$ peak absolute
magnitudes are $-17.2$, $-17.8$, $-18.0$, and $-18.1$. The $V$ limit
is the best determined.

As noted by others \citep{Des16,Kha19}, Arnett's rule does not work
well for Type Ib and Ic supernovae (\Sect{analytic}). \citet{Des16}
found that the $^{56}$Ni mass inferred for a given peak luminosity
overestimated the actual value by a factor, on the average, of $1.41
\pm 0.072$. Here, possibly because of different assumptions about
mixing explosion energy, and opacity, we find a smaller discrepancy
$1.30 \pm 0.095$ (\Sect{arnett}).  Starting with the analytic treatment
of \citet{Kha19}, we derive a more accurate way of determining the
$^{56}$Ni mass using Khatami's characteristic parameter, $\beta$, to
describe a given model set (\Sect{analytic}).

While the deficiency of bright events is interesting, it should not be
overlooked that our models do agree with the observed properties
of many observed events. Detailed comparisons have been made with six of
them: SN 2007Y, SN 2007gr, SN 2009jf, LSQ13abf, SN 2008D, and SN
2010X, and are largely successful. The models do display deficiencies
in the ultraviolet and infrared that increase with time 10 to 20 days
post-peak. We confirm and explain, on the basis of our models
(\Sect{indices}), the pinch seen in the color index ($V-R$) 10 days
after $V$ peak. \citet{Dro11} measured a mean value of $0.26 \pm 0.06$
mag for this quantity. \citet{Des16} calculated a mean value of $0.33 \pm
0.035$ mag. We calculate a median value here of 0.34 mag for our standard
$f_{\rm WR} = 1$ models and 0.29 mag for the models with greater mass
loss, $f_{\rm WR} = 1.5$. These and other predictions for similar pinches in
many other color indices are given in \Tab{imf} and \Fig{color}. Given
the robustness of this number and other indices as a measure of
temperature at a specific epoch, they may prove useful in calculating
the host-galaxy extinction \citep{Str18b}. We also note that it is not
just the colors, but the spectrum itself that appears to converge on a
standard value 10 days after the explosion (\Fig{spectra}).

\Fig{he4.5} and \Fig{he8.0} show the light curves and spectra of two
typical models from our survey, Models He4.50 and He8.00x2.  Despite a
very different presupernova evolution, these stars ejected similar
amounts of mass and, by construction, had the same $^{56}$Ni masses
and explosion energies. Even with very different compositions,
their bolometric light curves are almost identical. Except at early
times their colors and spectra are also nearly indistinguishable.
While some variation in $^{56}$Ni synthesis and explosion energy is
naturally expected, the supernova depends much more on the
presupernova mass than how the star got there. Unless one is
interested in colors at early times, surveys with variable mass loss
rates may not be necessary in certain situations. The natural chaos in
presupernova core structure \citep{Suk18} may dominate over the
smaller effects of variable composition. There are two clear
exceptions to this: 1) the weighting assigned to the event when
computing averages depends on the zero age main sequence mass which
will be different for different mass loss prescriptions; and 2) loss
of most of the helium envelope will make it difficult to produce a
supernova that is spectroscopically Type Ib.

In addition to ordinary Type Ib and Ic supernovae, we also studied the
properties of several explosions for masses that lay outside our
standard range ($M_{\rm preSN}$ = 2.2 - 5.6 \Msun). Models with radius
expansion, either due to the low mass of the star or a silicon flash a
few weeks before core collapse, can produce events similar to LSQ13abf
(\Fig{LSQ}), SN 2008D (\Fig{sn2008d}), and SN 2010X (\Fig{he2.5} and
\Fig{he2.6}). Interestingly, the available $r$-band data is, by
itself, consistent with two solutions for SN 2010X. For Model
He2.50, the light curve has a peak lasting about two weeks resulting
from the release of shock deposited energy in a blue supergiant
  envelope by helium recombination. On the peak, radioactivity plays
little role. This is a solution also explored by \citet{Kle18a}. Ours
differs in that the large radius is due to a silicon flash 19 days
prior to iron core collapse. Even so, a good fit to duration and
luminosity required an artificial, though not unrealistic increase in
the explosion kinetic energy.  One attractive feature of this model is
that a slight variation of the silicon flash energetics might also
produce other supernovae with rapid rises and declines and weak
radioactive tails, e.g., SN 2002bj. Other similar double-peaked light
curves have been reported \citep{Gor14,Tad16} and attributed to radius
expansion.

A second solution, similar to that suggested by \citet{Kas10}
  themselves and also studied by \citet{Tau13}, \citet{Tau15}, and
  \citet{Mor17}, has SN 2010X illuminated at peak by radioactive
  decay. Our standard Model He2.60 (\Fig{he2.6}) produces a
  distinctive double-peaked light curve in the $r$-band that should
  exist in nature, but does not look like SN 2010X. Several
  modifications to the ejected mass, explosion energy, and $^{56}$Ni
  mass are required to achieve a good fit, but they are again
  reasonable modifications. The most successful Model He2.60m differs
  from previous studies \citep[e.g.,][]{Mor17} by happening in a
  progenitor with much larger radius and is therefore brighter early
  on. Such a large presupernova radius would require a much wider
  binary system, but loss of the hydrogen envelope when a red
  supergiant forms near helium ignition and a blue supergiant
  structure at the presupernova stage are not incompatible.

Relative to SN 2010X the overall $r$-band light curve is better
  fit by the recombination model, especially at late times
  (\Fig{he2.5} and \Fig{he2.6}). The low nickel mass ($\ltaprx0.01$
  \Msun) and explosion energy are also consistent with expectations
  for a star of such small mass. On the other hand, the velocity is a
  bit low unless the ejected mass is further artificially reduced, and
  the properties of the silicon flash are uncertain. The high
  luminosity and very blue spectrum prior to peak may also be
  difficult to reconcile with observations of SN 2010X. The
  radioactive model requires a very low ejected mass that contains a
  substantial amount of $^{56}$Ni ($\sim0.03$ \Msun). This may be
  difficult to achieve in a self-consistent neutrino-powered explosion
  \citep{Ert20}, but for now we favor this model for SN 2010X based on
  its spectrum near peak light \citep[\Fig{2010xspec} here and Fig. 2
    of][]{Kas10}, but the issue needs further study.

Very massive explosions produce faint, red, long transients
(\Fig{he12.0} and \Fig{he19.0}).  SN 2013bb is not a bad match to Model
He12.00 (\Sect{heavy}) and such faint, long transients could have been
selected against in past surveys. Given their high mass, they would be
relatively rare. It is also possible that stars of such great mass
have explosions augmented by non-neutrino sources \citep{Ert20}.

All explosion models, spectra, light curves and tables in this paper
are available on request.

\section*{Acknowledgments}

We have benefited greatly, both in understanding and in privately
supplied data, from interactions with several observers. Max
Stritzinger provided directions for using the Carnegie data set and
the data used in \Fig{taddia}, \Fig{stritz}, \Fig{carncolor},
\Fig{2007Y}, and \Fig{LSQ}. Simon Prentice helped us on many occasions
with insights and data. He provided the data in \Fig{prentice},
\Fig{2007gr} and \Fig{2009jf}.  Mansi Kasliwai provided data for
\Fig{he2.5} and \Fig{he2.6}. Stefano Valenti provided infrared data
for SN 2007gr. Mark Phillips, Francesco Taddia, Maria Drout, Isaac 
Shivvers, and Alex Filippenko provided valuable assistance in 
understanding and using their works. Christopher Kochanek helped us 
to understand the complicated landscape of photometric systems. At 
Santa Cruz this work was partly supported by NASA NNX14AH34G.  TS was 
supported by NASA through a NASA Hubble Fellowship grant \#60065868 
awarded by the Space Telescope Science Institute, which is operated 
by the Association of Universities for Research in Astronomy, Inc., 
for NASA, under contract NAS5-26555. The authors acknowledge use of 
the LUX supercomputer at UC Santa Cruz, funded by NSF MRI grant AST 
1828315.


\newpage

\appendix
  \setcounter{table}{0}
  \setcounter{figure}{0}

\section{Analytic Fits to the Bolometric Light Curve and Arnett's Rule}
\lSect{analytic}

\subsection{$^{56}$Ni and the Peak Luminosity}

\citet{Kha19} have recently derived analytic approximations for the
light curves of stripped-envelope supernova. They find that
recombination and the spatial distribution of heating modify the peak
of the light curve and that these effects can be accounted for by
varying a single dimensionless parameter, $\beta$, which is a measure
of mixing. The parameter $\beta$ also captures some of the
  effects of recombination on the opacity. In particular, they find
for Type I supernovae heated by the decay of $^{56}$Ni and $^{56}$Co
\begin{equation}
 L_{\rm peak} \ = \ 2 \epsilon_{\rm Ni} M_{\rm Ni}  F(t_{\rm peak},\beta)
 \lEq{Lkhat}
\end{equation}
where
\begin{equation}
  F(t_{\rm peak},\beta) \ =  \frac{k_1 \, g(\tau_{\rm Ni}) \, + \, k_2 \, g(\tau_{\rm Co})}
  {\beta^2 \tau_{\rm Ni}^2}
\end{equation}
and
\begin{align}
k_1 \ &= \ 1  - r \,
  \frac{\epsilon_{\rm Co}}{\epsilon_{\rm Ni}} \\
  k_2 \ &= \ r \left(\frac{\epsilon_{\rm Co}}{\epsilon_{\rm Ni}}\right)
   \left(\frac{t_{\rm Co}}{t_{\rm Ni}}\right)^2 \\
  r \ &= \  \frac{t_{\rm Co}}{t_{\rm Co} - t_{\rm Ni}} \\
g(\tau_i) \ &= \ 1 - (1 + \beta \tau_i) e^{-\beta \tau_i}.
\end{align}
Here $M_{\rm Ni}$ is the initial mass of $^{56}$Ni; $t_{\rm Co}$ = 111.3 days
and $t_{\rm Ni}$ = 8.77 days are the mean lives of $^{56}$Co and
$^{56}$Ni; $\tau_{\rm Co} = t_{\rm peak}/t_{\rm Co}$ and $\tau_{\rm Ni} = t_{\rm
  peak}/t_{\rm Ni}$ where $t_{\rm peak}$ is the time of bolometric light
curve peak; and $\epsilon_{\rm Ni} = 3.90 \times 10^{10}$ erg g$^{-1}$
s$^{-1}$ and $\epsilon_{\rm Co} = 6.78 \times 10^9$ erg g$^{-1}$ s$^{-1}$
are the energy yields from $^{56}$Ni and $^{56}$Co decay. Typos in the
values \citet{Kha19} published for $k_1$, $k_2$, and $g(\tau_{\rm Ni})$
have been repaired. Evaluating fixed constants gives $k_1$ = 0.811 and
$k_2$ = 30.4. For comparison, the energy from radioactive decay at peak, if
all energy were trapped, is
\begin{equation}
L_{\rm heat} \ = \ M_{\rm Ni} \left[(\epsilon_{\rm Ni} - r \epsilon_{\rm Co})
  e^{-\tau_{\rm Ni}} \ + r \epsilon_{\rm Co} e^{-\tau_{\rm Co}}\right].
\end{equation}  

Based on models, \citet{Kha19} approximated $\beta \approx 4/3 (1 +
r_s^4)$ where $r_s$ is the ratio of the radius of the homogeneously
mixed region containing $^{56}$Ni to the ``radius of the ejecta'',
which is the maximum velocity of the ejecta times the age. $r_s$ = 0
corresponds to no mixing and $r_s = 1$ to complete mixing. While a
mathematically useful simplification, there are several difficulties
in applying this definition to an actual model. First the radius of
the ejecta is poorly defined when a shock wave in the outer layers can
result in extremely high velocity in a small amount of mass. We can
attempt to get around that by defining a cutoff at say, 99\% of the
mass. Second, as \citet{Kha19} discuss, recombination and other
alterations to the opacity can change the value of $\beta$ and its
interpretation. In particular, recombination can reduce
$\beta$. \citet{Des16} discuss the complicated time and spatial
dependence of the opacity, even in the absence of
recombination. Finally, prescriptions for mixing are not so
simple as homogenizing the composition inside a given velocity. There
is an appreciable gradient in $^{56}$Ni even in the mixed region
(\Fig{mix}).

If we examine a typical model, He6.00, in some detail, the velocity at
the 99\% mass point is 13,900 km s$^{-1}$. At the 95\% point it is
9890 km s$^{-1}$. The velocity containing 99\% of the $^{56}$Ni in the
mixed model is 6300 km s$^{-1}$, but the velocity containing
50\% of the $^{56}$Ni is only 3570 km s$^{-1}$. The dimensionless radius
would thus be somewhere between 0.26 and 0.64 and $\beta$ between 1.34
and 1.55, without corrections for recombination. Probably the 50\% mixing
radius is more appropriate, so values of $\beta$ less than about 1.4 are
preferred. While some might regard our models as ``well mixed'', the
$^{56}$Ni actually remains centrally concentrated. In a version of
He6.00 with no mixing at all (\Fig{mix}), the velocity at the outer
edge of the $^{56}$Ni was already 2700 km s$^{-1}$, or 75\% of the
radius of the half-nickel point in the mixed production model.

The average $\beta$ for the \citep{Des16} data set is 1.17
\citep[close to the 9/8 cited by][]{Kha19}. For our new models the
average $\beta$ is 1.36. There is a tendency for the lighter models
with thick helium shells to have a smaller value of $\beta$, around
1.3 in the new set, while the heavier ones have $\beta \approx 1.4$.

For this range of $\beta$ and a range of $t_{\rm peak}$ from 15 to 40 days,
$F(t_{\rm peak},\beta)$ in \Eq{Lkhat} can be fit to an accuracy of 2\% by
\begin{equation}
F(t_{\rm peak},\beta) \ = 0.1515 \left(\frac{\beta}{1.33} \frac{t_{\rm
      peak}}{20 \ {\rm days}}\right)^{-0.75},
\end{equation}
so that the luminosity at peak is
\begin{equation}
L(t_{\rm peak}) \ = \ 2.35 \times 10^{42} \frac{M_{Ni}}{0.10
  \ \Msun} \left(\frac{\beta}{1.33} \ \frac{t_{\rm peak}}{20 \ {\rm
    days}}\right)^{-0.75} \ {\rm erg \ s^{-1}}.
\lEq{lumcorr}
\end{equation}

\begin{figure}
\includegraphics[width=0.48\textwidth]{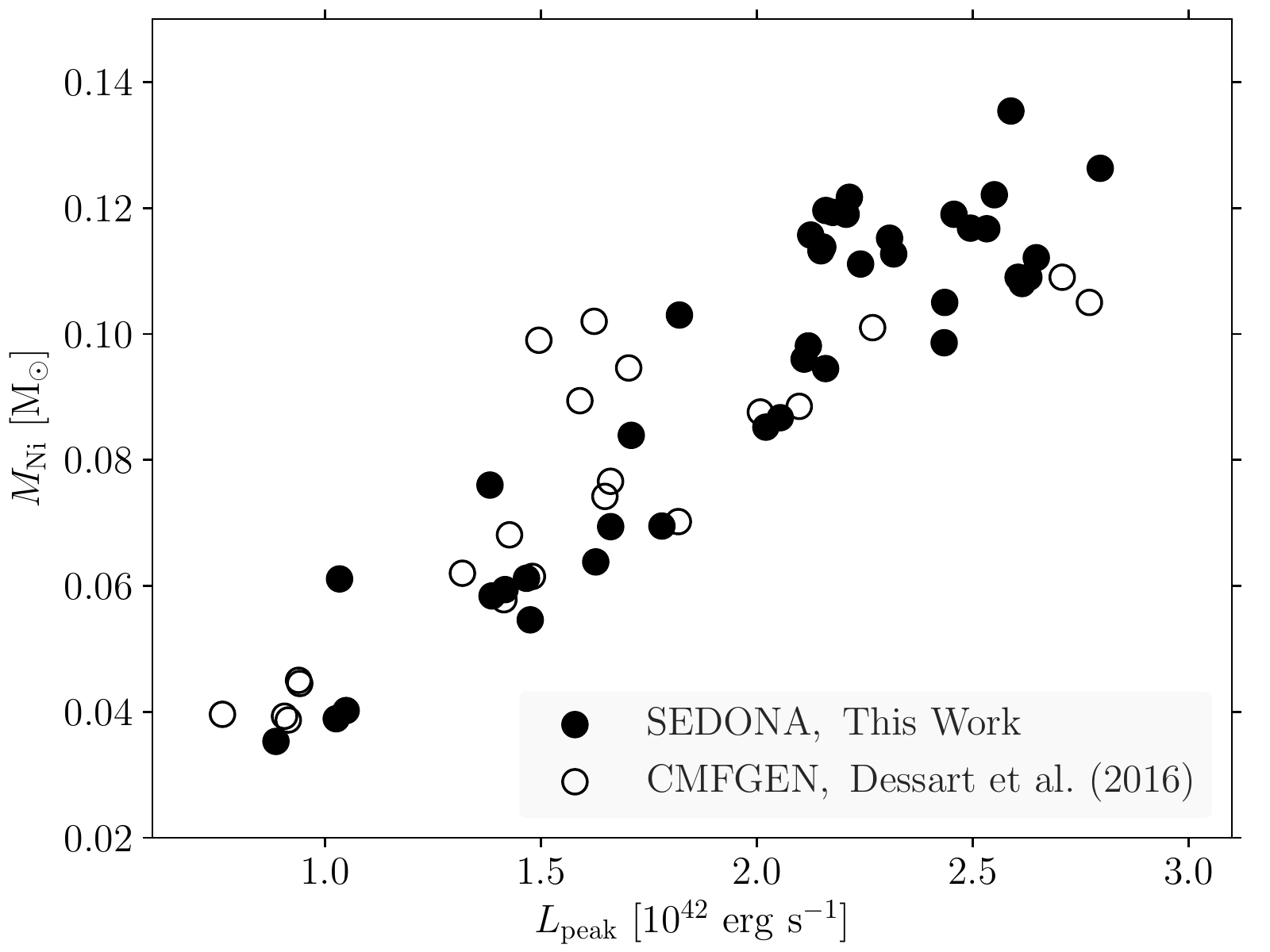}
\includegraphics[width=0.48\textwidth]{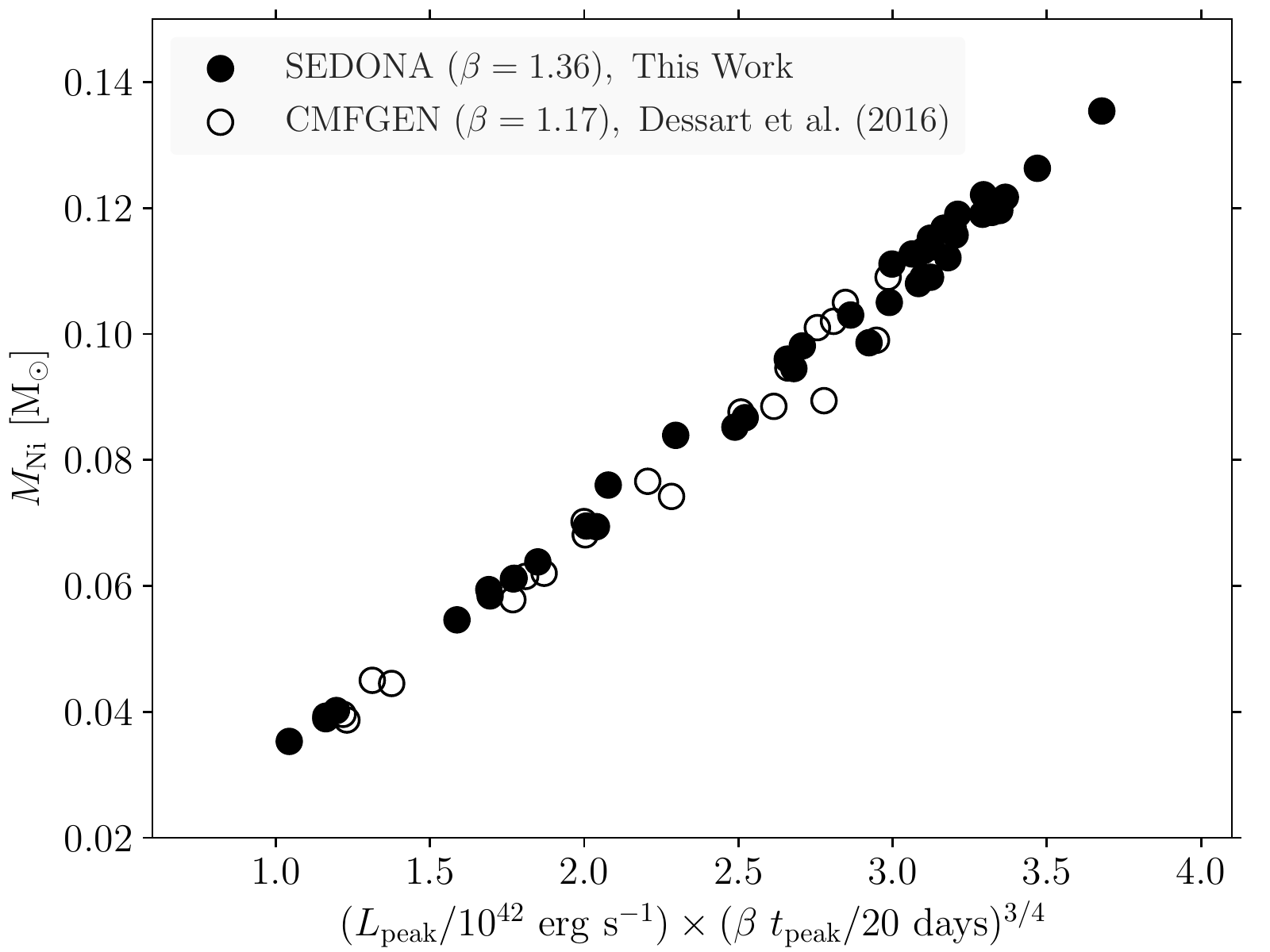}
\caption{Bolometric luminosity at peak vs the initial mass of
  $^{56}$Ni produced in the explosion. Filled circles are from the
  present study and open ones are from \citet{Des16}. (top:) The
  unmodified dependence. (bottom:) The comparison is greatly improved
  if the luminosity at peak is scaled by $(\beta t_{\rm peak})^{3/4}$
  where $\beta$ = 1.17 for the Dessart set and 1.36 for the present
  work. Four of the 27 models calculated by Dessart are not included
  because they produced more than 0.2 \Msun \ of $^{56}$Ni and are not
  regarded as physical. \lFig{Khat}}
\end{figure}

Because $t_{\rm peak}$ is not always easy to determine
observationally, it is possible to substitute other similar variables
such as the decline rate post-maximum or the width of the
peak. \Fig{tpeakv15} shows that for a given value of $\beta$, a
measurement of $\Delta M_{15}$ in the visual or red band can constrain
the peak of the bolometric light curve for the present models to about
a day, or $\sim5$\%. Taken to a power 0.75, this means the $^{56}$Ni
mass can be determined by the peak luminosity to about 4\%. The scaled
peak, $\beta t_{\rm peak}$ is a measure of when the energy deposited by
radioactive decay balances the light emitted by the star, i.e., the
time derivative of the internal energy is zero \citep{Kha19}. It is a
better choice for comparison with the decay time. Without the
multiplication by $\beta$ the \citet{Des16} data is systematically
higher than for the new models. Part of the remaining scatter results
from the choice of a single $\beta$ to characterize the two data
sets. The models of \citet{Des16} spanned a larger range of explosion
energies and mixing prescriptions than the present study. The larger
range also resulted from a broad choice of uncertain parameters. For
example, we do not consider the higher energy, large
$^{56}$Ni-producing events, since our neutrino-powered explosions do
not produce them.  Mixing here also has a more physical, though still
uncertain basis (\Sect{mix}). A relation between $t_{\rm peak}$
bolometric and $\Delta M_{15}$ in the $B$ band shows greater scatter,
perhaps reflecting inaccuracies in our $B$-band determination, and is
not recommended. We recommend that observers assume a value $\beta =
1.3$ when using these figures. This gives slightly greater weight to
our models ($\beta = 1.36$) than those of Dessart et al ($\beta =
1.17$).

All these approximations are derived from fitting models for Type Ib
and Ic supernovae whose light curves are powered by the decay of
$^{56}$Ni. Without further study, they are not directly applicable to other
situations.

\begin{figure}
\includegraphics[width=0.48\textwidth]{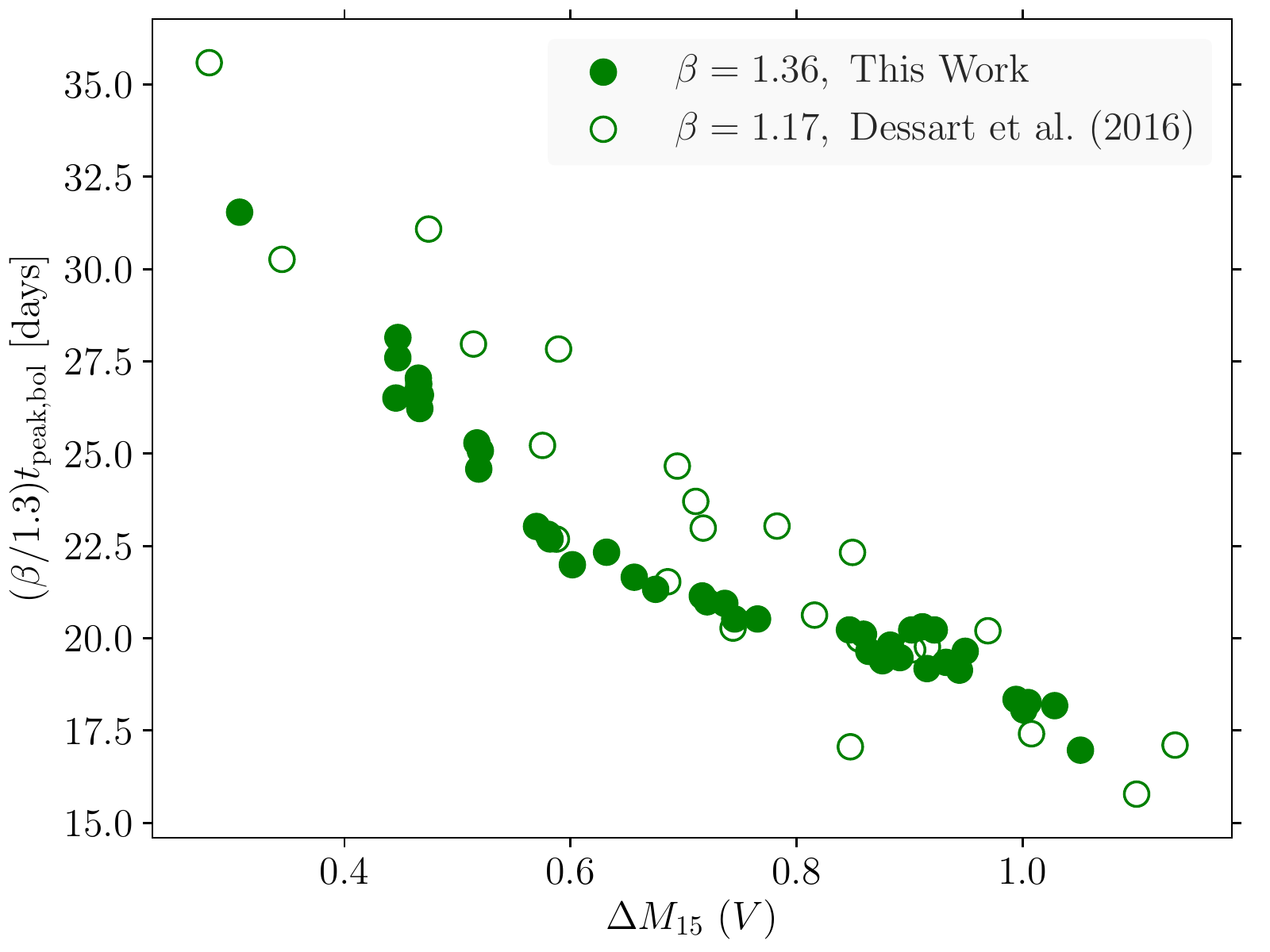}
\includegraphics[width=0.48\textwidth]{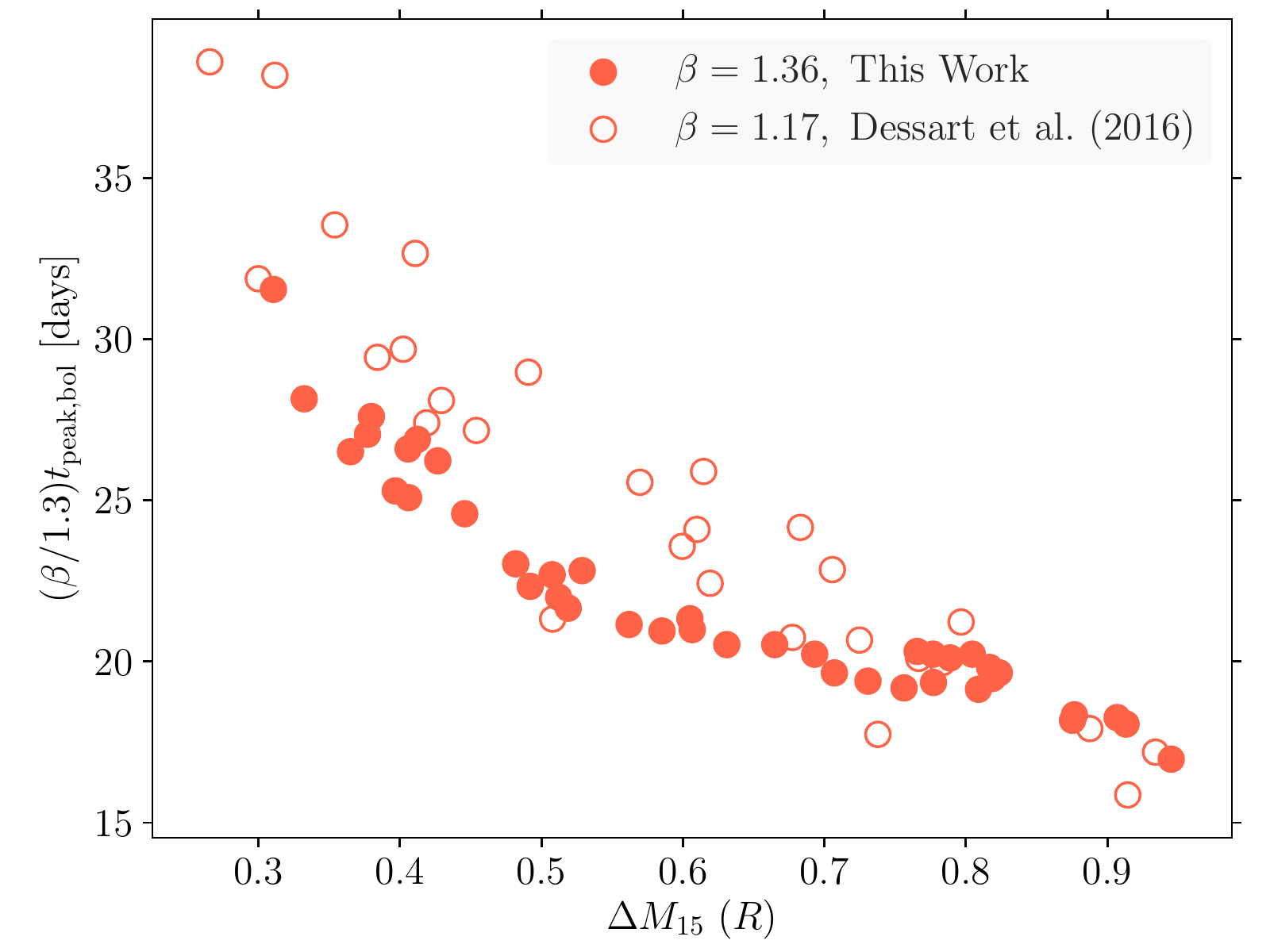}
\caption{The scaled time of bolometric peak, $\beta/1.3$ times
  $t_{peak}$ is plotted against the decline in the $V$ magnitude 15
  days after $V$-band peak (top) and decline in $R$-band (bottom). For the
  Dessart et al. data set $\beta$ = 1.17; for the current models
  $\beta = 1.36$. A larger spread is seen in the Dessart et al. models
  which included a broader range of mixing prescriptions and explosion
  energies than the current study. \lFig{tpeakv15}}
\end{figure}

\subsection{Arnett's Rule}
\lSect{arnett}

Originally derived with Type Ia supernovae in mind, but since often
applied to all manner of Type I supernovae, Arnett's Rule
\citep{Arn82} says the the bolometric luminosity, at peak, is equal to
the current power being generated by the decay of $^{56}$Ni and
$^{56}$Co. This radioactive power includes all radiation, trapped and
untrapped. The Rule is often used to infer the mass of $^{56}$Ni made
in an explosion whose bolometric luminosity at peak has been
determined \citep[e.g.,][]{Dro11}.  While a very good approximation
for Type Ia supernovae \citep[e.g.,][]{Str06}, previous studies have
shown that the rule leads to a substantial overestimate of the
$^{56}$Ni actually required to power Type Ib and Ic supernovae
\citep{Des15,Des16}. The reasons for deviations and discussion of the
restrictions for the rule's use have been discussed in the context of
a more general analytic derivation by \citet{Kha19}. The rule breaks
down because of violations of the assumptions in its original
variation, namely constant opacity and self-similarity of the energy
density profile.

\citet{Des16} also examined the relation between luminosity and radioactive
power at peak for a grid of Type Ib and Ic supernova light curves
calculated from current models of the explosion using the CMFGEN code
\citep{Hil12}.

\begin{figure}
\begin{center}
\includegraphics[width=0.48\textwidth]{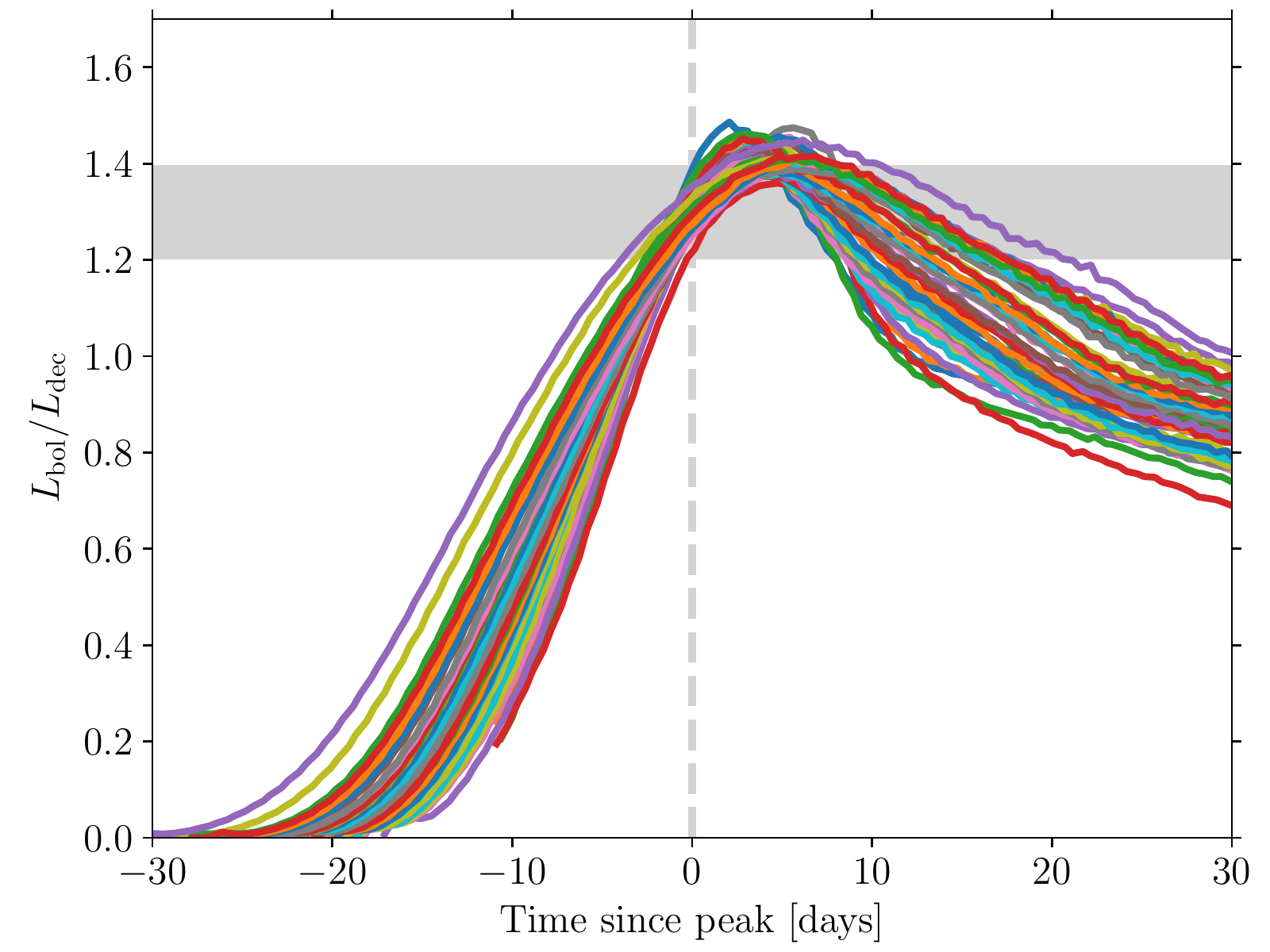}
\caption{Ratio of the bolometric luminosity to current total power
  from the decay of $^{56}$Ni and $^{56}$Co as a function of
  time. Time zero is adjusted here to be the peak of the bolometric
  light curve. The average ratio and error bar, $1.30 \pm 0.095$, are
  indicated by the gray band. The ratio is sensitive to the timing of
  the the peak, but ``Arnett's Rule'' significantly overestimates the
  amount of $^{56}$Ni required to explain the observed light curves of
  Type Ib and Ic supernovae. A similar study by \citet{Des16} found a
  ratio at peak of $1.41 \pm 0.072$.  \lFig{Arnett}}
\end{center}
\end{figure}

  \setcounter{table}{0}
  \setcounter{figure}{0}

\section{Model Properties}
\lSect{tables}

Here we provide additional details for all of the standard Type Ib and
Ic models that used the W18 central engine \citep{Ert20} and the
standard \citet{Yoo17} mass loss rates with $f_{\rm WR}$ = 1.  The
peak luminosities and magnitudes (listed in \Tab{peak_main}) and their
corresponding peak times (\Tab{tscales_main}) are measured using a
third degree polynomial fit to the brightest 5 points in the light
curve.  The rise ($t_{-1/2}$) and decline ($t_{+1/2}$) times are
measured between peak and 50\% of the peak for luminosity, and between
peak and peak plus 0.75 magnitude for the broad band filters
  \citep[e.g.,][]{Ho20}. The decline rates $\Delta M_{15}$
(\Tab{decline}) are measured between peak and 15 days post peak. The
bolometric and multiband light curves for our entire set of models are
available on request.

\begin{deluxetable*}{ccccccccccccc}
\tablecaption{Peak Luminosities and Magnitudes ($f_{\rm WR}=1$)}
\tablehead{ \colhead{$M_{\rm He,i}$}   &
            \colhead{$\log\ L_{\rm bol}$}  &
            \colhead{$\log\ L_{\rm 4-10k}$} &
            \colhead{$U_{\rm peak}$}  &
            \colhead{$B_{\rm peak}$}  &
            \colhead{$V_{\rm peak}$}  &
            \colhead{$R_{\rm peak}$}  &
            \colhead{$I_{\rm peak}$}  &
            \colhead{$u_{\rm peak}$}  &
            \colhead{$g_{\rm peak}$}  &
            \colhead{$r_{\rm peak}$}  &
            \colhead{$i_{\rm peak}$}  &
            \colhead{$z_{\rm peak}$}
            \\
            \colhead{[\Msun]}  &
            \colhead{[$\rm erg\ s^{-1}$]}  &
            \colhead{[$\rm erg\ s^{-1}$]}  &
            \colhead{}  &
            \colhead{}  &
            \colhead{}  &
            \colhead{}  &
            \colhead{}  &
            \colhead{}  &
            \colhead{}  &
            \colhead{}  &
            \colhead{}  &
            \colhead{}
            }\\
\startdata
  2.70  &  41.89  &  41.78  &  -16.26  &  -15.85  &  -16.41  &  -16.53  &  -16.73  &  -15.33  &  -16.12  &  -16.39  &  -16.23  &  -16.46 \\
  2.80  &  41.87  &  41.76  &  -15.81  &  -15.74  &  -16.35  &  -16.52  &  -16.70  &  -14.87  &  -16.03  &  -16.38  &  -16.21  &  -16.48 \\
  2.90  &  42.02  &  41.92  &  -16.23  &  -16.27  &  -16.79  &  -16.86  &  -17.02  &  -15.19  &  -16.53  &  -16.74  &  -16.51  &  -16.80 \\
  3.20  &  42.21  &  42.11  &  -16.59  &  -16.77  &  -17.30  &  -17.34  &  -17.50  &  -15.51  &  -17.05  &  -17.22  &  -16.99  &  -17.38 \\
  3.30  &  42.17  &  42.07  &  -16.50  &  -16.63  &  -17.16  &  -17.24  &  -17.40  &  -15.42  &  -16.90  &  -17.10  &  -16.91  &  -17.26 \\
  3.40  &  42.01  &  41.91  &  -15.86  &  -16.14  &  -16.73  &  -16.87  &  -17.03  &  -14.76  &  -16.42  &  -16.73  &  -16.57  &  -16.88 \\
  3.50  &  42.02  &  41.92  &  -15.79  &  -16.13  &  -16.75  &  -16.90  &  -17.07  &  -14.65  &  -16.43  &  -16.76  &  -16.61  &  -16.91 \\
  3.60  &  41.94  &  41.84  &  -15.48  &  -15.89  &  -16.55  &  -16.73  &  -16.89  &  -14.31  &  -16.20  &  -16.58  &  -16.46  &  -16.72 \\
  3.70  &  42.21  &  42.11  &  -16.54  &  -16.70  &  -17.25  &  -17.39  &  -17.54  &  -15.46  &  -16.98  &  -17.25  &  -17.06  &  -17.40 \\
  3.80  &  42.25  &  42.15  &  -16.66  &  -16.80  &  -17.35  &  -17.49  &  -17.63  &  -15.58  &  -17.07  &  -17.35  &  -17.16  &  -17.50 \\
  3.90  &  42.14  &  42.04  &  -16.30  &  -16.48  &  -17.05  &  -17.25  &  -17.40  &  -15.21  &  -16.74  &  -17.09  &  -16.96  &  -17.24 \\
  4.00  &  42.16  &  42.06  &  -16.44  &  -16.55  &  -17.11  &  -17.29  &  -17.45  &  -15.36  &  -16.81  &  -17.14  &  -16.99  &  -17.30 \\
  4.10  &  42.14  &  42.04  &  -16.19  &  -16.44  &  -17.04  &  -17.23  &  -17.39  &  -15.07  &  -16.71  &  -17.08  &  -16.94  &  -17.24 \\
  4.20  &  42.22  &  42.12  &  -16.45  &  -16.65  &  -17.25  &  -17.43  &  -17.58  &  -15.35  &  -16.92  &  -17.27  &  -17.13  &  -17.44 \\
  4.30  &  42.31  &  42.21  &  -16.79  &  -16.92  &  -17.50  &  -17.66  &  -17.80  &  -15.71  &  -17.19  &  -17.51  &  -17.33  &  -17.65 \\
  4.40  &  42.31  &  42.20  &  -16.74  &  -16.88  &  -17.47  &  -17.64  &  -17.78  &  -15.66  &  -17.15  &  -17.48  &  -17.32  &  -17.64 \\
  4.50  &  42.39  &  42.29  &  -16.95  &  -17.09  &  -17.70  &  -17.83  &  -17.98  &  -15.87  &  -17.37  &  -17.69  &  -17.48  &  -17.83 \\
  4.62  &  42.42  &  42.32  &  -17.07  &  -17.18  &  -17.78  &  -17.92  &  -18.06  &  -16.01  &  -17.47  &  -17.78  &  -17.57  &  -17.91 \\
  4.75  &  42.42  &  42.32  &  -17.03  &  -17.16  &  -17.77  &  -17.91  &  -18.05  &  -15.95  &  -17.45  &  -17.77  &  -17.56  &  -17.91 \\
  4.88  &  42.42  &  42.31  &  -16.95  &  -17.14  &  -17.77  &  -17.91  &  -18.05  &  -15.87  &  -17.42  &  -17.77  &  -17.56  &  -17.91 \\
  5.00  &  42.42  &  42.32  &  -17.02  &  -17.16  &  -17.78  &  -17.93  &  -18.07  &  -15.94  &  -17.44  &  -17.78  &  -17.58  &  -17.93 \\
  5.13  &  42.39  &  42.28  &  -16.82  &  -17.02  &  -17.67  &  -17.84  &  -17.97  &  -15.73  &  -17.32  &  -17.68  &  -17.50  &  -17.85 \\
  5.25  &  42.33  &  42.23  &  -16.61  &  -16.86  &  -17.53  &  -17.71  &  -17.85  &  -15.49  &  -17.16  &  -17.55  &  -17.39  &  -17.73 \\
  5.38  &  42.45  &  42.34  &  -17.06  &  -17.19  &  -17.83  &  -17.99  &  -18.12  &  -15.98  &  -17.47  &  -17.84  &  -17.66  &  -18.00 \\
  5.50  &  42.40  &  42.30  &  -16.72  &  -16.99  &  -17.70  &  -17.91  &  -18.03  &  -15.59  &  -17.29  &  -17.74  &  -17.60  &  -17.91 \\
  5.63  &  42.32  &  42.22  &  -16.53  &  -16.81  &  -17.50  &  -17.69  &  -17.83  &  -15.40  &  -17.11  &  -17.53  &  -17.39  &  -17.70 \\
  5.75  &  42.33  &  42.22  &  -16.55  &  -16.81  &  -17.50  &  -17.69  &  -17.82  &  -15.41  &  -17.12  &  -17.53  &  -17.39  &  -17.71 \\
  5.88  &  42.40  &  42.29  &  -16.83  &  -17.03  &  -17.68  &  -17.87  &  -17.99  &  -15.73  &  -17.31  &  -17.71  &  -17.56  &  -17.88 \\
  6.00  &  42.23  &  42.12  &  -16.05  &  -16.50  &  -17.24  &  -17.46  &  -17.61  &  -14.86  &  -16.83  &  -17.29  &  -17.20  &  -17.45 \\
  6.13  &  42.40  &  42.30  &  -16.81  &  -17.03  &  -17.69  &  -17.89  &  -18.02  &  -15.68  &  -17.32  &  -17.73  &  -17.58  &  -17.89 \\
  6.25  &  42.39  &  42.28  &  -16.72  &  -16.98  &  -17.65  &  -17.85  &  -17.98  &  -15.60  &  -17.27  &  -17.69  &  -17.55  &  -17.85 \\
  6.38  &  42.36  &  42.25  &  -16.59  &  -16.89  &  -17.58  &  -17.79  &  -17.92  &  -15.45  &  -17.20  &  -17.63  &  -17.50  &  -17.79 \\
  6.50  &  42.35  &  42.24  &  -16.48  &  -16.83  &  -17.54  &  -17.77  &  -17.90  &  -15.31  &  -17.14  &  -17.60  &  -17.49  &  -17.76 \\
  6.63  &  42.36  &  42.25  &  -16.56  &  -16.88  &  -17.57  &  -17.79  &  -17.92  &  -15.40  &  -17.19  &  -17.63  &  -17.51  &  -17.78 \\
  6.75  &  42.13  &  42.02  &  -15.64  &  -16.23  &  -16.97  &  -17.22  &  -17.38  &  -14.41  &  -16.57  &  -17.04  &  -16.98  &  -17.21 \\
  6.88  &  42.33  &  42.22  &  -16.38  &  -16.77  &  -17.48  &  -17.71  &  -17.84  &  -15.19  &  -17.09  &  -17.54  &  -17.44  &  -17.71 \\
  7.00  &  42.33  &  42.22  &  -16.33  &  -16.76  &  -17.48  &  -17.72  &  -17.85  &  -15.13  &  -17.07  &  -17.55  &  -17.45  &  -17.70 \\
  7.13  &  42.41  &  42.30  &  -16.73  &  -17.01  &  -17.68  &  -17.91  &  -18.04  &  -15.57  &  -17.30  &  -17.74  &  -17.63  &  -17.92 \\
  7.25  &  42.26  &  42.14  &  -15.94  &  -16.52  &  -17.27  &  -17.52  &  -17.68  &  -14.72  &  -16.85  &  -17.34  &  -17.28  &  -17.51 \\
  7.38  &  42.34  &  42.23  &  -16.33  &  -16.77  &  -17.49  &  -17.74  &  -17.87  &  -15.14  &  -17.09  &  -17.56  &  -17.48  &  -17.72 \\
  7.50  &  42.33  &  42.21  &  -16.24  &  -16.71  &  -17.45  &  -17.69  &  -17.84  &  -15.02  &  -17.04  &  -17.52  &  -17.44  &  -17.68 \\
  7.63  &  42.34  &  42.22  &  -16.25  &  -16.74  &  -17.47  &  -17.72  &  -17.85  &  -15.04  &  -17.06  &  -17.54  &  -17.47  &  -17.70 \\
  7.75  &  42.34  &  42.23  &  -16.27  &  -16.75  &  -17.49  &  -17.74  &  -17.89  &  -15.05  &  -17.08  &  -17.56  &  -17.49  &  -17.72 \\
  7.88  &  42.33  &  42.21  &  -16.17  &  -16.70  &  -17.46  &  -17.71  &  -17.86  &  -14.95  &  -17.03  &  -17.54  &  -17.47  &  -17.69 \\
  8.00  &  42.01  &  41.88  &  -14.85  &  -15.68  &  -16.59  &  -16.91  &  -17.10  &  -13.58  &  -16.09  &  -16.71  &  -16.73  &  -16.87 \\
\\
  12.0  &  42.06  &  41.92  &  -14.23  &  -15.61  &  -16.69  &  -17.03  &  -17.23  &  -12.72  &  -16.10  &  -16.83  &  -16.86  &  -17.00 \\
  19.0  &  42.12  &  41.99  &  -14.35  &  -15.80  &  -16.82  &  -17.17  &  -17.40  &  -12.87  &  -16.28  &  -16.98  &  -17.01  &  -17.19
\enddata  
\tablecomments{ $L_{\rm 4-10k}$ covers only 4000 -- 10000 \AA. $UBVRI$
  and $ugriz$ magnitudes are based on Vega-- and AB--systems
  respectively. For the peak, rise, and decline times for each model
  see \Tab{tscales_main}.}  \lTab{peak_main}
\end{deluxetable*}

\begin{deluxetable*}{crrrrrcrrrrrcrrrrr}
\tablecaption{Rise, Decline, and Peak Times ($f_{\rm WR}=1$)}
\tablehead{ \colhead{$M_{\rm He,i}$}   &
            \multicolumn{5}{c}{$t_{\rm -1/2}$} &
            \colhead{} &
            \multicolumn{5}{c}{$t_{\rm peak}$} &
            \colhead{} &
            \multicolumn{5}{c}{$t_{\rm +1/2}$}
            \\
            \colhead{[\Msun]}  &
            \multicolumn{5}{c}{[days]} &
            \colhead{} &
            \multicolumn{5}{c}{[days]} &
            \colhead{} &
            \multicolumn{5}{c}{[days]}
            }\\
\startdata
 & $L_{\rm bol}$ & $B$ & $V$ & $R$ & $I$ & & $L_{\rm bol}$ & $B$ & $V$ & $R$ & $I$ & & $L_{\rm bol}$ & $B$ & $V$ & $R$ & $I$ \\
 \cline{2-6} \cline{8-12} \cline{14-18}\\
  2.70 & 15.5 & 12.4 &  8.4 &  7.9 &  7.7 && 15.7 & 13.1 & 16.2 & 16.5 & 16.5 && 10.0 &  8.4 &  8.7 & 10.2 & 11.7\\
  2.80 & 15.9 & 13.3 &  9.3 &  8.8 &  8.3 && 16.1 & 14.0 & 16.5 & 17.2 & 16.8 && 11.1 &  8.6 &  9.7 & 10.2 & 12.3\\
  2.90 & 15.6 &  7.6 &  8.6 &  9.3 &  9.6 && 15.8 & 13.8 & 16.0 & 17.4 & 17.8 && 10.6 &  8.0 &  9.1 & 10.0 & 12.5\\
  3.20 & 15.0 &  7.5 &  8.6 & 10.9 & 10.6 && 16.1 & 13.8 & 16.4 & 19.2 & 19.0 && 12.0 &  9.1 & 10.0 & 10.5 & 14.0\\
  3.30 &  9.3 &  7.4 &  9.1 & 11.5 & 10.0 && 17.3 & 15.0 & 17.5 & 20.6 & 19.0 && 12.7 &  8.9 & 10.3 & 10.3 & 15.5\\
  3.40 & 10.2 &  7.9 & 10.1 & 10.7 & 11.1 && 18.6 & 15.8 & 18.7 & 19.8 & 20.2 && 13.3 &  9.5 & 10.4 & 12.4 & 15.6\\
  3.50 &  9.7 &  7.8 & 10.0 & 10.6 & 10.9 && 18.4 & 16.0 & 18.8 & 19.8 & 20.0 && 14.1 &  9.8 & 10.8 & 12.7 & 16.4\\
  3.60 & 10.1 &  8.1 & 10.2 & 11.7 & 11.0 && 19.2 & 16.8 & 19.3 & 21.2 & 20.4 && 15.5 & 10.4 & 11.8 & 13.0 & 18.0\\
  3.70 & 10.1 &  7.1 &  9.5 & 11.7 & 13.3 && 18.6 & 15.3 & 18.4 & 21.5 & 22.8 && 14.0 & 10.0 & 11.5 & 11.4 & 13.7\\
  3.80 & 10.4 &  6.7 &  9.3 & 11.7 & 10.4 && 18.8 & 14.8 & 18.0 & 21.3 & 19.9 && 13.5 & 10.2 & 11.6 & 11.1 & 16.5\\
  3.90 & 10.9 &  7.6 & 10.5 & 12.7 & 12.9 && 20.0 & 16.4 & 19.6 & 22.8 & 22.8 && 16.8 & 11.0 & 13.2 & 13.2 & 17.9\\
  4.00 & 12.2 &  7.5 & 11.9 & 12.3 & 12.4 && 21.0 & 15.8 & 20.9 & 22.3 & 22.4 && 14.9 & 11.1 & 11.3 & 13.2 & 17.9\\
  4.10 & 12.5 &  7.6 & 11.9 & 12.2 & 12.4 && 21.7 & 16.5 & 21.2 & 22.3 & 22.4 && 15.1 & 11.3 & 11.9 & 14.1 & 18.3\\
  4.20 & 11.9 &  7.4 & 11.3 & 12.7 & 12.1 && 21.0 & 16.0 & 20.4 & 22.8 & 22.1 && 15.2 & 11.1 & 12.0 & 12.9 & 17.6\\
  4.30 & 11.7 &  7.8 & 12.3 & 11.9 & 10.9 && 20.8 & 16.3 & 21.6 & 22.4 & 21.2 && 15.3 & 10.7 & 11.0 & 13.7 & 19.8\\
  4.40 & 11.8 &  8.1 & 12.1 & 11.9 & 11.9 && 20.9 & 16.5 & 21.2 & 22.3 & 22.3 && 15.6 & 10.7 & 11.6 & 13.9 & 18.9\\
  4.50 & 10.8 &  7.4 & 10.5 & 10.9 & 10.1 && 19.5 & 15.3 & 19.4 & 20.8 & 20.0 && 14.3 & 10.2 & 11.3 & 13.1 & 18.2\\
  4.62 & 10.9 &  7.3 & 10.4 & 11.3 & 10.4 && 19.7 & 15.3 & 19.3 & 21.4 & 20.4 && 15.1 & 10.7 & 12.2 & 13.4 & 19.0\\
  4.75 & 10.8 &  8.3 & 10.5 & 10.7 & 10.1 && 19.5 & 16.2 & 19.3 & 20.7 & 20.0 && 15.2 &  9.9 & 12.1 & 14.2 & 19.9\\
  4.88 & 11.1 &  8.0 & 10.6 & 11.5 & 10.4 && 20.0 & 16.0 & 19.6 & 21.6 & 20.4 && 15.5 & 10.5 & 12.4 & 13.8 & 20.5\\
  5.00 & 11.2 &  8.6 & 10.8 & 10.9 & 10.4 && 20.1 & 16.6 & 19.8 & 21.1 & 20.5 && 16.0 & 10.4 & 12.7 & 15.0 & 20.8\\
  5.13 & 11.5 &  8.2 & 10.7 & 12.2 & 10.8 && 20.6 & 16.3 & 19.8 & 22.6 & 21.0 && 16.8 & 11.6 & 14.0 & 14.5 & 22.6\\
  5.25 & 11.8 &  8.9 & 12.0 & 11.5 & 11.0 && 21.0 & 17.2 & 21.2 & 21.8 & 21.1 && 17.4 & 11.5 & 13.6 & 16.2 & 22.9\\
  5.38 & 12.1 &  8.5 & 11.9 & 11.8 & 12.5 && 21.2 & 16.6 & 21.0 & 22.3 & 22.8 && 18.0 & 12.1 & 14.1 & 16.7 & 23.6\\
  5.50 & 12.6 &  9.3 & 11.3 & 11.8 & 12.2 && 22.0 & 17.6 & 20.5 & 22.5 & 22.4 && 18.5 & 12.1 & 15.7 & 17.4 & 26.2\\
  5.63 & 12.6 &  8.6 & 11.9 & 13.2 & 12.3 && 22.0 & 17.2 & 21.3 & 23.6 & 22.4 && 18.3 & 12.3 & 14.6 & 16.0 & 23.5\\
  5.75 & 12.7 &  8.6 & 11.7 & 13.2 & 12.0 && 22.3 & 17.3 & 21.4 & 23.8 & 22.2 && 19.3 & 12.9 & 15.7 & 17.0 & 25.9\\
  5.88 & 12.5 &  8.3 & 12.2 & 13.4 & 12.4 && 22.0 & 17.0 & 21.7 & 24.2 & 22.7 && 19.8 & 12.9 & 15.5 & 17.0 & 27.0\\
  6.00 & 13.6 &  9.3 & 13.4 & 13.8 & 14.1 && 24.0 & 18.8 & 23.6 & 24.8 & 24.9 && 21.6 & 14.1 & 16.8 & 19.8 & 29.0\\
  6.13 & 12.9 &  8.9 & 11.7 & 13.4 & 14.7 && 22.7 & 17.9 & 21.6 & 24.4 & 25.2 && 20.5 & 13.0 & 17.1 & 18.2 & 26.8\\
  6.25 & 13.2 &  8.8 & 13.1 & 13.3 & 12.9 && 23.2 & 18.0 & 23.1 & 24.3 & 23.4 && 20.7 & 13.4 & 16.1 & 19.0 & 29.6\\
  6.38 & 13.5 &  9.1 & 12.5 & 13.5 & 13.5 && 23.7 & 18.6 & 22.6 & 24.6 & 24.0 && 20.7 & 13.1 & 17.2 & 19.0 & 29.5\\
  6.50 & 13.4 &  7.9 & 13.2 & 14.0 & 16.0 && 24.0 & 17.8 & 23.6 & 25.4 & 26.8 && 21.5 & 14.4 & 17.2 & 19.1 & 28.1\\
  6.63 & 13.5 &  8.8 & 12.2 & 14.1 & 13.0 && 24.2 & 18.8 & 22.8 & 25.6 & 24.0 && 21.8 & 13.7 & 18.4 & 19.4 & 32.3\\
  6.75 & 14.8 &  8.9 & 14.6 & 17.2 & 16.6 && 27.0 & 20.8 & 26.7 & 29.8 & 29.3 && 28.4 & 17.2 & 21.7 & 23.7 & 35.8\\
  6.88 & 14.3 &  9.3 & 13.5 & 15.4 & 16.0 && 25.8 & 20.2 & 24.8 & 27.6 & 27.8 && 24.6 & 15.0 & 19.7 & 21.4 & 32.5\\
  7.00 & 13.6 &  8.7 & 13.4 & 15.8 & 13.6 && 25.2 & 19.6 & 25.0 & 28.1 & 25.5 && 25.1 & 16.0 & 19.6 & 21.0 & 35.2\\
  7.13 & 13.9 &  8.6 & 13.6 & 15.0 & 13.4 && 25.2 & 19.3 & 24.8 & 27.2 & 25.0 && 24.0 & 15.2 & 19.3 & 21.2 & 35.0\\
  7.25 & 16.8 & 10.0 & 14.7 & 16.3 & 17.1 && 29.9 & 22.4 & 27.7 & 30.0 & 30.6 && 27.3 & 17.9 & 23.2 & 26.1 & 38.3\\
  7.38 & 14.9 &  9.9 & 13.2 & 15.9 & 17.0 && 26.9 & 21.2 & 25.1 & 28.6 & 29.2 && 25.1 & 15.6 & 21.4 & 22.2 & 34.7\\
  7.50 & 14.4 &  8.8 & 14.7 & 16.2 & 16.4 && 26.6 & 20.3 & 26.8 & 29.2 & 29.0 && 26.4 & 17.0 & 20.4 & 22.7 & 35.5\\
  7.63 & 14.8 &  9.7 & 14.2 & 17.1 & 14.3 && 27.1 & 21.4 & 26.4 & 30.2 & 27.1 && 26.8 & 16.3 & 21.4 & 22.0 & 37.4\\
  7.75 & 15.5 &  9.0 & 14.8 & 16.0 & 16.2 && 28.0 & 20.8 & 27.2 & 29.2 & 29.2 && 25.9 & 17.1 & 20.7 & 23.6 & 35.2\\
  7.88 & 15.0 &  9.5 & 13.6 & 16.7 & 15.7 && 27.9 & 21.6 & 26.4 & 30.2 & 29.0 && 26.7 & 17.1 & 22.4 & 23.1 & 36.6\\
  8.00 & 17.6 & 11.0 & 15.3 & 17.5 & 19.0 && 32.6 & 25.2 & 29.8 & 32.6 & 34.4 && 30.2 & 20.3 & 25.5 & 27.9 & 40.0\\
\\
 12.00 & 19.6 & 12.8 & 19.4 & 20.4 & 23.1 && 39.2 & 32.8 & 39.2 & 40.2 & 43.2 && 40.6 & 27.3 & 31.3 & 37.1 & 52.0\\
 19.00 & 24.5 & 16.6 & 23.4 & 28.2 & 27.9 && 47.4 & 39.9 & 46.5 & 51.5 & 51.7 && 43.1 & 29.6 & 35.4 & 37.5 & 59.4
 \enddata
\tablecomments{The rise ($t_{-1/2}$) and decline ($t_{+1/2}$)
  times are the intervals required for a 0.75 change in magnitude (i.e.,
  50\% change in flux) relative to their peak values.}
\lTab{tscales_main}
\end{deluxetable*}

\begin{deluxetable}{ccccccccc}
\tablecaption{$\Delta M_{15}$ Decline Rates ($f_{\rm WR}=1$)}
\tablehead{ \colhead{$M_{\rm He,i}$}   &
            \colhead{$\Delta B_{15}$}  &
            \colhead{$\Delta V_{15}$}  &
            \colhead{$\Delta R_{15}$}  &
            \colhead{$\Delta I_{15}$}  &
            \colhead{$\Delta g_{15}$}  &
            \colhead{$\Delta r_{15}$}  &
            \colhead{$\Delta i_{15}$}  &
            \colhead{$\Delta z_{15}$}
            \\
            \colhead{[\Msun]}  &
            \colhead{}  &
            \colhead{}  &
            \colhead{}  &
            \colhead{}  &
            \colhead{}  &
            \colhead{}  &
            \colhead{}
            }\\
\startdata
2.70  &  1.35  &  1.15  &  1.10  &  0.92  &  1.15  &  1.16  &  0.99  &  0.62 \\
2.80  &  1.30  &  1.10  &  1.11  &  0.86  &  1.12  &  1.18  &  0.98  &  0.61 \\
2.90  &  1.45  &  1.17  &  1.10  &  0.87  &  1.24  &  1.18  &  1.00  &  0.59 \\
3.20  &  1.49  &  1.12  &  1.07  &  0.78  &  1.28  &  1.15  &  0.93  &  0.64 \\
3.30  &  1.47  &  1.09  &  1.05  &  0.73  &  1.26  &  1.09  &  0.90  &  0.64 \\
3.40  &  1.37  &  1.05  &  0.92  &  0.74  &  1.15  &  0.98  &  0.88  &  0.59 \\
3.50  &  1.33  &  1.04  &  0.89  &  0.70  &  1.15  &  0.98  &  0.90  &  0.59 \\
3.60  &  1.26  &  0.95  &  0.88  &  0.64  &  1.07  &  0.90  &  0.85  &  0.54 \\
3.70  &  1.37  &  0.99  &  0.97  &  0.77  &  1.19  &  1.05  &  0.83  &  0.60 \\
3.80  &  1.34  &  0.99  &  1.00  &  0.68  &  1.20  &  1.08  &  0.83  &  0.62 \\
3.90  &  1.24  &  0.86  &  0.85  &  0.66  &  1.02  &  0.93  &  0.75  &  0.54 \\
4.00  &  1.22  &  0.97  &  0.86  &  0.66  &  1.02  &  0.94  &  0.76  &  0.56 \\
4.10  &  1.20  &  0.93  &  0.81  &  0.63  &  0.99  &  0.89  &  0.73  &  0.55 \\
4.20  &  1.19  &  0.94  &  0.86  &  0.65  &  1.01  &  0.88  &  0.73  &  0.55 \\
4.30  &  1.24  &  0.99  &  0.82  &  0.60  &  1.03  &  0.89  &  0.68  &  0.54 \\
4.40  &  1.23  &  0.96  &  0.82  &  0.62  &  1.02  &  0.83  &  0.67  &  0.51 \\
4.50  &  1.29  &  1.00  &  0.87  &  0.63  &  1.13  &  0.96  &  0.73  &  0.55 \\
4.63  &  1.25  &  0.95  &  0.85  &  0.61  &  1.08  &  0.94  &  0.60  &  0.54 \\
4.75  &  1.28  &  0.95  &  0.82  &  0.57  &  1.11  &  0.91  &  0.56  &  0.54 \\
4.88  &  1.22  &  0.91  &  0.83  &  0.56  &  1.02  &  0.92  &  0.54  &  0.51 \\
5.00  &  1.24  &  0.89  &  0.75  &  0.55  &  1.02  &  0.82  &  0.54  &  0.50 \\
5.13  &  1.12  &  0.80  &  0.78  &  0.52  &  1.00  &  0.84  &  0.62  &  0.49 \\
5.25  &  1.14  &  0.84  &  0.67  &  0.49  &  0.97  &  0.77  &  0.58  &  0.47 \\
5.38  &  1.07  &  0.80  &  0.66  &  0.50  &  0.94  &  0.68  &  0.49  &  0.46 \\
5.50  &  1.06  &  0.70  &  0.62  &  0.45  &  0.91  &  0.65  &  0.55  &  0.40 \\
5.63  &  1.02  &  0.78  &  0.70  &  0.46  &  0.85  &  0.69  &  0.58  &  0.45 \\
5.75  &  0.97  &  0.71  &  0.64  &  0.41  &  0.83  &  0.63  &  0.52  &  0.42 \\
5.88  &  0.97  &  0.72  &  0.65  &  0.39  &  0.88  &  0.62  &  0.48  &  0.41 \\
6.00  &  0.84  &  0.65  &  0.52  &  0.39  &  0.77  &  0.56  &  0.48  &  0.34 \\
6.13  &  0.95  &  0.64  &  0.58  &  0.45  &  0.76  &  0.59  &  0.46  &  0.37 \\
6.25  &  0.91  &  0.69  &  0.54  &  0.37  &  0.83  &  0.58  &  0.42  &  0.37 \\
6.38  &  0.94  &  0.64  &  0.54  &  0.37  &  0.77  &  0.58  &  0.50  &  0.35 \\
6.50  &  0.79  &  0.63  &  0.54  &  0.43  &  0.72  &  0.59  &  0.51  &  0.32 \\
6.63  &  0.86  &  0.57  &  0.52  &  0.32  &  0.71  &  0.53  &  0.47  &  0.36 \\
6.75  &  0.59  &  0.47  &  0.41  &  0.31  &  0.51  &  0.45  &  0.32  &  0.29 \\
6.88  &  0.75  &  0.52  &  0.47  &  0.34  &  0.61  &  0.52  &  0.40  &  0.28 \\
7.00  &  0.68  &  0.52  &  0.48  &  0.27  &  0.56  &  0.53  &  0.44  &  0.28 \\
7.13  &  0.73  &  0.54  &  0.47  &  0.26  &  0.70  &  0.49  &  0.37  &  0.31 \\
7.25  &  0.54  &  0.41  &  0.35  &  0.29  &  0.49  &  0.39  &  0.35  &  0.22 \\
7.38  &  0.71  &  0.46  &  0.43  &  0.33  &  0.59  &  0.48  &  0.38  &  0.29 \\
7.50  &  0.60  &  0.50  &  0.41  &  0.30  &  0.63  &  0.46  &  0.39  &  0.27 \\
7.63  &  0.64  &  0.47  &  0.45  &  0.24  &  0.56  &  0.47  &  0.41  &  0.24 \\
7.75  &  0.58  &  0.48  &  0.41  &  0.32  &  0.52  &  0.35  &  0.37  &  0.22 \\
7.88  &  0.59  &  0.41  &  0.41  &  0.26  &  0.57  &  0.42  &  0.37  &  0.24 \\
8.00  &  0.43  &  0.29  &  0.28  &  0.26  &  0.35  &  0.30  &  0.32  &  0.21 \\
\\
12.00  &  0.23  &  0.22  &  0.19  &  0.19  &  0.22  &  0.18  &  0.24  &  0.13 \\
19.00  &  0.21  &  0.16  &  0.17  &  0.12  &  0.17  &  0.19  &  0.15  &  0.09
\enddata
\tablecomments{$UBVRI$ and $ugriz$ magnitudes are based on Vega-- and
  AB--systems respectively. The peak magnitudes are given in
  \Tab{peak_main}, and peak, rise, and decline times for each model
  are given in \Tab{tscales_main}.}  \lTab{decline}
\end{deluxetable}

\clearpage

\setcounter{table}{0}
\setcounter{figure}{0}

\section{Filters and Zeropoints}
\lSect{offset}

Observers quantify their measurements using various combinations of
filters and zeropoint systems \citep[e.g.,][]{Bes05}.  To accommodate
the historical conventions as well as the specific types of data
employed in this study, we have used three sets of filters: the
generic Johnson--Cousins $UBVRI$ set \citep{Bes83}, the standard Sloan
$ugriz$ set (SDSS), and the $uBgVri$ set employed by the Carnegie
Supernova Project (CSP). All filter transmission characteristics were
obtained from \url{svo2.cab.inta-csic.es/theory/fps}, and a comparison
of their normalized transmission functions are given in \Fig{filters}.
For the most part CSP filters have a very similar response to the
equivalent filters of Johnson--Cousins and SDSS sets. There are,
however, small but noticeable differences generally with the CSP
filters having slightly higher transmission at longer wavelengths.

These three filter sets have been paired with three different
zeropoint systems. The Johnson--Cousins set was always normalized to
the Vega system, with zeropoint fluxes of 1564, 4024, 3563, 3028, 2458
Jy for $U$, $B$, $V$, $R$, and $I$ filters respectively (also obtained
from \url{svo2.cab.inta-csic.es/theory/fps}). The SDSS filter
magnitudes were always computed in the AB system \citep{Oke83}, with a
constant zeropoint of 3631 Jy for all filters. Finally, the CSP filter
magnitudes were computed in their ``natural'' system according to
\citet{Kri17}, using their Eq.(45) and Table 15.

\Tab{offset} lists offsets between various choices of filters and
zeropoint systems used in our study, measured at the bolometric peak of our
typical model He4.10 and 10 days after peak. There is a constant
  offset when only the zeropoint for a given set of filters is
  changed. Those are listed on top where only the zeropoints from
  natural to AB are being varied for CSP filters and Vega to AB in
  Johnson-Cousins filters.  The offsets in Johnson-Cousins $UBVRI$
  expressed in Vega vs. AB zeropoint systems are equivalent to the
  differences in their corresponding zeropoint fluxes
  \citep[e.g.,][]{Bla07}. When both filters and zeropoints are changed
  the offsets are time dependent. These changes are illustrated in the
  remainder of the table for a typical model, He4.10, at bolometric
  peak and 10 days after.

The difference between CSP filter magnitudes in natural vs. AB system
are negligible for our purposes, except for $B$ and $u$. These
values are also closely consistent with the published offsets from
\citet[][see their Table 16]{Kri17}. CSP natural magnitudes in $B$ are
essentially identical to Johnson $B$ magnitude expressed in Vega
system, while the same comparison for $V$ yields a difference of 0.07-0.08
mag. CSP natural magnitudes in $ugri$ are also quite close to SDSS
filter magnitudes in AB system, but there is about 0.09 mag offset in
$g$ at late times. Overall, the offsets do not exceed 0.1 mag within the 
time range in which SEDONA provides reliable results.

\begin{deluxetable*}{rclccccccccc}
\tablecaption{Various Magnitude offsets at the peak of He4.10}
\tablehead{
            \colhead{}    &
            \colhead{}    &
            \colhead{}    &
            \colhead{$U$} &
            \colhead{$B$} &
            \colhead{$V$} &
            \colhead{$R$} &
            \colhead{$I$} &
            \colhead{$u$} &
            \colhead{$g$} &
            \colhead{$r$} &
            \colhead{$i$}
            }\\
\startdata
CSP$_{\rm nat}$ &--& CSP$_{\rm AB}$     &       &  0.12 &  0.02 &       &       &  0.06 &  0.02 &  0.01 & -0.00 \\
\vspace{3mm}
JC$_{\rm Vega}$ &--& JC$_{\rm AB}$      & -0.91 &  0.11 & -0.02 & -0.20 & -0.41 &       &       &       &       \\

&& & \multicolumn{9}{c}{Peak}\\
\cline{4-12}\\
CSP$_{\rm nat}$ &--& JC$_{\rm Vega}$    &       & -0.00 &  0.07 &       &       &       &       &       &       \\
CSP$_{\rm nat}$ &--& SDSS$_{\rm AB}$    &       &       &       &       &       &  0.02 & -0.05 &  0.02 &  0.02 \\
&& & \multicolumn{9}{c}{Peak + 10 days}\\
\cline{4-12}\\
CSP$_{\rm nat}$ &--& JC$_{\rm Vega}$    &       & -0.02 &  0.08 &       &       &       &       &       &       \\
CSP$_{\rm nat}$ &--& SDSS$_{\rm AB}$    &       &       &       &       &       &  0.00 & -0.09 &  0.01 &  0.04 \\
\enddata
\tablecomments{The peak is bolometric.}
\lTab{offset}
\end{deluxetable*}

\begin{figure}
\begin{center}
\includegraphics[width=0.48\textwidth]{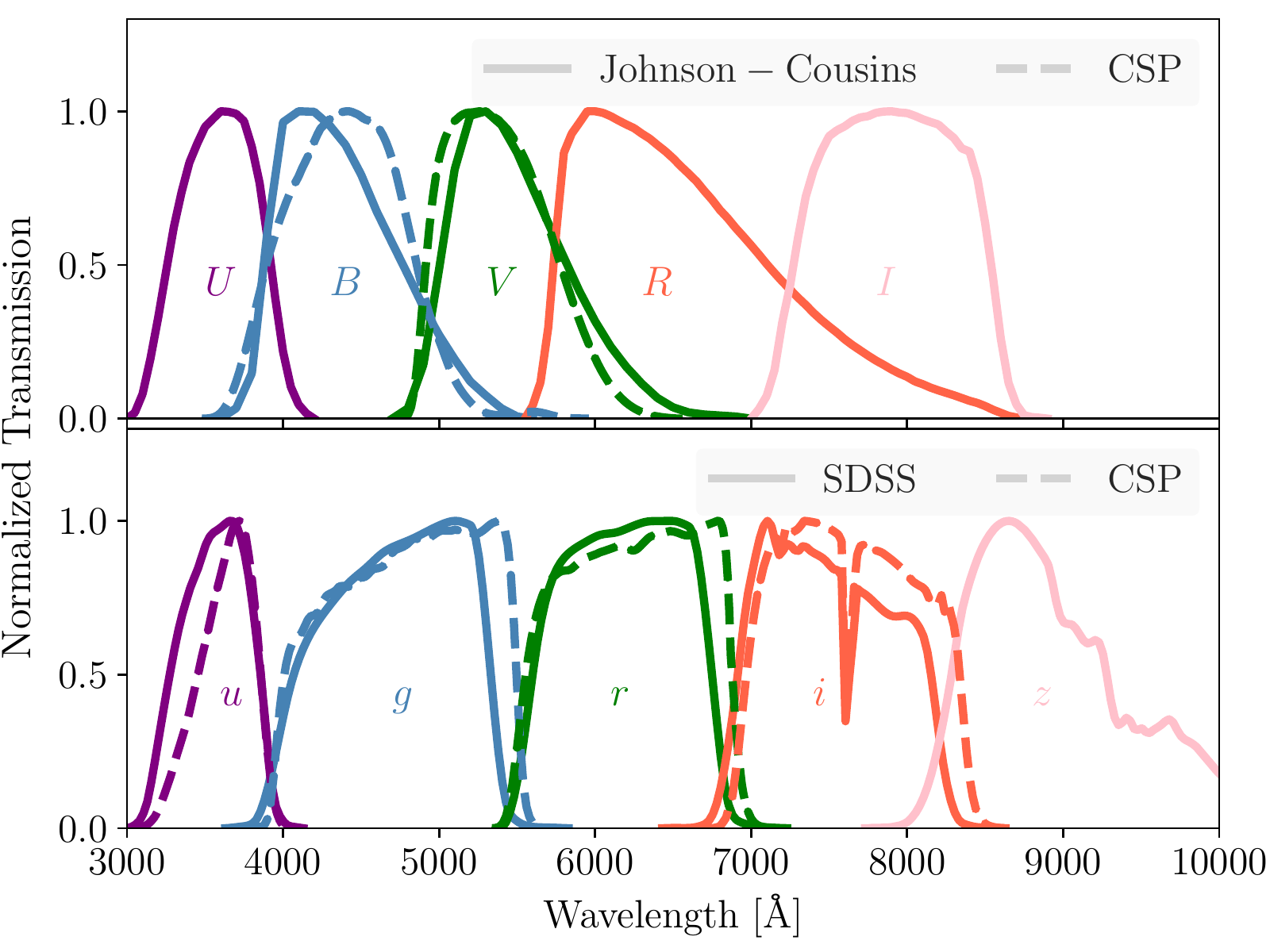}
\caption{Normalized transmission functions for the three filter sets
  used in this study. (top:) Johnson-Cousins $UBVRI$ (solid) compared
  with CSP $BV$ (dashed). (bottom:) SDSS $ugriz$ (solid) compared with
  CSP $ugri$ (dashed).  \lFig{filters}}
\end{center}
\end{figure}
\clearpage


\end{document}